%% file: main.tex
\PassOptionsToPackage{mathlines}{lineno}
\documentclass[twocolumn]{aastex631}

\usepackage{array}
\usepackage{graphicx}
\usepackage{comment}
\usepackage{natbib}
\usepackage{amsmath}
\usepackage{graphics}
\usepackage{tabularx}
\usepackage{multirow}
\usepackage{longtable}

\usepackage{ulem}
\usepackage{CJKutf8}            % Chinese name
\usepackage{listings}
\lstdefinestyle{arxivcode}{
    basicstyle=\ttfamily\footnotesize,
    breaklines=true,
    breakatwhitespace=true,
    frame=lines,
    framesep=2mm,
    xleftmargin=0pt,
    xrightmargin=0pt,
    keepspaces=true,
    showstringspaces=false,
    columns=fullflexible
}

\newcommand{\noop}[1]{}

\def\bv{\hbox{$B\!-\!V$}}

\newcommand{\elixer}{\textsc{ELiXer}}
\newcommand{\OII}{[\ion{O}{2}]} 
\newcommand{\OIII}{[\ion{O}{3}]}
\newcommand{\oiis}{[O\,\textsc{ii}] emitters}
\newcommand{\hbeta}{H$\beta$}
\newcommand{\hgamma}{H$\gamma$}
\newcommand{\hdelta}{H$\delta$}
\newcommand{\HeII}{\ion{He}{2}}

\newcommand{\lya}{Ly$\alpha$}
\newcommand{\contdetcount}{682,067}
\newcommand{\linedetcount}{31,961,716}

\newcommand{\nsource}{1{,}107{,}763} 
\newcommand{\nstar}{150{,}608 }
\newcommand{\nlae}{426{,}654}
\newcommand{\noii}{491{,}411}
\newcommand{\nagn}{18{,}303}
\newcommand{\nlzg}{19{,}457}
\newcommand{\ndettable}{3,296,101}
\newcommand{\skycoverage}{86.67}
\newcommand{\zhet}{\texttt{z\_hetdex}}

\newcommand{\sdss}{SDSS}
\newcommand{\hetg}{$g_\mathrm{HETDEX}$}
\newcommand{\nhetdexobs}{6778}
\newcommand{\nifu}{431,713}
\newcommand{\namps}{5{,}180{,}556}
\newcommand{\nlzgtooiiswitch}{32,031}
\newcommand{\fluxden}{\ensuremath{10^{-17}\,\mathrm{erg\,s^{-1}\,cm^{-2}\,\mathring{A}^{-1}}}}

\begin{document}
\nolinenumbers

\title{HETDEX Public Data Release 1: Source Catalog 2 and \\ Data Cubes from $\sim90$~deg$^2$ of Integral-Field Optical Spectroscopy}

%\footnote{Based on observations obtained with the Hobby-Eberly Telescope, which is a joint project of the University of Texas at Austin, the Pennsylvania State University, Ludwig-Maximilians-Universit\"at M\"unchen, and Georg-August-Universit\"at G\"ottingen.  The HET is named in honor of its principal benefactors, William P.~Hobby and Robert E.~Eberly.}}

\author[0000-0002-2307-0146]{Erin {Mentuch Cooper}}
\affiliation{Department of Astronomy, The University of Texas at Austin, 2515 Speedway Boulevard, Austin, TX 78712, USA}
\email{erin.hetdex@gmail.com}

\author[0000-0002-8433-8185]{Karl Gebhardt}
\affiliation{Department of Astronomy, The University of Texas at Austin, 2515 Speedway Boulevard, Austin, TX 78712, USA}

\author[0000-0002-8925-9769]{Dustin Davis}
\affiliation{Department of Astronomy, The University of Texas at Austin, 2515 Speedway Boulevard, Austin, TX 78712, USA}

\author[0000-0001-5561-2010]{Chenxu Liu\begin{CJK*}{UTF8}{gkai} (刘辰旭) \end{CJK*}}
\affiliation{South-Western Institute for Astronomy Research, Key Laboratory of Survey Science of Yunnan Province, Yunnan University, Kunming, Yunnan 650500, People's Republic of China}

\author[0000-0001-7010-7637]{Barbara G. Castanheira}
\affiliation{Department of Physics and Astronomy, Baylor University, Waco, TX 76798-7316}

\author[0000-0002-0304-5701]{Owen Chase}
\affiliation{Department of Astronomy, The University of Texas at Austin, 2515 Speedway Boulevard, Austin, TX 78712, USA}
\affiliation{Weinberg Institute, The University of Texas at Austin, 2515 Speedway Boulevard, Austin, TX 78712, USA}

\author[0000-0003-2332-5505]{\'{O}scar A. Ch\'{a}vez Ortiz}
\affiliation{Department of Astronomy, The University of Texas at Austin, 2515 Speedway Boulevard, Austin, TX 78712, USA}

\author[0000-0002-1328-0211]{Robin Ciardullo}
\affiliation{Department of Astronomy \& Astrophysics, The Pennsylvania State University, University Park, PA 16802, USA}
\affiliation{Institute for Gravitation and the Cosmos, The Pennsylvania State University, University Park, PA 16802, USA}

\author[0000-0002-0212-4563]{Olivia Curtis}
\affiliation{Department of Astronomy \& Astrophysics, The Pennsylvania State University, University Park, PA 16802, USA}
\affiliation{Institute for Gravitation and the Cosmos, The Pennsylvania State University, University Park, PA 16802, USA}

\author[0000-0002-5223-8315]{Delaney A. Dunne}
\affiliation{California Institute of Technology, 1200 E.~California Blvd., Pasadena, CA 91125, USA}

\author[0000-0001-5175-1777]{Neal J. Evans II}
\affiliation{Department of Astronomy, The University of Texas at Austin,
2515 Speedway, Stop C1400, Austin, Texas 78712-1205, USA}

\author[0000-0003-2575-0652]{Daniel J. Farrow}
\affiliation{E. A. Milne Centre for Astrophysics
University of Hull, Cottingham Road, Hull, HU6 7RX, UK}
\affiliation{Centre of Excellence for Data Science,
Artificial Intelligence \& Modelling (DAIM),
University of Hull, Cottingham Road, Hull, HU6 7RX, UK}

\author[0000-0002-7025-6058]{Maximilian Fabricius}
\affiliation{Max Planck Institute for Extraterrestrial Physics, Giessenbachstr. 1, 85748 Garching, Germany}
\affiliation{Universit\"ats-Sternwarte M\"unchen, Fakult\"at f\"ur Physik, Ludwig-Maximilians-Universit\"at M\"unchen, Scheinerstrasse 1, 81679 M\"unchen, Germany}

\author[0000-0001-8519-1130]{Steven L. Finkelstein}
\affiliation{Department of Astronomy, The University of Texas at Austin, 2515 Speedway Boulevard, Austin, TX 78712, USA}
\affiliation{Cosmic Frontier Center, The University of Texas at Austin, Austin, TX, USA}

\author[0000-0001-6842-2371]{Caryl Gronwall}
\affiliation{Department of Astronomy \& Astrophysics, The Pennsylvania
State University, University Park, PA 16802, USA}
\affiliation{Institute for Gravitation and the Cosmos, The Pennsylvania State University, University Park, PA 16802, USA}

\author[0009-0003-7103-9076]{Nathaniel J. Hamme}
\affiliation{Department of Astronomy \& Astrophysics, The Pennsylvania
State University, University Park, PA 16802, USA}

\author[0000-0001-6717-7685]{Gary J. Hill}
\affiliation{McDonald Observatory, The University of Texas at Austin, 2515 Speedway Boulevard, Austin, TX 78712, USA}
\affiliation{Department of Astronomy, The University of Texas at Austin, 2515 Speedway Boulevard, Austin, TX 78712, USA}

\author[0000-0002-1496-6514]{Lindsay R. House}
\affiliation{Data Science Institute, The University of Chicago, 5460 S University Ave, Chicago, IL 60615, USA}
\affiliation{NSF-Simons AI Institute for the Sky (SkAI), 172 E. Chestnut St., Chicago, IL 60611, USA}

\author[0000-0001-7039-9078]{Matt J. Jarvis}
\affiliation{Astrophysics, Department of Physics, University of Oxford, Keble Road, Oxford, OX1 3RH, UK}
\affiliation{Department of Physics and Astronomy, University of the Western Cape, Robert Sobukwe Road, 7535 Bellville, Cape Town, South Africa}

\author[0000-0002-8434-979X]{Donghui Jeong}
\affiliation{Department of Astronomy \& Astrophysics, The Pennsylvania State University, University Park, PA 16802, USA}
\affiliation{Institute for Gravitation and the Cosmos, The Pennsylvania State University, University Park, PA 16802, USA}
\affiliation{School of Physics, Korea Institute for Advanced Study, Seoul 02455, Korea}

\author[0000-0002-2196-4699]{Andreas Kelz}
\affiliation{Leibniz-Institut for Astrophysik Potsdam (AIP), An der Sternwarte 16, 14482 Potsdam, Germany}

\author[0000-0002-0136-2404]{Eiichiro Komatsu}
\affiliation{Max-Planck-Institut f\"{u}r Astrophysik, Karl-Schwarzschild-Str. 1, 85741 Garching, Germany}
\affiliation{Ludwig-Maximilians-Universität München, Schellingstr. 4, 80799 München, Germany}
\affiliation{Kavli Institute for the Physics and Mathematics of the Universe (WPI), The University of Tokyo Institutes for Advanced Study (UTIAS), The University of Tokyo, Chiba 277-8583, Japan}

\author[0009-0003-1893-9526]{Mahan Mirza Khanlari}
\affiliation{Department of Astronomy, The University of Texas at Austin, 2515 Speedway Boulevard, Austin, TX 78712, USA}

\author[0000-0001-5610-4405]{Hasti Khoraminezhad}
\affiliation{Institute for Multi-messenger Astrophysics and Cosmology, Department of Physics, Missouri University of Science and Technology, 1315 N. Pine St., Rolla MO 65409, USA}

\author[0000-0002-0417-1494]{Wolfram Kollatschny}
\affiliation{Institut f\"{u}r Astrophysik, Universit\"{a}t G\"{o}ttingen, Friedrich-Hund-Platz 1, 37077 G\"{o}ttingen, Germany}

\author[0000-0002-6907-8370]{Maja Lujan Niemeyer}
\affiliation{Max-Planck-Institut f\"{u}r Astrophysik, Karl-Schwarzschild-Str. 1, 85741 Garching, Germany}
\affiliation{Ludwig-Maximilians-Universität München, Schellingstr. 4, 80799 München, Germany}

\author[0000-0002-3559-5310]{Hanshin Lee}
\affiliation{McDonald Observatory, The University of Texas at Austin, 2515 Speedway Boulevard, Austin, TX 78712, USA}

\author[0009-0006-7054-0100]{Phillip MacQueen}
\affiliation{McDonald Observatory, The University of Texas at Austin, 2515 Speedway Boulevard, Austin, TX 78712, USA}

\author[0000-0002-1350-019X]{Deeshani Mitra}
\affiliation{Institute for Multi-messenger Astrophysics and Cosmology, Department of Physics, Missouri University of Science and Technology, 1315 N. Pine St., Rolla MO 65409, USA}

\author[0000-0003-3823-8279]{Shiro Mukae}
\affiliation{Department of Astronomy, The University of Texas at Austin, 2515 Speedway Boulevard, Austin, TX 78712, USA}
\affiliation{MIRAI Technology Institute, Shiseido Co., Ltd., 1-2-11, Takashima, Nishi-ku, Yokohama, Kanagawa, 222-0011, Japan}

\author[0000-0002-1049-6658]{Masami Ouchi}
\affiliation{National Astronomical Observatory of Japan, 2-21-1 Osawa, Mitaka, Tokyo 181-8588, Japan}
\affiliation{Institute for Cosmic Ray Research, The University of Tokyo, 5-1-5 Kashiwanoha, Kashiwa, Chiba 277-8582, Japan}
\affiliation{Department of Astronomical Science, SOKENDAI (The Graduate University for Advanced Studies), Osawa 2-21-1, Mitaka, Tokyo, 181-8588, Japan}
\affiliation{Kavli Institute for the Physics and Mathematics of the Universe (WPI), The University of Tokyo Institutes for Advanced Study (UTIAS), The University of Tokyo, Chiba 277-8583, Japan}

\author{Jennifer Poppe}
\affiliation{Department of Astronomy, The University of Texas at Austin, 2515 Speedway Boulevard, Austin, TX 78712, USA}

\author[0000-0003-2284-8603]{Meredith C. Powell}
\affiliation{Leibniz-Institut for Astrophysik Potsdam (AIP), An der Sternwarte 16, 14482 Potsdam, Germany}

\author[0000-0001-7066-1240]{Mahdi Qezlou}
\affiliation{Department of Astronomy, The University of Texas at Austin, 2515 Speedway Boulevard, Austin, TX 78712, USA}

\author[0000-0002-6186-5476]{Shun Saito}
\affiliation{Institute for Multi-messenger Astrophysics and Cosmology, Department of Physics, Missouri University of Science and Technology, 1315 N. Pine St., Rolla MO 65409, USA}
\affiliation{Kavli Institute for the Physics and Mathematics of the Universe (WPI), The University of Tokyo Institutes for Advanced Study (UTIAS), The University of Tokyo, Chiba 277-8583, Japan}

\author[0000-0001-7240-7449]{Donald P. Schneider}
\affiliation{Department of Astronomy \& Astrophysics, The Pennsylvania State University, University Park, PA 16802, USA}
\affiliation{Institute for Gravitation and the Cosmos, The Pennsylvania State University, University Park, PA 16802, USA}

\author[0000-0002-4974-1243]{Laurel Weiss}
\affiliation{Department of Astronomy, The University of Texas at Austin, 2515 Speedway Boulevard, Austin, TX 78712, USA}

\author[0000-0003-2977-423X]{Lutz Wisotzki}
\affiliation{Leibniz-Institut for Astrophysik Potsdam (AIP), An der Sternwarte 16, 14482 Potsdam, Germany}

\author[0000-0003-2307-0629]{Gregory R. Zeimann}
\affiliation{Hobby-Eberly Telescope, University of Texas at Austin, Austin, TX
78712, USA}

%\collaboration{50}{The HETDEX collaboration}

%\end{comment}

\begin{abstract}

The Hobby-Eberly Telescope Dark Energy Experiment (HETDEX) is a wide-field, integral-field spectroscopic survey designed to map the large-scale distribution of Lyman-$\alpha$–emitting galaxies (LAEs) at $1.88 < z < 3.52$ and constrain dark energy at cosmic noon. Using the 10-m Hobby-Eberly Telescope and the Visible Integral-Field Replicable Unit (IFU) Spectrograph, HETDEX obtains $>35{,}000$ spectra per exposure over 3500–5500 \AA\ at $R\sim800$ with $\sim1\farcs8$ image quality, enabling an untargeted census of emission-line galaxies across 540 deg$^2$. We present HETDEX Public Data Release 1 (PDR1), comprising \nifu\ IFU observations covering \skycoverage\ deg$^2$ of noncontiguous sky in the Spring ($13^{\rm h}, +51^\circ$) and Fall ($1.5^{\rm h}, 0^\circ$) fields, along with legacy regions (COSMOS, GOODS-N, NEP, SA22). PDR1 includes the HETDEX Public Source Catalog 2 (HPSC2), an expanded and reprocessed version of \citet{Cooper2023} incorporating four additional years of data, improved quality control, and new machine learning classifiers. HPSC2 contains \nlae\ LAEs, \noii\ [O II] emitters, \nlzg\ low-$z$ galaxies, \nagn\ active galactic nuclei, and \nstar\ stars, providing coordinates, redshifts or stellar velocities, and 1D spectra for each source. Because the data cubes use local sky subtraction optimized for faint emission-line detection, they are not suited for absolute surface-brightness measurements or very extended nearby galaxies. Appendix materials include the full detection catalog, the 1.6 million–candidate LAE sample, and raw detection databases. All products are publicly accessible through the HETDEX data portal (\url{https://hetdex.org/data-results/}), including access to a public \textsc{JupyterLab}. HPSC2 is also publicly available via Zenodo (\href{https://doi.org/10.5281/zenodo.19581262}{DOI: 10.5281/zenodo.19581262}).

\end{abstract}


\section{Introduction} 
\label{sec:intro}

Once considered a simple celestial sphere, the sky is now understood in full three dimensions thanks to wide-area spectroscopic surveys. Over the last two decades, surveys such as the Sloan Digital Sky Survey \citep[\sdss;][]{sdss2000}, the SDSS-III Baryon Oscillation Spectroscopic Survey \citep[BOSS;][]{boss2013}, the SDSS-IV extended BOSS \citep[eBOSS;][]{eboss2016}, and the Dark Energy Spectroscopic Instrument \citep[DESI;][]{desi2016, desi2022} have revealed the 3D, large-scale structure of the Universe. Galaxy spectra encode the distance measurements that allow us to map the 3D positions of millions of galaxies and quantify the evolving cosmological state of the Universe.

Astronomical spectra provide both precise distance measurements and detailed diagnostics of a source's astrophysical nature. Galaxy properties derived from spectra—including star-formation rate, stellar mass, and metallicity—statistically reveal how galaxies evolve as a population. Spectra from stars reveal their temperature, surface gravity, metallicity and atmospheric structure, just to name some of the vast range of content encapsulated in a source's spectrum.

Most spectroscopic surveys such as SDSS, eBOSS and DESI, select their targets based on multiwavelength photometric imaging. Targets are chosen to maximize detection and ensure enough signal at a range of wavelengths such that multiple spectroscopic features can secure a spectroscopic redshift with high confidence. This approach generally requires long exposures, often many hours. These surveys position hundreds of fibers -- or thousands in DESI's case -- and obtain spectra of previously detected objects over a large field of view (FOV) in order to maximize the area surveyed. They are optimized for brighter galaxies and only obtain a single spectrum per source target.

The Hobby-Eberly Telescope Dark Energy Experiment \citep[HETDEX;][]{Gebhardt2021} takes a different approach by using an array of up to 78 integral field units (IFUs) to perform an untargeted search of the sky. Designed to detect strong line emission from $1.9 < z < 3.5$ Lyman-$\alpha$ Emitting galaxies (LAEs), HETDEX trades continuum sensitivity for sky coverage. The strong \lya\ line emission allows for detection over a wide range of stellar masses (e.g., \citealt{Shapley2003, HuCowie2006}) and redshifts for objects generally too faint for detection in broadband images \citep{Hagen2016, Oyarzun2017, Santos2020}. See \citet{Ouchi2020} and references therein for a thorough review. 

\begin{figure*}
    \includegraphics[width=0.9\textwidth]{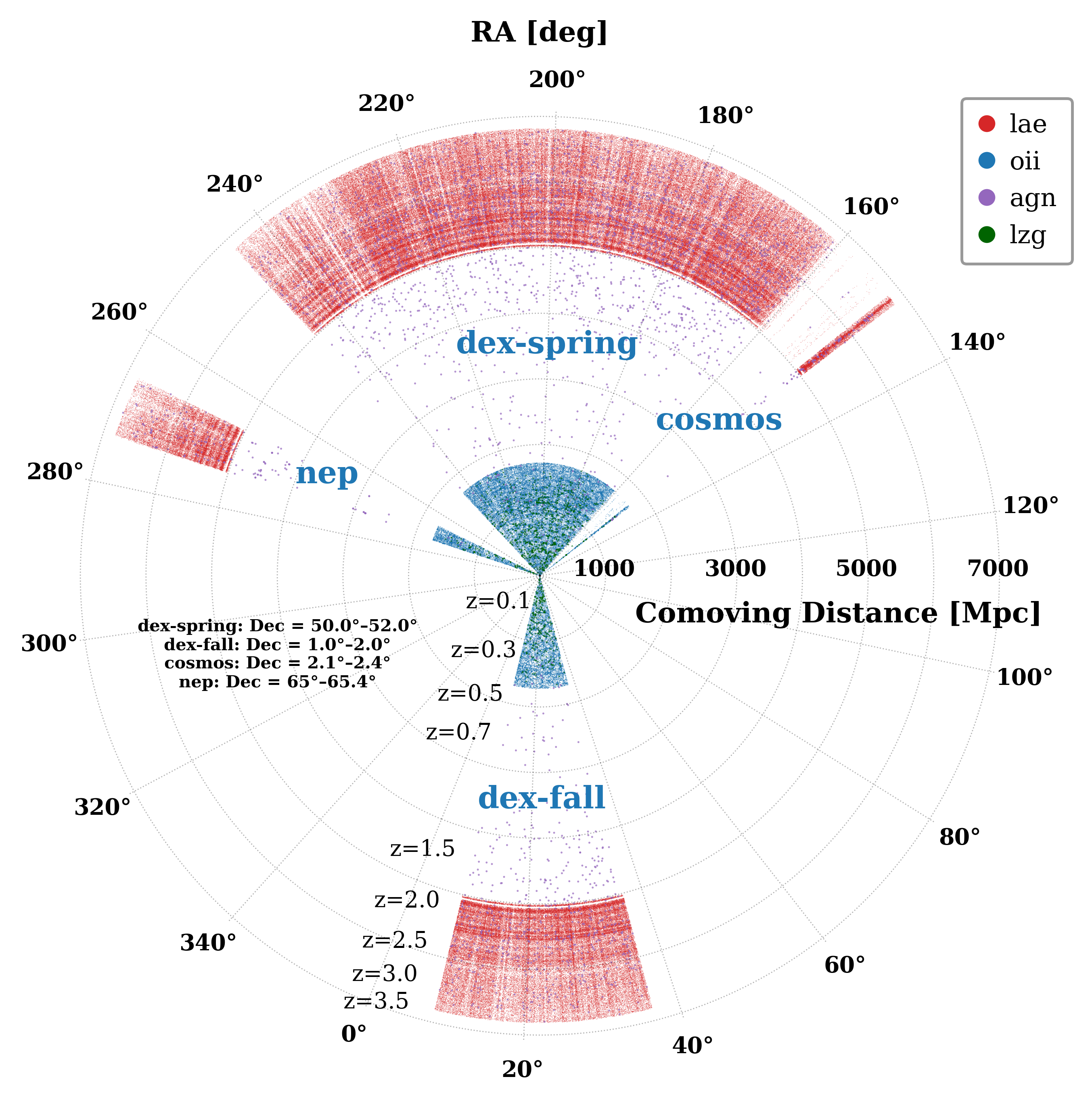}
    \caption{Projected distribution of \lya\ emitters (`lae', red), active galactic nuclei (`agn', purple), \oiis\ (`oii', blue) and low-redshift, non-emission-line galaxies (`lzg', green) in comoving space, collapsed in decl. Radial labels indicate comoving distance in Mpc, with corresponding redshift ticks annotated along the wedge. Clustering is evident at low redshift among \oiis\ and passive low-z galaxies, while the distribution of the higher-redshift LAEs appears relatively smooth. The variation in LAE number density reflects the effect of coverage variation, wavelength-dependent flux sensitivity, and an intrinsic decline toward higher redshift. Only AGNs are found in the region between $0.5<z<1.9$, identified from their bright continuum emission and broad spectral lines in \citet{liu2025}.}
    \label{fig:polarwedge}
\end{figure*}

HETDEX is performed with the Visible IFU Replicable Unit Spectrograph  \citep[VIRUS;][]{Hill2021}, on the 10\,m Hobby-Eberly Telescope (HET, \citealt{Ramsey1998,Hill2021}). VIRUS obtains~$\approx$~35,000 spectra simultaneously, each covering the wavelength range $3470 \mathrm{\AA} \le \lambda \le 5540 \mathrm{\AA}$ with a corresponding spectral resolving power $750\lesssim R\lesssim 950$. HETDEX uses LAEs as a (biased) tracer of dark matter density; by measuring their clustering, HETDEX characterizes the Universe's dark energy density and tests for potential evolution \citep{Shoji2009}. HETDEX aims to measure the Hubble parameter, $H(z)$, and the angular diameter distance, $D_{A}(z)$, to better than 1\% accuracy in the redshift range $1.9 < z < 3.5$. To achieve this precision, approximately one million LAE positions must be mapped over 540 deg$^2$ of sky, or $10.9$~Gpc$^3$ in the targeted redshift range. 

HETDEX is unlike traditional LAE narrowband surveys that identify LAE candidates via a narrowband minus broadband color excess (e.g., \citealt{cowie1998,rhoads2000, Gronwall2007a, Ouchi2008, Konno2016, Sobral2018, Spinoso+2020, ono2021, benitez+2014}). These narrowband LAE surveys can effectively cover large areas at a slice of redshift space, but lack line-of-sight volume information; although multiple narrowband filters can increase the survey volume and/or redshift resolution \citep{eriksen2019,Bonoli+2021, odin2024}. With an IFU, not only is the redshift coverage much wider and continuous, but the effective bandpass for emission-line detection is generally 10 to 50 times narrower, enabling deeper LAE detections in shorter exposures and with higher-redshift precision. Searches for LAEs have proved successful at fainter sensitivities \citep{vanB2005, Bacon2015, Adams2011a, Urrutia2019} but not over large volumes. HETDEX trades line flux sensitivity for volume.

With only two to three LAEs detected in a typical VIRUS IFU observation, they make up a minuscule fraction of the total spectral data obtained by HETDEX. For example, a typical LAE's signal is spread across 3-15 fibers, depending on observational quality, and the spectral signature is typically only 4-8\,\AA\, or 3-5 pixels, in the spectral dimension. An IFU consists of 1344 fiber spectra with spectral coverage from 3500 to 5500\,\AA; thus LAEs are less than 0.01\% of the signal in an HETDEX IFU data cube, with HETDEX IFUs observing far more empty sky than LAEs. The remainder of the IFU fiber coverage is filled with hundreds of thousands of other astronomical sources, along with statistical signatures of large-scale unresolved \lya-emission \citep{maja2025} and absorption \citep{laurel2025, mahan2025}.

%figure generated in work/stampede2/pdr1/mosaic-full-survey.ipynb
\begin{figure*}[th]
    \centering
    \includegraphics[width=\linewidth]{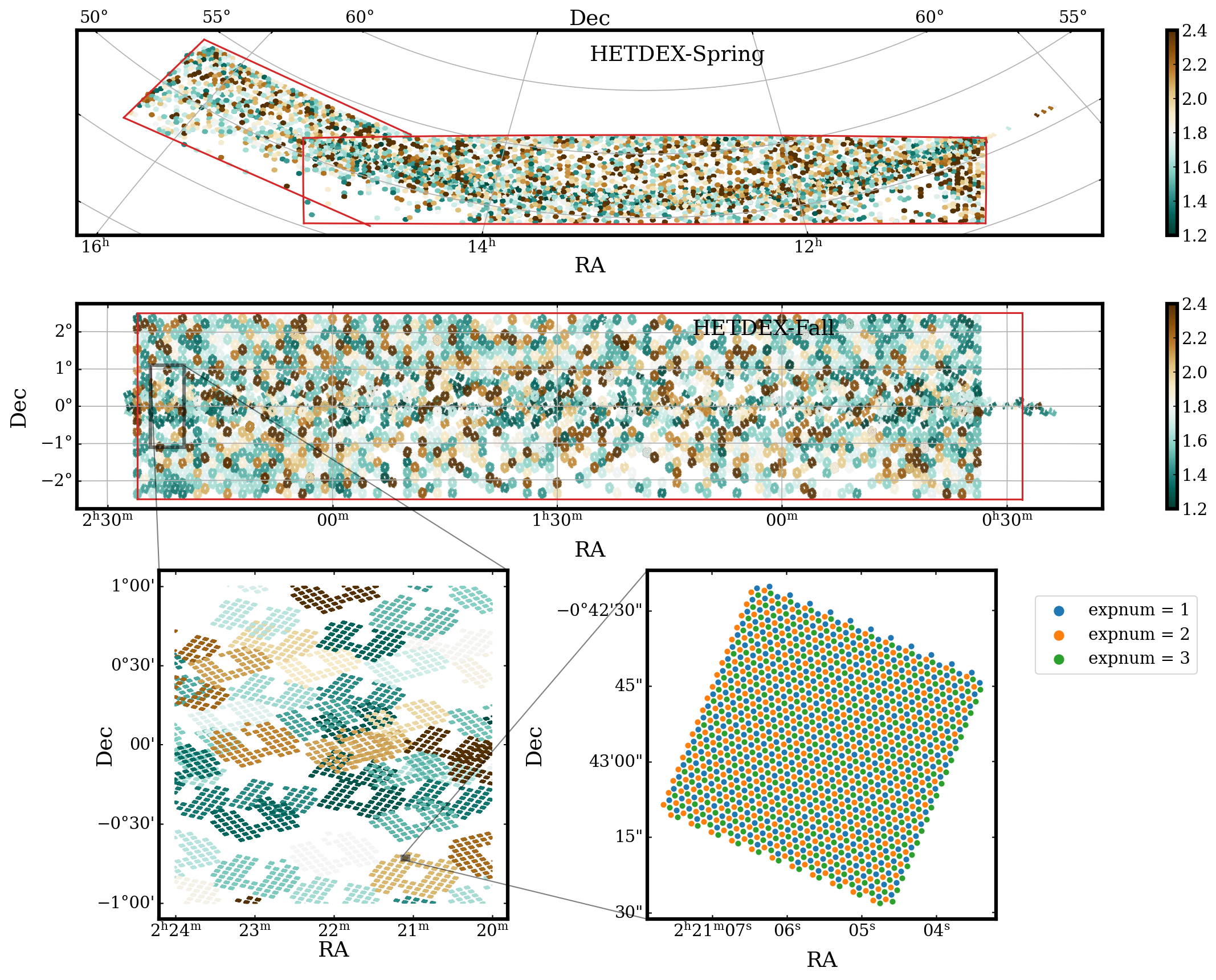}%made in pdr1/mosaic.ipynb
    \caption{Outline of the two main HETDEX science fields (in red) and the footprint of IFU data cubes in this data release. Colors indicate the image quality of individual VIRUS IFU pointings, quantified as the FWHM of the point-spread-function (PSF) in arcseconds. Fields are (1)~the high declination Spring Field (top), which is centered at (13.5$^{\rm h}$, +51$^{\circ}$) and covers $\sim390$~deg$^2$ of the sky, and (2)~the equatorial Fall Field (middle), which is centered at (1.5$^{\rm h}$, 0$^{\circ}$), and covers $\sim150$~deg$^2$ of the sky. Each IFU has a FOV of $51\arcsec \times 51\arcsec$, which means that the full VIRUS IFU array has a 0.22 fill factor in the HET's~22$'$-diameter FOV. The expanded inset on the bottom-left presents a zoomed-in area on a typical 2~deg$^2$ region in the HETDEX-Fall field. The expanded inset on the bottom right shows a zoom in on a singular IFU indicating the array of fiber spectra, each 1\farcs5 in diameter. Each colour represents one of the three dithered exposures. A single HETDEX observation can contain up to 78 IFU observations, consisting of 936 CCD exposures and 104,832 fiber spectra. An individual IFU three-dither observation of 1344 ($448\times3$) fibers is used to create a single HETDEX data cube.}
    \label{fig:coverage}
\end{figure*}

% figure generated in work/stampede2/pdr1/coverage_mosaic-legacy.ipynb
\begin{figure*}[t]
    \centering
    \includegraphics[width=\textwidth]{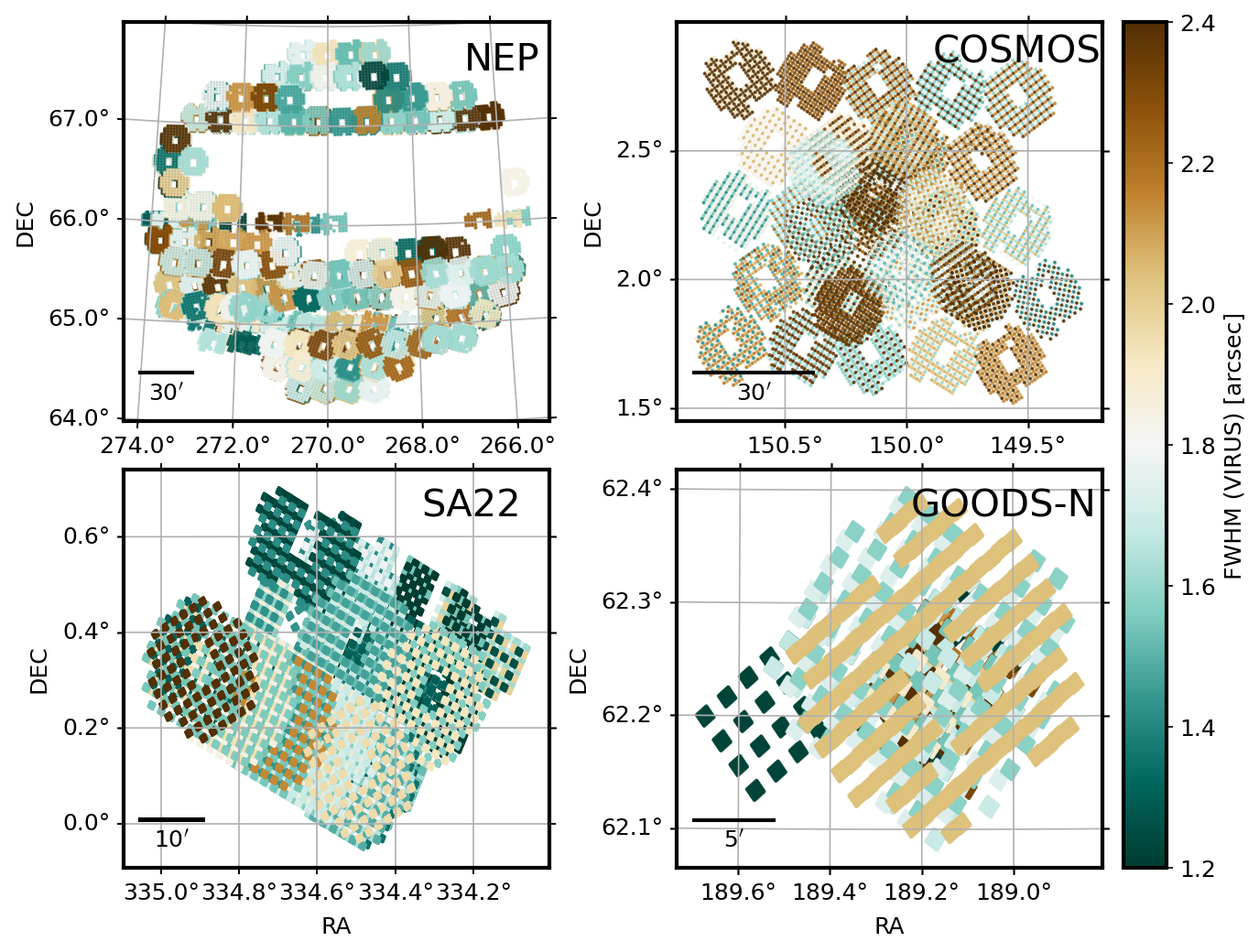}
    \caption{Coverage of the four legacy fields included in this release: NEP, COSMOS, SA22, and GOODS-N. 
    Colors indicate the image quality of individual VIRUS IFU pointings (FWHM in arcseconds). The bar in each panel indicates the image scale. As in Figure~\ref{fig:coverage}, the IFU layout creates gaps in each observation that are most notable in the higher-resolution panels on the bottom row.}
    \label{fig:legacycoverage}
\end{figure*}

In the context of the number of spectra collected, HETDEX is currently the largest spectroscopic sky survey ever conducted. The main survey was completed in July 2024; its database consists of over 600\,million observed fiber spectra of the night sky obtained from 2500 dark sky observing hours. After being interpolated and cleaned for various quality issues, this dataset comprises \nhetdexobs\ observations and \nifu\ HETDEX IFU data cubes. Figure \ref{fig:polarwedge} shows the distribution of HETDEX sources as a function of R.A. and cosmological comoving distance, collapsed in decl. The diagram illustrates the survey’s ability to map large-scale structure across several Gpc, with high-redshift \lya\ emitters (red) dominating the volume at $z\gtrsim2$, while \OII\ emitters (blue), active galactic nucleai (AGN, purple), and luminous low-redshift galaxies (green) trace the foreground. 

We present the first full public data release (PDR1) from the HETDEX main survey, providing access to over 430,000 science-quality IFU data cubes across the survey footprint. The paper describes the PDR1 data model, IFU data products, and methods for accessing and manipulating data cubes, with example tutorial notebooks. This paper also serves as a reference for the second HETDEX public source catalog (HPSC2) for HETDEX, which is an expansion of the early HETDEX Public Source Catalog 1 (HPSC1) described in \citet{Cooper2023}. Future data releases may consist of updated mask modeling, revised flux calibration, and/or expanded data from post main survey operations.

The outline of this paper is as follows. Section~\ref{sec:observations} describes the HETDEX observations, the generation of the IFU data cubes and details about the data masking. Section\,\ref{sec:catalog} describes the catalog accompanying this release, HETDEX Public Source Catalog 2.
Section~\ref{sec:datamodel} outlines the data model for the public data release and describes the format of the index and data cube files. In Section~\ref{sec:access}, a description of access options is provided including a list of tutorial Jupyter notebooks, although access is not limited to the Python computing language as the data cubes follow standard FITS file format \citep{fits1981, fits2010}. Section\,\ref{sec:caveats} describes the limitations of this data release.

HETDEX public data products are found at the data mount hosted by the Texas Advanced Computing Center (TACC) at The University of Texas at Austin at \url{https://web.corral.tacc.utexas.edu/hetdex/HETDEX/pdr/}. Additional information and data hosting mirrors are provided at \url{https://hetdex.org/data-results/}. Example access using Python can be found at \url{https://github.com/HETDEX/dexcube}. Notebooks within this repository demonstrate remote access for users who wish to download data cubes of interest and provide tutorials showing how to access source catalogs, extract spectra, and generate narrowband images directly from the data cubes.

All positions reported in this paper are in the International Celestial Reference System (ICRS; epoch J2000)\null.  We adopt the flat $\Lambda$-cold-dark-matter cosmology with $H_0=67.7\,\mathrm{km}\,\mathrm{s}^{-1}\,\mathrm{Mpc}^{-1}$ and $\Omega_{\mathrm{m},0}=0.31$ measured by \citet{Planck2018}. All magnitudes are expressed in the AB system \citep{oke1983}. We assume a rest-frame vacuum wavelength of $\lambda=1215.67$~\AA\ for Ly$\alpha$ and rest-frame air wavelength of $\lambda=3727.8$~\AA\ for the \OII\ doublet, integrated to our instrumental resolution. Observed wavelengths expressed in this paper and associated data products are as measured in air. %All redshifts are appropriately calculated for any differences between air and vacuum wavelengths using the standard in \cite{Morton1991}.
Redshifts are calculated from rest-frame wavelengths with conversions from air to vacuum \citep{Greisen2006} and include a radial-velocity correction to the Solar System barycenter \citep{astropy:2013}.
All spectral data are provided in air in the wavelength range from 3470\,\AA\ to 5540\,\AA\ in steps of 2\,\AA.

\section{Observations}
\label{sec:observations}

\subsection{Survey Overview}
% table stats generated in work/stampede2/pdr1/coverage_mosaic-legacy.ipynb
\begin{table*}[th]
    \centering
    \small
    \setlength{\tabcolsep}{4pt}
    \caption{Catalog Release Survey Statistics}
    \begin{tabular}{l c r r r r r r r r}

        \hline
    field ID & Center (deg, deg) & N(IFU) & Area (deg$^2$) & N(source) & N(LAE) & N(OII) & N(AGN) & N(LzG) & N(star) \\
    \hline\hline
    dex-spring & $(195.00^\circ,\ +51.00^\circ)$ & 244{,}176 & 49.00 & 640{,}603 & 253{,}458 & 287{,}017 & 11{,}093 & 10{,}807 & 77{,}522 \\
    dex-fall   & $(22.50^\circ,\ +0.00^\circ)$   & 136{,}829 & 27.46 & 305{,}915 & 114{,}419 & 140{,}133 & 5{,}531  & 6{,}586  & 38{,}886 \\
    nep        & $(270.00^\circ,\ +66.00^\circ)$ & 34{,}269  & 6.88  & 100{,}872 & 25{,}649  & 45{,}035  & 1{,}232  & 1{,}381  & 27{,}359 \\
    cosmos     & $(150.12^\circ,\ +2.21^\circ)$  & 11{,}271  & 2.26  & 24{,}662  & 7{,}794   & 11{,}521  & 431     & 498     & 4{,}390 \\
    ssa22      & $(336.50^\circ,\ +0.00^\circ)$  & 4{,}393   & 0.88  & 10{,}905  & 3{,}975   & 4{,}036   & 146     & 93      & 1{,}930 \\
    goods-n    & $(189.18^\circ,\ +62.24^\circ)$ & 775       & 0.16  & 2{,}172   & 1{,}025   & 877      & 42      & 27      & 195 \\
    \hline
    \textbf{total} & $(\text{—},\ \text{—})$ & 431{,}713 & 86.64 & 1{,}085{,}129 & 406{,}320 & 488{,}619 & 18{,}475 & 19{,}392 & 150{,}282 \\
    \hline
    \end{tabular}

    \raggedright
    \vspace{0.25ex}
    Note: Listed is the number count of IFU observations observed and the count included after observation quality inclusion criteria. Reported area coverage includes repeat observations and therefore does not represent unique sky coverage. Field centers are given in equatorial coordinates (ICRS; epoch J2000).
    \label{tab:summary}
\end{table*}

The data cubes and catalogs released in PDR1 consist of the full main HETDEX survey \citep{Gebhardt2021}, drawn from the fifth internal data release (HDR5) from the HETDEX Collaboration. It contains \nhetdexobs\ observations obtained starting in January 2017, when the VIRUS IFU assembly contained just 16 operational IFUs, and concluding on 31~July~2024 with the full 78 IFUs installed within the VIRUS array in July~2021.

The HETDEX footprint consists of two primary fields that allow for full-year surveying as shown in Figure~\ref{fig:coverage}. The HETDEX-Spring Field, labeled as the \texttt{dex-spring} field throughout this paper and in the associated catalog, covers 390\,deg$^2$ of high-decl. ($\delta \sim 51^{\circ}$) sky while the HETDEX-Fall Field, labeled as \texttt{dex-fall} in the catalog, covers 150\,deg$^2$ along the celestial equator (see \citealt{Gebhardt2021} for full details on field selection). The project does not need complete coverage within this sky area to accomplish its scientific goals, as discussed by \citet{Chiang2013}; a fill factor of 0.22 (1/4.6), which optimizes the number of IFUs given the area of the focal plane of the HET, is sufficient. 

In addition to the two main survey fields, this HETDEX data release also includes coverage from multiple legacy fields, as presented in Figure~\ref{fig:legacycoverage}. The most extensive coverage comes from collaborative observations with the Texas Euclid Survey for \lya\ \citep{tesla2023} of the North Ecliptic Pole (NEP\null). Nearly full field coverage of the central 1~deg$^2$ of the Cosmic Evolution Survey survey (COSMOS; \citealt{cosmos2007}) is included, as well as sparse coverage of the SA22 \citep{sa221998} and GOODS-N \citep{GOODSN} fields. Field coordinates, coverage area and the number of IFU observations contained in each field are summarized in Table~\ref{tab:summary}.

HETDEX uses VIRUS \citep{Hill2021}, the fiber-fed, multi-spectrograph instrument of the upgraded Hobby-Eberly Telescope \citep{Ramsey1998, Hill2021}. Each IFU connects to a pair of VIRUS spectrographs with 448~1\farcs5 diameter fibers positioned on a rectangular array with fiber center separations of~2\farcs5. In a single HETDEX observation, three exposures, each typically lasting 6-7 minutes (the exposure times range from 3.6 to 12 minutes, depending on observing conditions), are captured; the telescope is dithered in a triangular pattern to ensure complete fill factor for each of the \hbox{$51'' \times 51''$} IFU fields (see \citealt{Hill2021, Gebhardt2021}). The bottom-right panel in Figure\,\ref{fig:coverage} provides an example IFU fiber layout for this three-dither pattern. In a single IFU observation, 1344 fiber spectra are collected, providing full sky coverage of the IFU. At full completion of the VIRUS instrument, its 78 IFUs cover approximately~21.7\% (a factor of 4.6) of the HET's 18$'$-diameter FOV. The exact layout and numbering scheme for the array of IFUs is shown in Figure 3 in \citet{Gebhardt2021}.

HETDEX data processing is described in detail in \citet{Gebhardt2021}. Briefly, reduction begins with CCD-level calibrations, including bias subtraction, pixel flats, and twilight sky flats. These are followed by fiber extraction, wavelength calibration, and sky subtraction to produce a calibrated spectrum for each fiber in the array.

Sky subtraction is performed using a local sky model constructed at the amplifier level from fibers within each dithered exposure. This local sky subtraction is optimized for the detection of low signal-to-noise ($S/N$) emission lines, which are the primary targets of HETDEX.

In addition to this approach, the internal HETDEX data model also supports a full-frame sky estimate constructed using all IFUs across the VIRUS focal plane within a given observation. While this full-frame model can better capture large-scale sky structure, it introduces low-level systematics in the calibrated spectra that can degrade the flux calibration on small spatial scales. For this reason, the public data products are calibrated using the local sky model. As a consequence, flux measurements near bright or extended sources, or on spatial scales comparable to the IFU FOV, may be affected by oversubtraction of extended emission.

Astrometric calibrations are achieved by measuring the centroid of each field star using the signal between 4400\,\AA\ and 5200\,\AA\ and comparing their IFU positions to the stars’ equatorial coordinates in the Sloan Digital Sky Survey \citep[SDSS;][]{sdss2000} and \textit{Gaia} \citep{Gaia2018} catalogs.  This process typically results in global solutions that are accurate to $\sim 0\farcs 2$. 

Absolute flux calibrations are achieved using $g < 24$ \sdss\ field stars as \textit{in situ} flux standards. Their $ugriz$ colors \citep{Padmanabhan2008}, \textit{Gaia} parallaxes \citep{Gaia2018}, and foreground reddening estimates \citep{Schlafly2011} are used to determine their most likely spectral energy distribution in a grid of model spectra \citep{Cenarro2007, Falcon-Barroso2011}.  The final system throughput curve is derived from the most likely flux distribution of $\sim 20$ stars, and is generally accurate to $\sim 5\%$ \citep{Gebhardt2021}. Uncertainties in the calibrated fiber spectra are derived by propagating the photon noise through the reduction steps and tracking the uncertainties that come with each step.

Each dither in a HETDEX observation is flux calibrated independently, as there may be small differences in their relative throughput due to variations in the  observing conditions. For inclusion in the HETDEX survey, we require that a nominal throughput (assuming a 360\,s exposure time) exceeds 08, and that the relative throughput of each dithered exposure cannot differ by more than a factor of three. The most common reason for rejection by this criterion is a significant drop in transparency during the third dithered exposure when clouds drifted into the FOV.

The data are organized into a database of flux-calibrated, 1D fiber spectra each with their own corresponding sky coordinate. This forms the fundamental data product of the internal HETDEX data model. The internal data model stores not only the calibrated fiber data but also a number of intermediate data products including the raw and calibrated CCD images from each dithered observation. Internally, these intermediate data are used for quality control and artifact identification. Cutouts of these products are used in machine learning-based classification and vetting of detections \citep{house2023, house2024,Mukae2025}. 

\begin{table*}[th]
    \centering
    \caption{FITS HDU structure for HETDEX IFU data cubes}
    \begin{tabular}{cllc}
        \hline
        \hline
        HDU Index & EXTNAME & Description & Data Shape \\
        \hline
        0 & PRIMARY & Primary HDU with no data & None \\
        1 & DATA    & Flux data cube in units of $10^{-17}$ erg s$^{-1}$ cm$^{-2}$  per 2\AA\ spectral bins & (1036, 104, 104) \\
          &         & Flux is sampled from 3470\,\AA\ to 5540\,\AA\ in steps of 2\,\AA\null & \\
        2 & ERROR   & 1$\sigma$ uncertainty estimates for each spaxel & (1036, 104, 104) \\
        3 & MASK    & Bitmask cube indicating quality flags per spaxel & (1036, 104, 104) \\
        \hline
    \end{tabular}
    \label{tab:data cube_structure}
\end{table*}

Internally, HETDEX data are stored in HDF5 files \citep{hdf5}, with the primary data model constructed around a database of calibrated fiber spectra. Direct use of the fiber data is nontrivial: because the HET and is not equipped with an atmospheric dispersion corrector, each fiber samples a slightly different sky position as a function of wavelength. This atmospheric differential refraction (ADR) is corrected in the internal pipeline with dedicated software (\textsc{hetdex-api}\footnote{\url{https://github.com/HETDEX/hetdex_api}}), which is also required for subsequent interpolation and spectral extraction. 

To simplify data access and avoid imposing these software dependencies, the public release is distributed as interpolated IFU data cubes rather than the native fiber database. This format also allows users to download only data cubes of interest, each of modest size ($<20$\,MB), and analyze the data using standard astronomical software tools.

In contrast to the internal release, the public release excludes the raw and intermediate data products. This decision is motivated primarily by the large data volume and complexity of the internal data model: the full internal dataset, including raw CCD frames, calibrated products, and intermediate processing files, exceeds $\sim350$\,TB, while the processed internal data products alone comprise $\sim150$\,TB. By comparison, the public release of data cubes and catalogs requires less than $10$\,TB of storage. At present, we do not plan a full public release of the intermediate data products. However, we are happy to provide access to these data upon request for users interested in reproducing specific aspects of the internal processing or investigating instrumental effects in greater detail. The publicly released data cubes and catalogs are sufficient for the majority of scientific investigations with the HETDEX survey.

\subsection{Data Cube Generation}

The database of calibrated fibers is interpolated using custom software from \textsc{hetdex-api}\footnote{\url{https://github.com/HETDEX/hetdex_api/blob/master/hetdex_tools/create_dex_cube.py}} into a set of IFU data cubes in a format that is similar to other IFU survey data from Calar Alto Legacy Integral Field Area (CALIFA, \citealt{califa}), Mapping Nearby Galaxies at APO (MaNGA; \citealt{bundy2015, drory2015, law2015}), and MUSE \citep{Urrutia2019}. As VIRUS is composed of an array of many IFUs with 1\arcmin\ gaps between them, the natural unit of data cube storage is that of a single IFU. A given HETDEX exposure is thus composed of 16-78 IFU data cubes depending on how many units were active on the instrument at the time of observation. Each IFU is stored in an individual FITS \citep{fits1981} file with the format described in Table~\ref{tab:data cube_structure}. The spectral sampling of each cube is set to 2\AA, the same as the spectral sampling of the internal fiber spectra. Pixel spatial sampling is set to 0\farcs5. With a median image quality of FWHM=1\farcs8 this is sufficient to sample the point spread function (PSF) of HETDEX observations while keeping storage requirements manageable.

The IFU data cubes cover the same spectral range and resolution as the internally calibrated fibers, sampling a linear wavelength grid from 3470\,\AA\ to 5540\,\AA\ in steps of 2\,\AA\null. For each wavelength slice, the apparent sky position of every fiber is adjusted to account for ADR using the astrometric solution from the internal pipeline \citep{Gebhardt2021}. The resulting irregularly spaced fiber positions are then resampled onto a $52\arcsec \times 52\arcsec$ square grid with $0\farcs5$ spatial sampling. Interpolation is performed with the \texttt{griddata} routine from \textsc{SciPy} \citep{scipy}, which linearly interpolates the flux and associated uncertainties of all contributing fibers to the output coordinate grid. The WCS of each cube follows standard FITS conventions, with the CRPIXn values specifying 1-indexed reference pixel coordinates at pixel centers rather than pixel edges. The data cubes are oriented according to the position angle of the telescope and VIRUS focal plane to minimize any unnecessary data storage space. Thus, the projected orientation is not constrained to the conventional “north up, east left” display; users should rely on the WCS headers when visualizing the data.

The resulting data cube spaxels are per-bin flux values sampled in 2\,\AA\ wide spectral bins in units of $10^{-17}$\,erg\,s$^{-1}$\,cm$^{-2}$ per 2\,\AA\ bin. Flux values are stored in Header Data Unit 1 (HDU1) and their associated flux uncertainties are stored in HDU2. Spaxels that fall outside the region with valid fibers or within fully masked regions are assigned \texttt{NaN}. To prevent spurious interpolation across masked fibers, a Boolean coverage map is propagated by resampling the good-pixel indicator with nearest-neighbour interpolation and blanking any output spaxels where this indicator is false. 

In addition, a bitmask array, described in detail in Section~\ref{sec:bitmask}, is interpolated from the internal data model to the data cube wavelength slices following the same methodology as the flux interpolation except that, for the bitmask, the nearest neighbour method is used for interpolation.

\subsection{Masking Model}
\label{sec:bitmask}

The calibrated fiber spectra in the internal HETDEX data model include a single mask for problems automatically identified by the reduction pipeline. This includes detector artifacts, poor-quality fibers yielding unusable spectra, transient cosmic rays, and various calibration failures. Such cases are flagged through a fill value of 0.0 in the calibrated fiber error arrays. While the fiber spectrum itself may still report a flux value, it is considered invalid when flagged in the error array.

In practice, given the complexity and sheer volume of HETDEX data, a wide range of previously undiscovered issues must also be masked from the data. Over the years, these have been identified through an extensive manual effort. In particular, data reports generated by the Emission Line eXplorer (\elixer) tool developed by \citet{Davis2023} have proved invaluable in diagnosing artifact features that are clear in the view of the raw CCD science frames but may not be obvious in the interpolated fiber spectrum. This manual inspection led to the development of a number of automated procedures to remove spurious artifacts from our source catalogs and fiber spectral data.

For each flux-calibrated spectrum, these masks are stored as bitmask values for every spectral element in the array. In the public data model, these bitmasks are interpolated into HDU3 of the data cube FITS files as listed in Table~\ref{tab:data cube_structure}. The interpolation is performed such that the nearest bitmask value is stored in the cube spaxel.

Table~\ref{tab:maskdq} provides a reference for the bits used to flag various issues in the data. The bitmask format offers the compact representation for a large array of flagging options. A user can select which flags they wish to include, or simply use the default setting where any nonzero value in the mask is flagged. In this section we provide additional detail on each bitmask flag, in the order they are encoded into the model.

\begin{table*}[ht]

    \centering
    \caption{Masking Description}
    \begin{tabular}{lrrll}
    \hline
    \hline
Hex Value &Binary    & Integer Value    & Name        & Description \\
    \hline
0x00000000 &              0 & 0                & Good         & No flag   \\          
0x00000001 &              1 & 1                & MAIN         & Value flagged in reduction (calfibe$==0.0$) \\
0x00000002 &             10 & 2                & FTF          & Average fiber-to-fiber in spectrum is $>0.5$ \\
0x00000004 &            100 & 4                & CHI2FIB      & chi2fib $> 150$ \\
0x00000008 &           1000 & 8                & BADPIX       & On a bad pixel region\\
0x00000010 &          10000 & 16               & BADAMP       & On a bad amplifier \\
0x00000020 &         100000 & 32               & LARGEGAL     & Located within a large galaxy mask \\
0x00000040 &        1000000 & 64               & METEOR       & On a known meteor track \\
0x00000080 &       10000000 & 128              & BADSHOT      & In bad shot list \\
0x00000100 &      100000000 & 256              & THROUGHPUT   & Relative response at 4540\,\AA\ $< 0.08$ \\
0x00000200 &     1000000000 & 512              & BADFIB       & On a known bad fiber \\
0x00000400 &    10000000000 & 1024             & SAT          & On a known satellite track \\
0x00000800 &   100000000000 & 2048             & BADCAL       & Masking due to sky and calibration issues \\
0x00001000 &  1000000000000 & 4096             & PIXMASK      & Masking spectrum==0 in native spectrum counts \\
0x00002000 & 10000000000000 & 8192             & BADDET      & $5\times5\times5$~pixel mask where a detection has been flagged \\
\hline
    \end{tabular}
    \label{tab:maskdq}
\end{table*}

\subsubsection{MAIN (Pipeline Reduction Flags)}

The most basic mask in the data model is inherited from the HETDEX data reduction pipeline \citep{Gebhardt2021}. Any fiber with a calibrated fiber error array value of zero (calfibe$==0.0$ in HETDEX's internal data model) is flagged, indicating that the pipeline determined the data to be unusable. This flag may occur for cosmic rays, readout glitches, or fibers where the calibration is poor, causing a specific spectral region of the fiber to be flagged. These defects are directly marked in the error arrays and carried through to the bitmask as the \texttt{MAIN} flag. It is strongly recommended to always use this mask.

\subsubsection{FTF (Fiber-to-Fiber Outliers)}

This flag identifies spectra where the average deviation between neighbouring fibers is anomalously high. Such cases indicate strong flat-fielding residuals or improperly calibrated regions, which can mimic emission lines when interpolated into data cubes. Masking these features ensures a more uniform background. It is recommended to always use this mask.

\subsubsection{CHI2FIB (High \texorpdfstring{$\chi^2$}{chi2} Fibers)}

Spectral array elements in a fiber spectrum where the fiber profile is not fitted well by the fiber profile model are flagged by the CHI2FIB bitmask flag (indicated by $\chi^2_{\rm fib}>5$). This cut identifies fibers whose calibration may result in spurious detections if left unmasked. Pixel elements in the spectral direction are masked, but the rest of the fiber spectrum remains unflagged. It is recommended to always use this mask.

\subsubsection{BADPIX (Bad Pixel Regions)}

This bit marks all fibers intersecting detector regions with persistent defects, dust spots, or charge traps that cannot be corrected by flat-fielding alone. 
As described in \citet{Gebhardt2021} and further in \citet{Cooper2023}, several detectors have significant features, including large dust spots, many charge traps, and a “pox” contamination where the quantum efficiency of individual pixels can be suppressed by 10\%-40\%. Issues such as these were especially prevalent for data taken during the first two years of observations.  While the flat-field calibrations identify many of the worst features automatically, many low-count defects remain in the data and can produce false-positive line detections.  It is recommended to always use this mask.

\subsubsection{BADAMP (Amplifier Variability)}

Every VIRUS IFU is composed of 448 fibers, which are divided into two spectrograph channels.  Each channel has a \hbox{$2064 \times 2064$} CCD detector, which is read out by two amplifiers binned $2 \times$ in the spectral direction. Thus, each IFU consists of four amplifier (``amp'' for short) channels, labeled `RU', `RL', `LL', `LU', each with 112 fiber spectra that are individually stored in separate \hbox{$1032 \times 1032$} raw data arrays and processed independently within the HETDEX reduction pipeline. Our standard three-dithered observation set generates 936 FITS files (3 dithered observations consisting of 312 Amplifier Readouts for the 78 IFUs). Although the IFU spectrographs are designed to be identical, the "R" in VIRUS stands for "replicable" after all; in practice, there are a number of issues that alter the consistency and quality of the fibers on any given amplifier. Some examples include low photon counts, electronic interference, scattered light, and poor calibration due to bright stars and/or galaxies. 

The PDR1 dataset contains \namps\ amplifier images. Over time, a combination of statistics that are calculated either during the reduction process or in post-reduction provide a good assessment on whether to reject the data produced by an amplifier's readout. Please see Table 2 and related discussion in \citet{Cooper2023} for more details on criteria used to flag bad amplifiers. Roughly  92\% of the FITS files are of the necessary science quality to detect the faint signature of emission-line galaxies. Those that are not are flagged by the BADAMP bitmask.

A conservative approach is taken when assigning automated criteria to identify poor quality amplifiers.  This decision is based on our desire to mitigate false positives in the LAE sample. It is likely that a small fraction of flagged amplifiers are of science quality for brighter sources. If a user plans to visually inspect their data then this flag can be ignored.

\subsubsection{LARGEGAL (Large Galaxy and Planetary Nebula Masks )}
\label{sec:galmask}

\input{HETDEX_pn}

We mask bright galaxies whose optical diameters $D_{25}$  (measured at the 25 mag~arcsec$^{-2}$ isophote) exceed 1\arcmin\null. Their locations and shapes are provided by the Third Reference Catalog of Bright Galaxies \citep[RC3;][]{rc3} and the Uppsala General Catalogue of Galaxies \citep[UGC;][]{1973ugcg.book.....N}.  The Spring field contains 644 such galaxies; the fall field, 447. All detections that fall within $1.5\times$ the $D_{25}$ scale of a bright galaxy's elliptical isophote are removed. This factor was determined by examining the HETDEX spectra at different scalings and ensuring that all detections related to the bright galaxy were encompassed in the aperture mask. These galaxy masks are consistently applied to the HETDEX Source Catalogs and the PDR1 data cubes.

\begin{figure*}[t]
    \centering
    \includegraphics[width=0.95\linewidth]{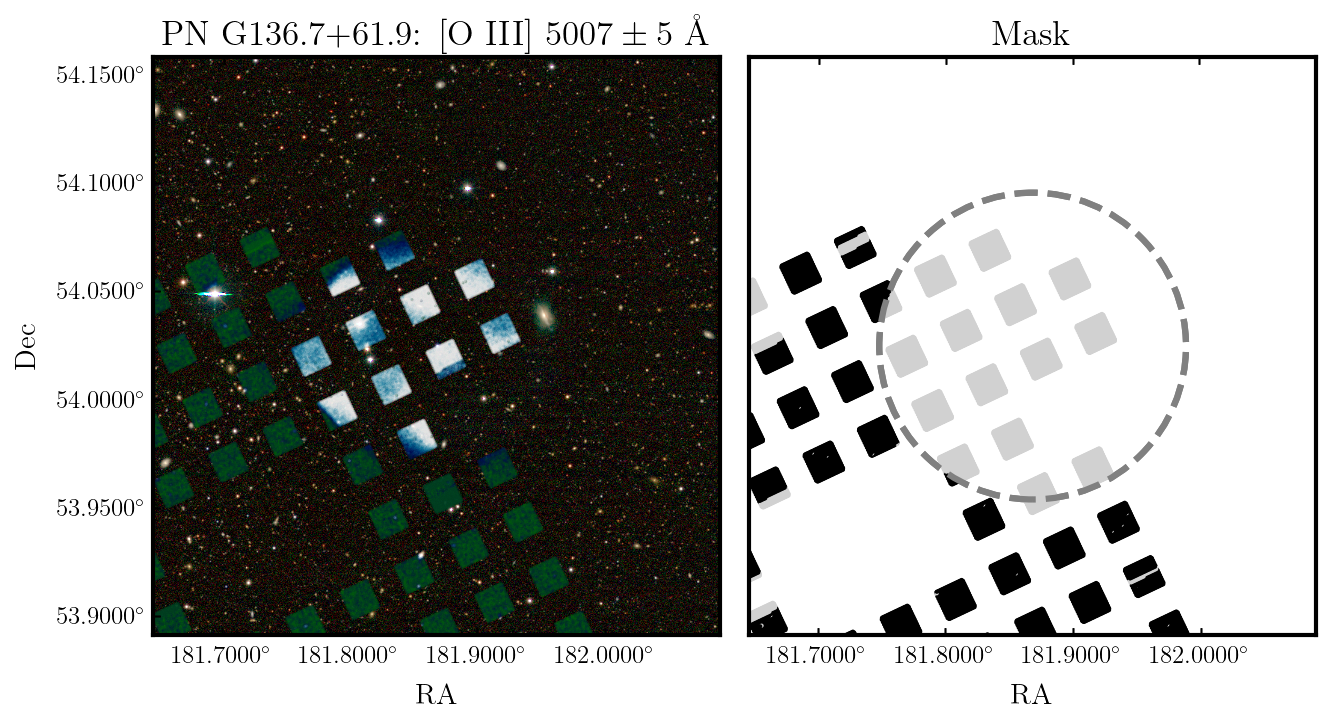}
    \caption{This example demonstrates how large extended sources, such as the planetary nebulae PN G136.7+61.9 displayed here, are identified and masked to prevent false emission-line detections in the catalog and data cubes. Left panel: collapsed data cube highlighting the \OIII\ $5007 \pm 5$ \AA\ emission line flux over an 8\,\arcmin\ field centered at the planetary nebula coordinates (RA: $181^\circ.8685$, Dec: $54^\circ.0248$ ). The RGB background image is constructed from Legacy Survey data \citep{Dey2019}. Right panel: visualization of the mask for the same region, where grey regions indicate masked HETDEX data, and black indicates nonmasked data. The gray dashed ellipse represents the sky region masked from the HETDEX fiber data when the \texttt{LARGEGAL} bitmask is set (as described in Section\,\ref{sec:galmask}), PNe are treated similar to large galaxies for masking purposes.}
    \label{fig:pn_example}
\end{figure*}

Multiple planetary nebulae (PNe) were identified in the dataset due to an excess of bright emission lines found in the emission-line detection search. An example of a resolved PN is shown in Figure~\ref{fig:pn_example}. Spanning several arcminutes in diameter, the PNG 136.7+61.9 is seen in emission at 5007\,\AA\  across several VIRUS IFUs. We mask these in a similar manner as the large galaxy mask. The masked regions are listed in Table~\ref{tab:pn_cat}, with their central coordinate information and angular sizes provided by the HASH PN catalog\footnote{\url{https://hashpn.space}} from \citep{parker2016}. Note that all three PNe extend over multiple HETDEX observations due to their large angular extent. Flagged detections are removed from our catalog and masked as circular regions similar to the large galaxy masking. 

Users interested in studying these masked galaxies or PNe can ignore this flag. Similarly, there are background sources that can be seen in the outskirts of these galaxies that are masked conservatively at $1.5\times$ the isophotal sizes. With visual inspection, this mask can be ignored, but the calibration may be off in the vicinity of bright sources.  Users ignoring this flag should proceed with caution.

%If you use this resource in a publication, please cite this paper: Parker, Bojičić & Frew 2016 and include the following acknowledgement: "This research has made use of the HASH PN database at hashpn.space”."

\subsubsection{METEOR (Meteors)}
\label{sec:meteor}
Meteors crossing the focal plane generate spatially extended line emission unassociated with fixed astronomical sources. These tracks are identified in two stages. Then for each shot, exhibiting many of these detections, we manually fit the locations with a linear fit to create a meteor mask.

The first search method comes from the \elixer\ classifying software tool. Strong line emission appearing in just a single dithered HETDEX observation, notably associated with Mg, Al, Ca, and Fe, is flagged as a meteor candidate. The second search method is performed using AI software from RAIC Labs,\footnote{\url{https://raiclabs.com/}} which uses Rapid Automatic Image Categorization (RAIC; \citealt{raiclabs}) software to quickly generate a labeled dataset that can be trained to classify test data. Streaks in emission-line maps generated by meteor detections are easily identified using this software.  We discuss this method in greater detail later in Section~\ref{sec:raic}. We use the meteor labels generated by both \elixer\ and RAIC to identify meteor detections in the HETDEX catalog.

For any observation with more than ten such detections, we visually inspect the data to confirm the presence of a meteor, then create a simple linear mask by fitting to the positions of the flagged detections. This mask extends $12\arcsec$ above and below the linear fit; while a narrower mask could suffice in some cases, this conservative width ensures that the brightest events are fully removed. The linear mask is consistently applied to both the line emission and continuum emission raw catalogs and is masked in the PDR1 data cubes at all wavelengths under the METEOR bitmask. 

In total, 96 meteors have been detected so far in HETDEX PDR1 observations, with the streaks often spanning several IFU data cubes. A list of the affected observations and their linear fits can be found in the \textsc{hetdex-api} github repository\footnote{\url{https://github.com/HETDEX/hetdex_api/blob/master/known_issues/hdr3/meteor.txt}}, but this information is not needed for PDR1 users, as meteors are easily identified using the METEOR bitmask. The bitmask is activated at all wavelengths in the fiber spectra for regions that contain meteors. We do note that in most cases, meteor emission affects only a limited region of our spectra.  However, the compromised wavelengths are not consistent and are dependent on a number of properties of the meteor itself (e.g. chemical composition, size, altitude) that is beyond the scope of this work.

While only a small amount of data is flagged by the meteor masking ($\sim0.04\%$), the affect on line-emission detections is significant. A single meteor can create several hundred emission-line candidates. Without the bitmask, a meteor streak without any associated continuum emission appear as very bright and strongly clustered LAE candidates. It is, thus, very important to remove these features for HETDEX cosmological analysis.

Users are recommended to always apply the METEOR bitmask unless they are either (1) studying the meteor spectra or (2) manually masking the affected emission lines.

\subsubsection{BADSHOT}

Entire exposures (``shots'') may be deemed unsuitable for analysis due to poor observing conditions, failed guiding, or instrumental issues. A list of such exposures is maintained internally, and fibers from these shots are flagged. This bitmask is not relevant for the public data release, as these observations are not included in the public data products. The bitmask flag remains for consistency with the internal data model structure.

\subsubsection{THROUGHPUT}

Fibers with an effective throughput response at 4540\,\AA\ below 0.08 (8\%) are flagged within the internal HETDEX data model. PDR1 data has all low throughput data removed prior to creation, so this is not applicable to the public data products. It is retained for consistency.

\subsubsection{BADFIB (Known Bad Fibers)}

% note to self. The bad fiber list comes from
%/corral-repl/utexas/Hobby-Eberly-Telesco/lib_calib/Fiber_Locations/current

Fiber spectra are deemed unfit for scientific use for multiple reasons. Some may have too low a throughput, while others fail fiber profile fitting. Additionally, a small number of fibers create a surplus of spurious detections at multiple wavelengths; these are are poorly handled by our noise model. These fibers are masked from the catalogs and are masked within the data cubes with the BADFIB flag. Less than 0.2\% of fibers are flagged overall, but these are important to remove, as some can create multiple emission-line artifacts that can introduce both spatial clustering between HETDEX tiled observations and line-of-sight source clustering in the spectral dimension.  It is recommended to always use this mask.

\subsubsection{SAT (Satellites)}
\label{sec:sat}

Satellites trails are a significant source of continuum contamination, with approximately one-third of each HETDEX observation being affected by a satellite in one of its dithers.  Though emission line contamination caused by satellites is less common, it can still occur.  Masking these regions and accounting for the lost survey volume is important, as these streaks and the increase in continuum they produce results in the loss and/or misidentification of LAE candidates in the areas of the sky they traverse.

Satellites are identified using classifications generated by RAIC Labs AI classification platform (see Section~\ref{sec:raic} for more details). In spectrally collapsed images at the locations of HETDEX continuum detections, satellite contamination appears as a streak. When more than five satellite detections are found, a linear model is fit to the positions of all point-source satellite detections in a HETDEX observation. All detections and fibers within $\pm6$\arcsec of the fitted line are masked.

A satellite will only contaminate a single dithered observation, but we mask data in all dithers, since HETDEX detections are based on the three-dither combined data. In PDR1, we mask a total of 546 satellite streaks. In some cases, some HETDEX observations contain two to three satellite streaks. A list is maintained in the \textsc{hetdex-api} github repository\footnote{\url{https://github.com/HETDEX/hetdex_api/blob/master/known_issues/hdr3/satellite_tracks.txt}}, but users can mask out the streaks using the SAT bitmask option. 

Several fainter satellites, beyond those currently masked, are still evident when visually inspecting spectrally collapsed IFU data cubes. These objects are missed, as they do not contain many continuum detections related to the streaks, and so are not found with the current detection algorithms.

The satellite mask is conservative, often masking more regions than is needed, so a user can consider not using this flag, provided that they include visual inspection.

\subsubsection{BADCAL (Calibration Residual Issues)}

This bit identifies wavelength regions where strong sky lines, calibration errors, or other systematic issues result in unreliable flux calibration. These wavelength-dependent masks reduce spurious emission-line detections in these contaminated regions. There are three spectral windows that are flagged with this bitmask. The latter two windows are not always flagged and are observation and amplifier dependent.

\begin{itemize}
    \item $3534-3556$\,\AA: this sky line is difficult to fully remove and is masked in every observation under the BADCAL bitmask.
    \item $5194-5197$\,\AA\ \& $5200-5205$\,\AA\: Although these two sky regions are often corrected for through flat-fielding, a sharp spectral feature often persists and can create false emission lines. The bitmask is set for this region for the shotid/amplifiers listed in the \textsc{hetdex-api} github repository\footnote{\url{https://github.com/HETDEX/hetdex_api/blob/master/known_issues/hdr3/flag5200.txt}}.
    \item $5456-5466$\,\AA: This sky line can be difficult to fully remove. The bitmask is set for this region for the amplifier/shotid combinations also contained on \texttt{hetdex\_api} repository\footnote{\url{https://github.com/HETDEX/hetdex_api/blob/master/known_issues/hdr3/flag5460.txt}}.
\end{itemize}

This bitmask can be ignored for analysis performed on bright objects like stars, nearby galaxies, most AGNs, meteors, and satellites. 

\subsubsection{PIXMASK (spectrum==0)}

Spectral regions in which the raw spectral data are set to zero are also flagged.  These wavelength-dependent pixel masks ensure that no spurious flux is interpreted from missing data. The bitmask is expanded $\pm1$ pixel in the spectral dimension to minimize low $S/N$ false positive detections. It is recommended to always use this wavelength-dependent bitmask.

\subsubsection{BADDET}
\label{sec:baddet}

In Section~\ref{sec:MLAI}, we describe targeted machine learning methods used to identify and flag catalog detections classified as artifacts. The majority of these artifacts are already masked in the HETDEX data cubes through the procedures described above; however, roughly 10\% (65,078 detections) remain unaccounted for.

To mitigate these remaining artifacts, we provide a cubic bitmask for detections flagged by machine learning methods or other manual inspection efforts. For these sources, we apply a $5\,{\rm pix} \times 5\,{\rm pix} \times 5\,{\rm pix}$ mask (corresponding to $2\farcs5 \times 2\farcs5 \times 10$~\AA\ in spatial and spectral dimensions) centered on the detection coordinates and emission-line wavelength (\texttt{RA\_det}, \texttt{DEC\_det}, \texttt{wave}).

These objects do not appear in the source catalog but can be found in the raw detection database (included in this release and described in Appendix~\ref{appendix:detections}). We recommend applying this mask when performing independent emission-line searches. However, if extracting spectra at known object positions and performing visual inspection, this mask can be ignored.

\subsection{Spectral Calibration Comparisons}

%made in pdr1/pipeline_comps.ipynb
\begin{figure*}[t]
    \centering
    \includegraphics[width=0.95\linewidth]{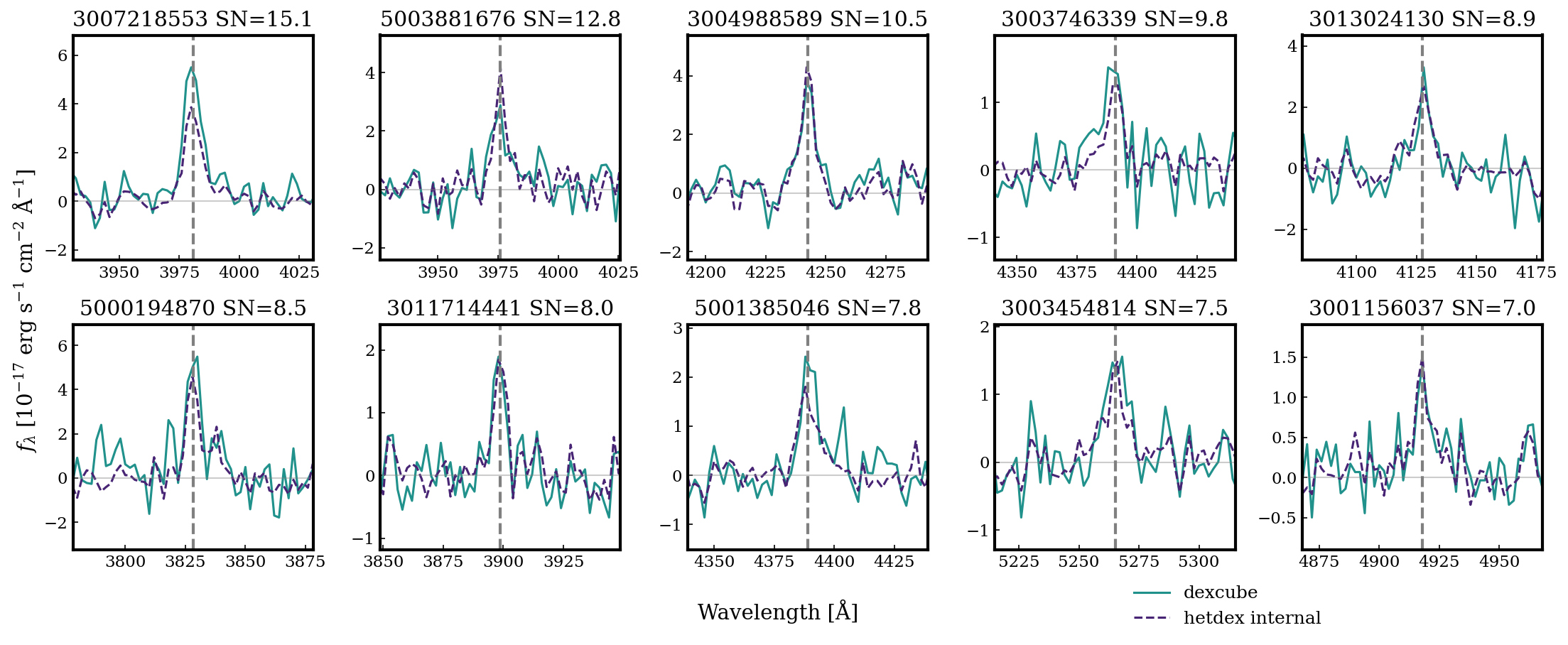}
    \caption{Comparison between aperture-corrected, extracted spectra of LAEs (with their emission wavelength indicated by the grey dashed vertical line) from the internal data model (shown in purple dashed line) and the public data model (shown in blue solid line). The internal spectra come from summing the flux in PSF-weighted aperture, while the dexcube spectral extractions are created from circular apertures with a radius of 3\farcs5. Interpolation and differences in aperture definitions account for minor difference between the two with a moderate 5-10\% difference in the blue region of the spectrum.}
    \label{fig:linecomps}
\end{figure*}

The internal HETDEX data model is shown to be well matched to SDSS spectra \citep{Gebhardt2021, Cooper2023}. In this section, we perform two tests to ensure that the extracted spectra from HETDEX data cubes are consistent with the internal model and with SDSS external spectral data. 

\subsubsection{Comparison with Internal Pipeline Spectra}

HETDEX pipeline spectra are composed of spectral extractions using several fibers and adding their flux values based on PSF-weighted extraction \citep{Horne1986, Gebhardt2021}. For point-source objects, this will be well approximated by a circular aperture extraction. In Figure~\ref{fig:linecomps}, we provide examples of LAE spectra, focusing on a 100\,\AA\ window surrounding their emission-line wavelength. These panels display examples for sources with decreasing in $S/N$ and compare spectra generated by the HETDEX pipeline in dashed purple to extracted circular (r=3\farcs5) aperture spectra generated by the data cubes. We see that for these emission-line sources, the spectra between the public and internal model are well matched. 

\subsubsection{Comparisons with SDSS}

\begin{figure}[t]
    \centering
    \includegraphics[width=\linewidth]{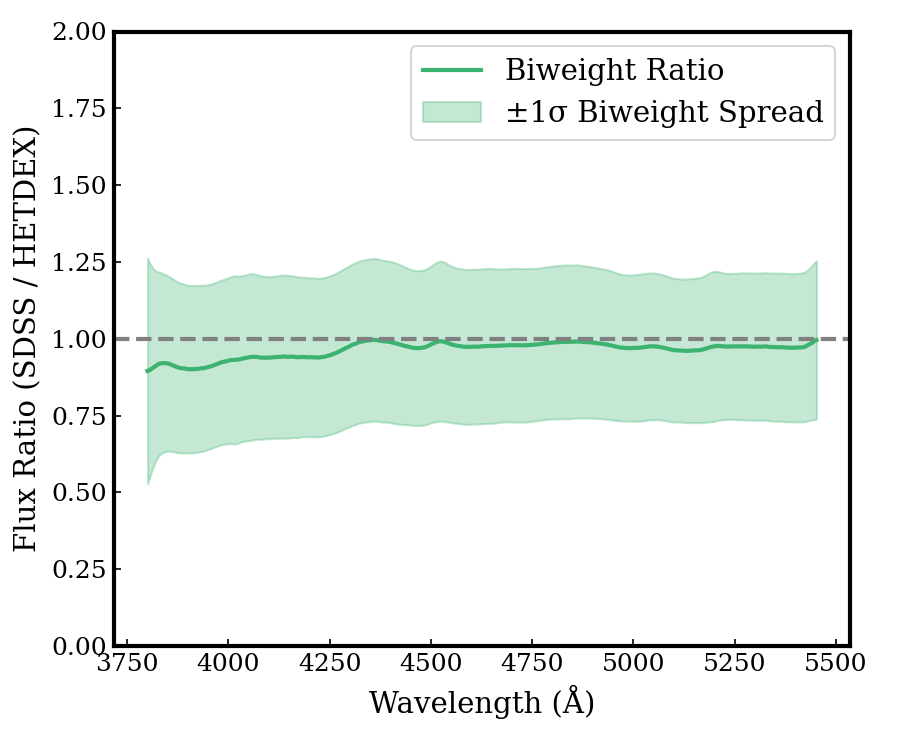}
    \caption{Biweight average flux ratio spectrum comparing SDSS and HETDEX observations of 8,076 AGN. Each spectrum was Gaussian-smoothed ($\sigma$ = 5 pixels) before computing the flux ratio SDSS/HETDEX at each wavelength. The solid green line shows the biweight location of the flux ratio across all matched sources, while the shaded region indicates the $\pm 1\sigma$ biweight spread after sigma-clipping outliers at each wavelength bin. The dashed line at ratio = 1 indicates perfect flux agreement. Overall, the SDSS and HETDEX spectra show good agreement.}
    \label{fig:fluxratio}
\end{figure}

We consider spectral comparisons between a sample of 8,076 AGN that are found in both HETDEX and SDSS in Figure~\ref{fig:fluxratio}. Spectral extractions are performed on the HETDEX IFU data cubes with 3\farcs5 circular apertures. The median SDSS/HETDEX flux ratio is consistent with unity with a scatter in the agreement of under 20\%. A modest offset of less than 10\% is evident in the blue region on the spectrum.

\section{Catalog}
\label{sec:catalog}

HETDEX Public Source Catalog 1 (HPSC1; \citealt{Cooper2023}) describes the generation of the first public source catalog taken from data that is a subset of this release (early data spanning from 2017-01-01 to 2020-06-26; internally referred to as HDR2.1). Included in PDR1 is an update to this catalog, titled HETDEX Public Source Catalog 2, which contains four additional years of data and incorporates several updates to masking, source selection and false-positive mitigation.  Hereafter, we refer to this catalog as HPSC2.

Details of generating the catalog are described in HPSC1, but brief descriptions are included here for ease and clarity. Source detection, data quality selections, redshift assignment, and the derivation of resolved \OII\ fluxes for low-redshift galaxies are applied consistently in both catalogs. Given the considerable updates to data quality assessment and detection masking (see Section \ref{sec:bitmask}), we outline the differences between HPSC1 and HPSC2 throughout this section.

As issues have been found in HETDEX data and the masking model has developed, we aimed to maintain consistent masking of both HETDEX detections and calibrated fiber data in our data model. However, we have also investigated and found success using multiple means of classifying and cleaning catalog sources through citizen science, also known as participatory science, and through multiple Machine Learning/Artificial Intelligence (ML/AI) classification approaches. We discuss these issues in Section~\ref{sec:MLAI}. 

Upon the identification of systematic issues through these methods, such as a pixel defect in the detector, a faulty fiber, or a meteor streak, we can reliably mask the problem at both the fiber spectrum (and interpolated data cube) and detection level. For transient features that could not be masked through an automated method, we mask at the detection level through the BADDET mask (described in Section\,\ref{sec:baddet}). These include horizontal readout artifacts caused by by cosmic rays, calibration problems due to scattered light, and residuals produced sky lines detected by the different classification methods. 

%Therefore, we advise future users against independently applying detection methods to the data cubes without collaborating with a HETDEX team member to gain access to intermediate data products.

\vspace{10pt}

Before getting into the details, we provide a few HETDEX catalog concepts to highlight:

\begin{itemize}
    \item HETDEX source detection (see Section~\ref{sec:detection}) is performed under a point-source assumption. Fluxes and extracted spectra are reported under this assumption unless otherwise stated. Detections that are within 3\arcsec\ of each other and 3\,\AA\ in spectral separation are reduced to a single detection in the HETDEX pipeline. For low-redshift, resolved galaxies, we supplement the pipeline fluxes and provide an aperture flux measurement for the \OII\ emission line (see Section~\ref{sec:aperflux}). 
    \item \texttt{detectid:} this is a unique integer identifier from the raw detection pipeline. Nine digits long, the 1st digit signals the internal release database of the detection. The 3rd digit indicates whether the detection is a emission-line detection (3rd digit=0), or a continuum detection (3rd digit=9). For example, the detection 5090045587 is from the internal HDR5 data release and the 9 in 509 indicates that it is a continuum detection. 
    \item \texttt{source\_id:} a single astronomical source can be comprised of many \texttt{detectids} if it is spatially resolved or if it has multiple emission lines. These detections are grouped together (as described in Section~\ref{sec:detgroup}) to make one source, identified by the integer \texttt{source\_id}. If the same astronomical source is observed in multiple observations, it is listed multiple times in the catalog and will have a different  \texttt{source\_id}. It is up to the user to combine the data from repeat observations into a single source.
    \item \texttt{z\_hetdex} and \texttt{source\_type}: We assign source classifications and redshifts using three methods: continuum template fitting with \textsc{Diagnose} \citep{debski2025,diagnose}, probabilistic emission-line identification via \elixer\ \citep{Davis2023}, and secure AGN redshifts from dedicated AGN catalogs \citep{liu2022, liu2025}. See Section~\ref{sec:classification} for full description. This approach achieves an overall redshift accuracy of $\sim$96\% compared to external spectroscopic catalogs. HETDEX sources are grouped into five \texttt{source\_types}: `lae', `agn', `oii', `lzg', `star' (see Section~\ref{sec:sourcetype} for definitions).
\end{itemize}

\subsection{Object Detection}
\label{sec:detection}
Two independent, but complementary, object-detection search techniques are performed as part of the main HETDEX reduction pipeline: one to identify emission lines, the other to detect continuum sources.  In the fifth internal data release for HETDEX (HDR5) a search was performed across more than $>6\times10^8$ flux-calibrated fibers. The search method details can be found in Section 7 of \citet{Gebhardt2021}. We briefly summarize the procedures here. 

During this process, no imaging pre-selection is used; the HETDEX data itself provides object detection. Both the emission-line and continuum detection algorithms are designed to identify point sources and account for the variable image quality, or PSF of each independent HETDEX three-dither observation. 

The raw detections are stored in HDF5 files. We include these files in the PDR1 data release and describe their content and access in Appendix~\ref{appendix:detections}.

\subsubsection{Emission-Line Detection}

Line emission is identified using a two-stage Gaussian fitting procedure. An initial grid search steps through spatial (0$\farcs$5) and spectral (8\,\AA) dimensions, fitting a fixed-width Gaussian ($\sigma=1.7$\,\AA) after subtracting the local continuum. Candidate emission lines with a $S/N>4$ and a reduced chi-square ($\chi^2$) $<3$ are retained and refit at higher ($0\farcs15$) spatial resolution, this time allowing $\sigma$ to vary. The raster location that maximizes $S/N$ defines the source centroid, and the best-fit amplitude yields the line flux. Duplicate detections within 3\arcsec\ and 3\,\AA\ are merged by selecting the highest-$S/N$ measurement. 

Final line catalog membership is determined by additional quality cuts: a detection must satisfy $S/N$, line width, and $\chi^2$ combinations that help to identify real galaxies from artifacts. Spurious features from negative continua, poor fiber fits, or detector artifacts are removed through these cuts but also through a number of automated methods described in Section~\ref{sec:detection_cleaning}.

The raw emission-line database consists of \linedetcount\ detections and reaches $\sim50\%$ completeness at $11\times10^{-17}$ erg s$^{-1}$ cm$^{-2}$, although this number varies with wavelength, image quality, and observing conditions.

\subsubsection{Continuum Detection}

Continuum sources are searched for independently by measuring fiber counts in two 200\,\AA\ wide spectral regions, one in the blue (3700–3900\,\AA) and one in the red (5100–5300\,\AA). If either region exceeds 50 counts per 2\,\AA\ pixel, the fiber is stored as a continuum detection. This translates to roughly \hetg~$\sim22.5$, where \hetg\ is a pseudo-magnitude derived from the HETDEX source spectrum by convolving it with the SDSS $g$-band transmission curve and integrating the resulting flux. A PSF model fit is then used to centroid the source with a $15\times15$ grid raster in 0$\farcs$1 steps, and a point-source spectrum is extracted at the best-fit position following the methodology of \citet{Horne1986}. As with line detections, quality checks remove sources located on low-quality detectors or those flagged as artifacts or transients. The raw continuum catalog contains \contdetcount\ detections, with sensitivity set by photon statistics and typically reaching \hetg~$\sim$~22.5 under typical HETDEX observing conditions. Continuum detections fainter than \hetg=22 are not included in the final catalog, as they do not provide enough sensitivity to determine a redshift.

\subsubsection{AGN Catalog Selection}

The HETDEX AGN catalog \citep{liu2022, liu2025}, derived from the same internal HDR5 release, is incorporated into our combined source catalog. While it shares the same raw detection databases, their selection criteria differ.  The AGN catalog results from targeted searches for broad emission lines, whereas the main catalog, as discussed in Section~\ref{sec:detection_cleaning}, excludes lines with large Gaussian widths ($\sigma > 14$\,\AA) to mitigate artifacts that often appear as broad features. The AGN classification method also identifies candidates through the presence of common spectral line pairs across the full emission-line database, enabling the identification of both Type~I and Type~II AGNs. Roughly a quarter of the AGN sample overlaps with our line and continuum catalogs, but the majority represent broad-line AGN that fail the Gaussian line-fit cuts due to their large line widths and high $\chi^2$ values. These sources are essential for capturing the diversity of AGNs and are retained after visual verification to mitigate contamination from artifacts. 

In the combined source catalog, AGN detections are grouped with any associated line or continuum emission according to the detection grouping method (Section~\ref{sec:detgroup}), with redshift assignments taken from the AGN catalog classifications of \citet{liu2025}.

We note that the HSPC2 source catalog applies stricter data quality cuts than the \citet{liu2025} AGN catalog.  The former removes frames with poor observing conditions or failed amplifiers, and eliminates objects that happen to lie on regions masked due to satellite trails, meteor emission, and the like.  The AGN catalog retains some of these sources through visual inspection. In addition, detection grouping in HPSC2 is done only by observation, so the same AGN may have multiple entries in HPSC2, while \citet{liu2025} opted to merge all observations of the same AGN into a single \texttt{agnid}. 

%These sources are essential for capturing the diversity of AGN and are retained after visual verification to mitigate contamination from artifacts. In the combined catalog, AGN detections are grouped with any associated line or continuum emission, with redshift assignments and classifications carried through the grouping process.

\subsection{Quality Control and Selection of Raw Detections}
\label{sec:detection_cleaning}

\input{obs_column_info.tex}

The internal HETDEX pipeline produced 32,643,783 raw detections for HDR3/4/5 line- and continuum- emission all combined. These files are included in this release and are described in Appendix \ref{appendix:detections}. This raw database is processed to a quality-controlled catalog (internal version ‘5.0.2’) of 4,627,812 detections. The outputs from the three object-detection methods (continuum, line, and AGN) are down-selected from the raw detection database through two primary steps: (1) artifact mitigation via the application of the mask model, and (2) filtering based on detection parameters.

\begin{enumerate}

    \item \textbf{ Detection Masking}

    Internal masking of HETDEX data is done at a detection level. For each issue described in Section~\ref{sec:bitmask}, the relevant detection parameters are searched to determine if the source candidate should be flagged. For example, for meteors and satellites, a combination of observation name (e.g. \texttt{shotid}) and sky coordinates are used to decide whether to exclude the detection. In contrast, for calibration, amplifier, or fiber issues, the relevant instrument info is queried for an affected time range.

    \item \textbf{ Detection Parameter Selections}

    For the line search, the following line-parameter ranges are required for catalog inclusion:

    \begin{equation}
    \left\{
    \begin{aligned}
    3510~\text{\AA} &\leq \lambda \leq 5496~\text{\AA} \\
    S/N &\geq 4.8 \\
    \chi^{2} &\leq 2.5 \\
    \texttt{continuum} &> -1 \\
    \chi^{2}_{\mathrm{fib}} &< 4.5 \\
    1.7~\text{\AA} &\leq \texttt{sigma} \leq 14~\text{\AA} \\
    \texttt{apcor} &\geq 0.45
    \end{aligned}
    \right.
    \end{equation}

    Here $S/N$ and $\chi^2$ are the signal-to-noise ratio and reduced quality of fit of the emission line, respectively, and \texttt{sigma} is the line width measured from the Gaussian line fit. 
    \texttt{continuum} measures the local continuum around the HETDEX emission-line detection. It is derived from a linear fit to two 50\,\AA\ spectral windows placed $3\times {\rm sigma}$ on either side of the detected line.   $\chi^{2}_{\mathrm{fib}}$ is the quality of fit of the fiber profile of the highest weighted fiber for the detection. High values of $\chi^{2}_{\mathrm{fib}}$ are associated with detection artifacts. We only include detections with an aperture correction factor, \texttt{apcor}, greater than 0.45. This factor is the fraction of the 3\farcs5 radius circular aperture with fiber coverage.

    For continuum sources, the only exclusion is based on \texttt{apcor} and \hetg, the magnitude derived from the detection spectrum by folding the detected spectrum through the SDSS-$g$ filter's bandpass and summing the flux.
    
    \begin{equation}
    \left\{
    \begin{aligned}
        g_\mathrm{HETDEX} &\leq 22 \\
        \texttt{apcor} &\geq 0.45 
    \end{aligned}
    \right.
    \end{equation}
    
%    Note: Detection masking often considers the three nearest fibers to the candidate object’s central position, so that each of these fibers is evaluated for possible masking. As a result, some detections that are not flagged by the initial masking criteria may still be flagged if a nearby secondary fiber falls on a bad fiber, a bad-pixel region, a faulty amplifier, or any of the other conditions described in Section \ref{sec:bitmask}. These masks are stored in a simplified bit, \texttt{flag\_best}, for internal HETDEX users. For the public catalog, we require \texttt{flag\_best == 0} and exclude all detections that do not satisfy this criterion.

\end{enumerate}

\subsection{Detection Grouping}
\label{sec:detgroup}

Both the line- and continuum-emission pipelines are optimized for point-source detection. For LAEs and many faint \OII\ emitters, this assumption is generally valid, although neither LAEs nor \OII\ emitters are truly point-like. As shown in \citet{MentuchCooper2026_HETDEX_LAN}, more than 50\% of LAEs with $S/N>6$ are better described by a two-component surface-brightness model consisting of a compact core and an extended exponential halo. Roughly 25\% of HETDEX sources, such as $z<0.5$ galaxies, AGNs, and stars, are composed of multiple detections, arising either from extended emission or from multiple emission lines. In addition, bright sources can be independently detected in both the line- and continuum-emission searches.

To robustly associate detections and mitigate confusion from extended low-$z$ \OII\ emission, we apply a 3D friends-of-friends (FOF) clustering algorithm. Using spatial and spectral linking lengths of 6\arcsec\ and 8\,\AA, we combine emission features into coherent source groups as described in \citet{Cooper2023}. This procedure ensures multiline galaxies are properly grouped, although in some blended cases, background sources may be merged into a foreground object. This will cause some background sources in gravitational lenses to be merged with the foreground object; although, these can be discovered through independent searches \citep{Laseter2022} within the expanded catalog described in Appendix~\ref{appendix:DetInfoTable}.

For each source, we designate a single representative detection, identified by the value in the \texttt{selected\_det} column of the expanded catalog \textit{Detection Information Table} (Appendix~\ref{appendix:DetInfoTable}). In the main catalog table, only this detection is printed for each source (including its exact detection position at \texttt{RA\_det}, \texttt{DEC\_det}), all other detection entries are removed. Non-LAEs are assigned the brightest \hetg\ member, while LAEs are assigned the highest–$S/N$ Ly$\alpha$ detection. This procedure produces a quality-controlled source catalog in which multiple detections are linked through the unique source identifier \texttt{source\_id}. In summary, \texttt{source\_id} provides a common identifier for all detections associated with the same astrophysical object, enabling the association of extended emission features, multiple emission lines, and the assignment of a consistent redshift and source classification.

\subsection{Source Classification and Redshift Assignment}
\label{sec:classification}

\begin{figure}[t]
    \centering
    \includegraphics[width=3.25in]{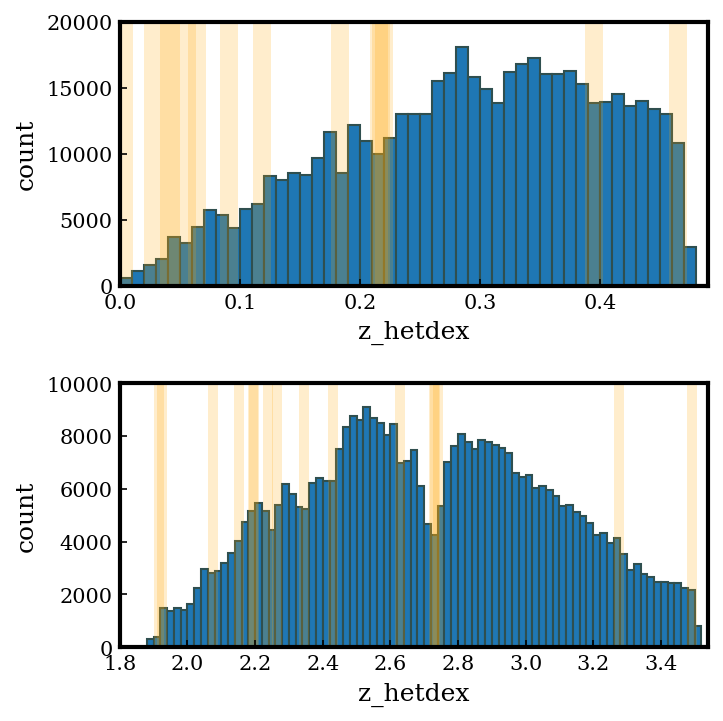}
    \caption{The redshift distribution of the low-$z$ ($\Delta z_{bin} = 0.01$) and high-$z$ ($\Delta z_{bin} = 0.02$) galaxy samples in the top and bottom panels, respectively. The low-$z$ sample is a combination of \OII\ emitters and LZGs; the high-$z$ dataset includes LAEs and AGNs. The brightest sky lines are marked by light-yellow vertical bars; these lines suppress the number counts in both distributions.}
    \label{fig:z_hist}
\end{figure}

Each HETDEX source group is assigned a classification and redshift using one of three methods depending on brightness and spectral features. For sources with $g < 22$, we apply \textsc{Diagnose} \citep{debski2025,diagnose}, a PCA-based template-fitting algorithm developed for the Hobby-Eberly Telescope VIRUS Parallel Survey (HETVIPS; \citealt{Zeimann2024}). \textsc{Diagnose} assigns classifications such as \texttt{star}, \texttt{galaxy}, or \texttt{qso} for bright sources, along with a redshift derived from spectral template matching. It recovers $\sim$97\% of classifications reliably at \hetg~$<22$. However, for fainter objects, confusion between \lya\ and low-$z$ lines demands we rely on alternative methods.

Objects with little to no continuum emission are classified using \elixer\ \citep{Davis2023}.  This program is based on the initial Bayesian work presented  in \citet{Leung2017}, incorporating equivalent widths, broadband imaging, and luminosity priors to infer line identity and redshift. Each detection is assigned a likelihood \texttt{P(\lya)} of being \lya, with a threshold of 0.4 chosen to balance recovery of true LAEs (96\%) and \OII\ contamination in the LAE candidate sample ($<$3\%). For high-redshift sources, \elixer\ assigns the redshift of the most central detection, considering clustering of lines and multiple line associations.

Redshifts and classifications for AGNs are adopted from the HETDEX AGN catalog \citep{liu2022, liu2025}, which provides redshift assignments for single broad-line detections (generally assuming the line is \lya), for emission-line pairs indicative of AGN activity, and for broad-line sources identified through cross-matching with SDSS DR14.

A logical sequence is implemented to assign each source a best redshift and object classification. Priority is given to AGN matches, followed by \textsc{Diagnose} redshifts for brighter objects (\hetg\ $<22$), and \elixer\ results otherwise. A histogram of spectroscopic redshifts is presented in Figure~\ref{fig:z_hist}.

\begin{figure}
\centering
    \includegraphics[width=0.45\textwidth]{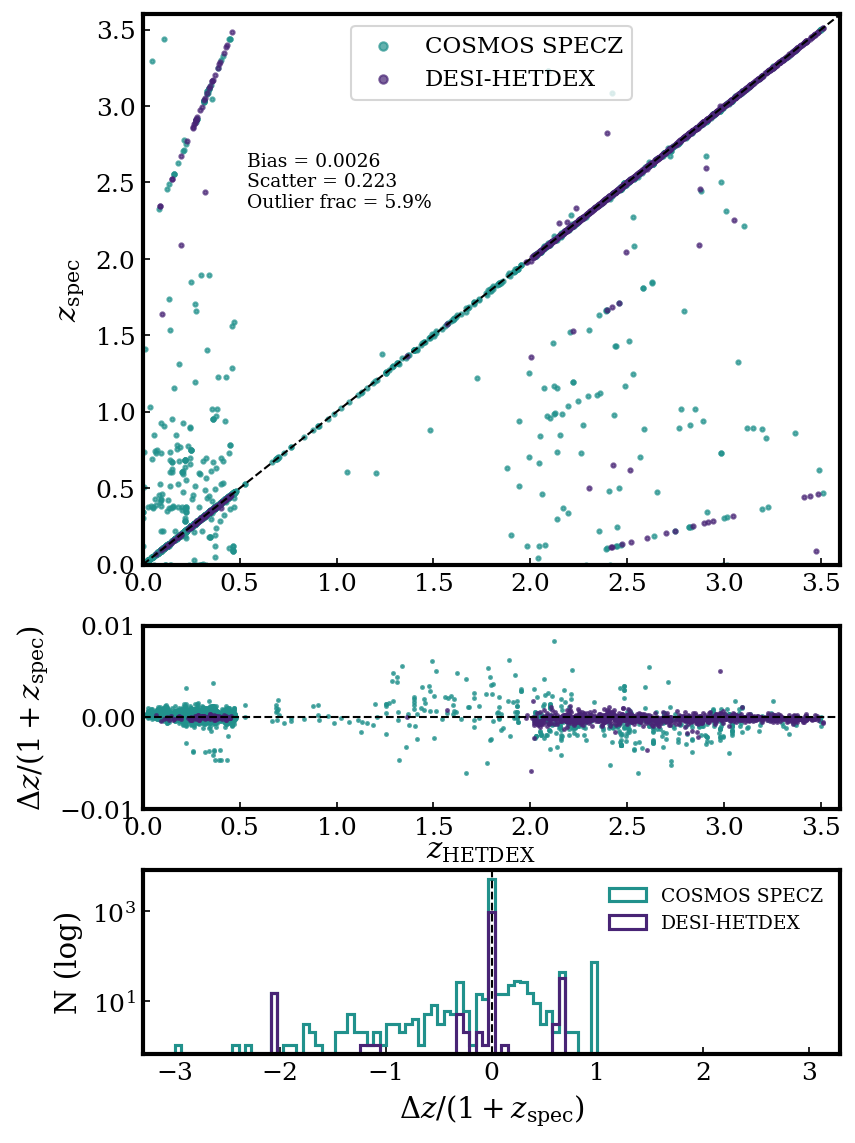}
    \caption{
Comparison of HETDEX redshifts with external spectroscopic measurements in the COSMOS legacy field compiled by \citet{khostovan2025} and visual spectroscopic confirmation from DESI–HETDEX \citep{landriau2025}. \textit{Top panel:} direct comparison between $z_{\mathrm{HETDEX}}$ and $z_{\mathrm{spec}}$, with the dashed line showing the one-to-one relation. 
\textit{Middle panel:} normalized residuals, defined as $(z_{\mathrm{spec}} - z_{\mathrm{HETDEX}}) / (1+z_{\mathrm{spec}})$, as a function of $z_{\mathrm{HETDEX}}$. 
%The horizontal dashed line marks zero offset, and vertical dotted lines in the histogram indicate the $\pm 0.02$ outlier threshold. 
\textit{Bottom panel:} distribution of residuals on a logarithmic scale, highlighting the extended tails. Across the full sample, we measure a small bias ($0.0026$), scatter ($\sigma = 0.22$), and an outlier fraction of 5.9\%.}
\label{fig:zcomps}
\end{figure}

We assess the accuracy of HETDEX redshifts through a comparison to two independent spectroscopic samples. The first is the COSMOS field spectroscopic redshift compilation created by \citet{khostovan2025}; this sample is skewed toward brighter sources, and, when restricted to objects with high-quality redshifts (\texttt{flag} $\ge 3$), the cross-matched sample (with a matching radius of 1\farcs0) consists of 395 LAEs, 303 AGNs, 347 low-redshift galaxies, and 4767 [\ion{O}{2}] emitting sources. We note that since COSMOS was used by HETDEX as a science verification field, the region was visited multiple times in the course of the survey.  The data acquired during each visit were reduced and analyzed independent of the other visits; as a result, a single COSMOS source can potentially have different HETDEX redshifts and object classifications.  The  numbers listed above therefore represent the total number of redshift comparisons, not the number of unique sources on the sky.  

The second sample comes from objects observed by the DESI-HETDEX survey \citep{landriau2025}.  This program concentrated on sources with HETDEX
classifications that straddled the decision boundary between LAE and [\ion{O}{2}] galaxy, and is composed of 851 LAEs, 67 AGNs, and 139  [\ion{O}{2}] galaxies, all with high quality (\texttt{VI\_QUALITY} $>3$) redshifts. Thus, our redshift comparison involves over 6800 measurements in both the local and high-redshift Universe.  
Note that both DESI and HETDEX measure the LAE redshifts assuming that it is a single Gaussian emission line profile, even though it is not uncommon to observe a double-peak feature \citep{landriau2025}. 
Using the mock LAE spectral profile in \citet{Khoraminezhad2025}, we have checked that the impact of this assumption is given by $\Delta z/(1+z)\sim 0.0005$ for DESI and $\Delta z/(1+z)\sim 0.0001$ for HETDEX, where HETDEX has a smaller bias due to its slightly poorer spectral resolution than DESI. 

Figure~\ref{fig:zcomps} shows the direct comparison between $z_{\mathrm{HETDEX}}$ and $z_{\mathrm{spec}}$, as well as the distribution of normalized residuals, defined as $(z_{\mathrm{spec}} - z_{\mathrm{HETDEX}}) / (1+z_{\mathrm{spec}})$. The ensemble statistics demonstrate that HETDEX redshifts are unbiased, with an average offset of $0.0026$ and a scatter of $\sigma = 0.22$.
The normalized median absolute deviation is consistent with zero bias (${\rm NMAD} = 0.0003$), and the catastrophic outlier fraction (defined by $|\Delta z|/(1+z) > 0.2$) is 5.9\%.

As noted above, the DESI–HETDEX program specifically targeted faint emission-line galaxies near the classification boundary between \lya\ and \OII\, and, therefore, consists primarily of objects with $g>23$. By design, there are no false positives or neighbor confusion. In contrast, the heterogeneous collection of object classes in the COSMOS spectroscopic compilation is biased toward brighter sources. Despite these differences, the catastrophic outlier fractions are nearly identical in the two samples: 5.97\% for COSMOS and 6.03\% for DESI–HETDEX.

In the COSMOS comparison, catastrophic outliers can arise from incorrect counterpart associations, including occasional HETDEX false positives or nearby on-sky neighbors within the matching radius. In contrast, the DESI–HETDEX sample is constructed from emission-line detections identified independently by both surveys, so catastrophic failures primarily result from incorrect emission-line identification. The most common case is confusion between \lya\ and [\ion{O}{2}], which produces the two distinct outlier tracks visible in Figure~\ref{fig:zcomps}. Overall, these results demonstrate that HETDEX redshifts are robust across the survey footprint, with only a modest tail of failures dominated by low-$S/N$ detections or occasional misclassifications.

%The COSMOS sample exhibits slightly tighter agreement than the DESI–HETDEX subset. As stated above, the DESI-HETDEX sample was selected to include faint line-emitting galaxies and provide better representation of the HETDEX emission line sample and consists primarily of objects at $g>23$. Conversely, the heterogeneous collection of object classes in the COSMOS sample leans towards brighter objects. Overall, the results demonstrate that HETDEX redshifts are robust across the survey footprint, with only a modest tail of low-$S/N$ failures caused primarily by incorrect counterpart matches, false positives or occasional source misclassifications.

\subsubsection{Source Classification}
\label{sec:sourcetype}

\begin{figure}[t]
    \centering
    \includegraphics[width=3.2in]{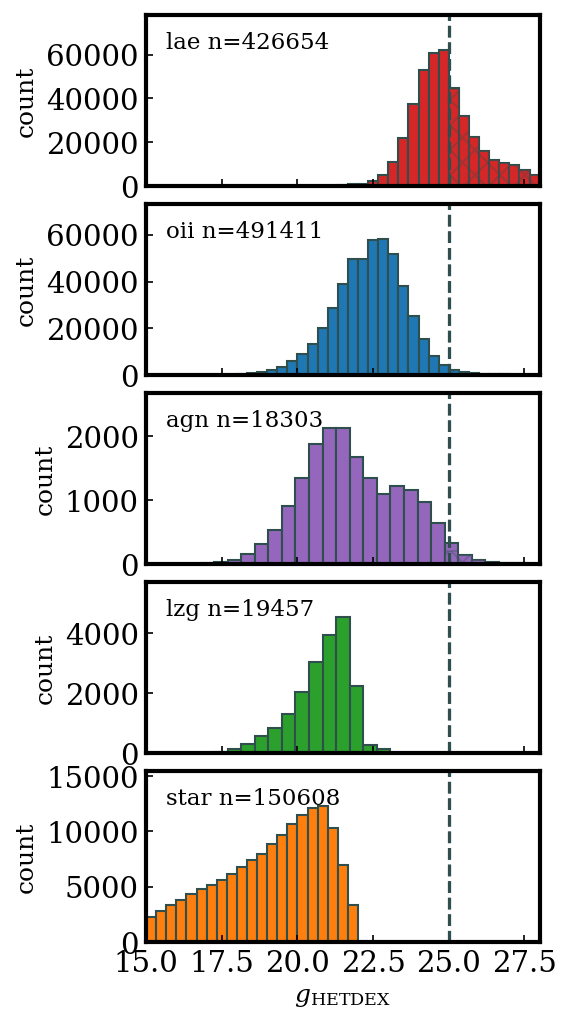}
    \caption{Histogram of \hetg\ magnitudes for each source type as measured by summing the 1D extracted spectra, weighted by the \sdss\ $g$-band response curve. If multiple detections exist for the source, the brightest detection is used. The vertical dashed line at \hetg=25 represents the HETDEX average sensitivity limit.}
    \label{fig:g_hist}
\end{figure}

HETDEX sources are broken into five \texttt{source\_types}. Source counts by field are provided in Table~\ref{tab:summary} and the magnitude distributions for each source type are shown in the histograms in Figure~\ref{fig:g_hist}.

\begin{itemize}
    \item \texttt{star} — Galactic stellar sources; classified by \textsc{Diagnose} and assigned $\zhet = 0$. 
    
    \item \texttt{agn} — Active Galactic Nuclei; identified via the HETDEX AGN catalog or \textsc{Diagnose}; redshift range: $0<\texttt{z\_hetdex}<4.6$.
    
    \item \texttt{lae} — Lyman-alpha emitters classified by \elixer; redshift range: $1.88 < \texttt{z\_hetdex} < 3.52$.
    
    \item \texttt{oii} — Low-redshift galaxies with detected \OII\ emission ($\lambda$3727\,\AA), classified via \textsc{Diagnose} or \elixer; redshift range: $0 < \texttt{z\_hetdex} < 0.48$.
    
    \item \texttt{lzg} — Low-redshift galaxies without detected emission lines classified by \textsc{Diagnose} using continuum template fits; redshift range: $0 < \texttt{z\_hetdex} < 0.5$.
\end{itemize}

\subsection{Spatially Resolved [\ion{O}{2}]}
\label{sec:aperflux}

\begin{figure*}
\centering
    \includegraphics[width=0.3\textwidth]{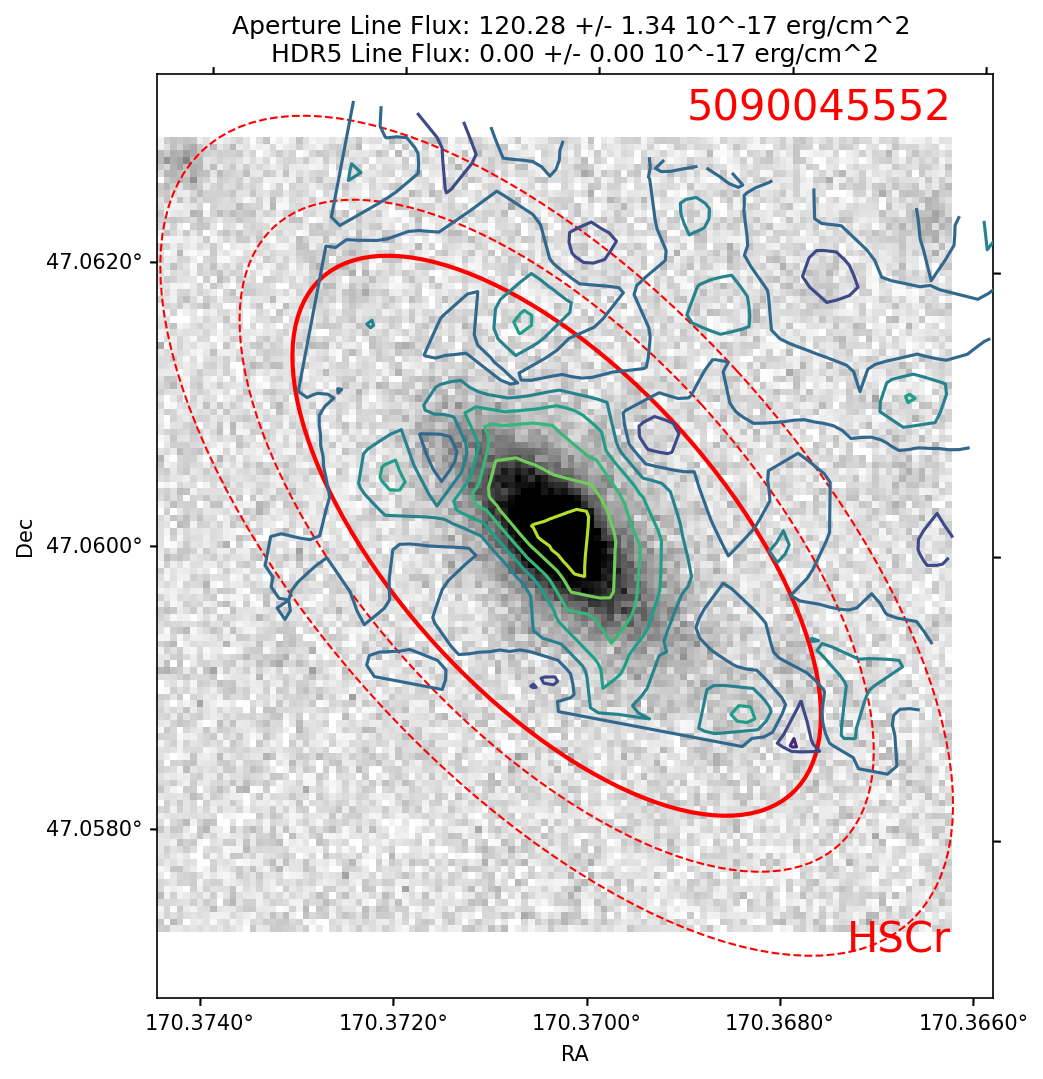}
    \includegraphics[width=0.3\textwidth]{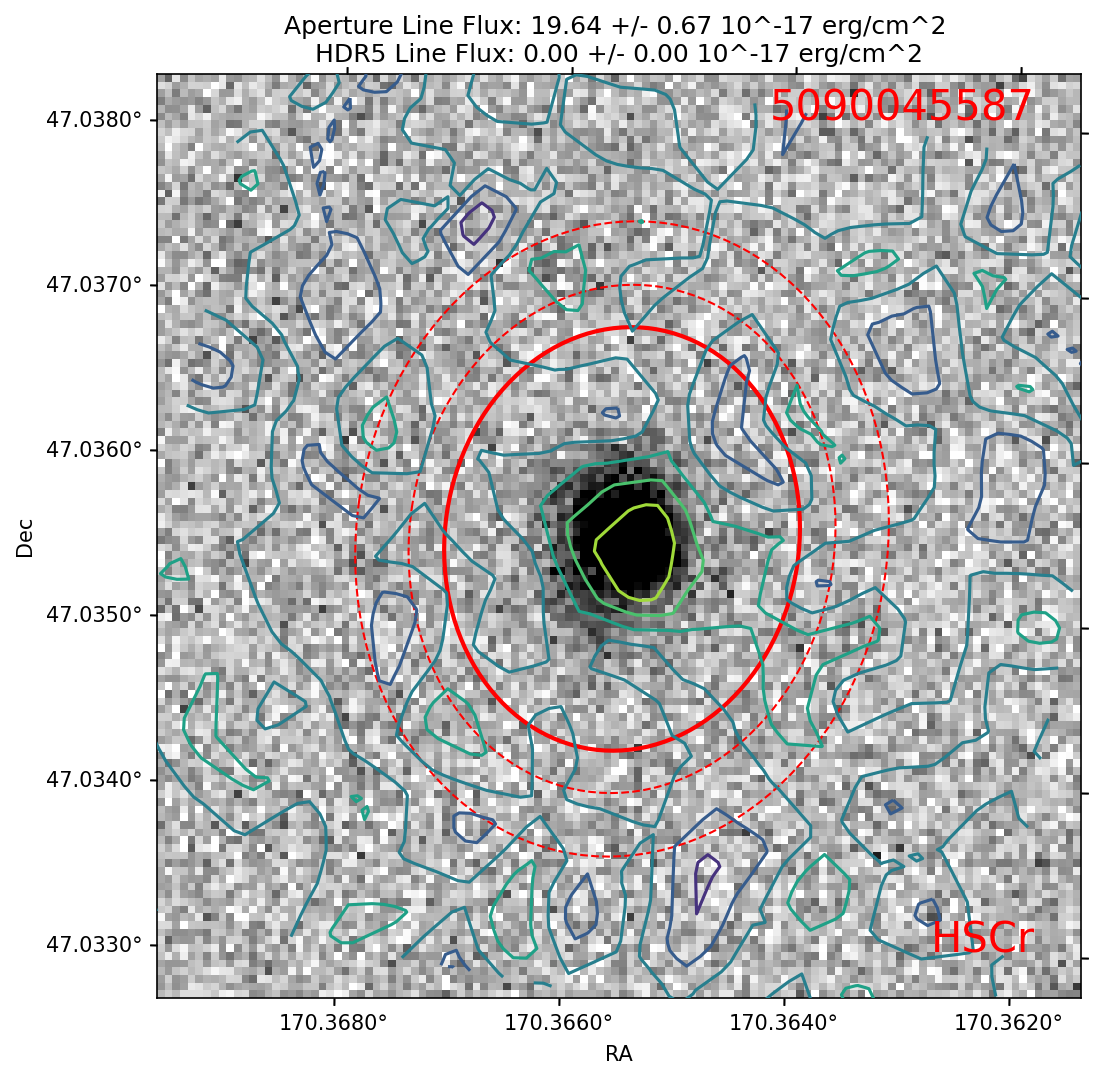}
    \includegraphics[width=0.2\textwidth]{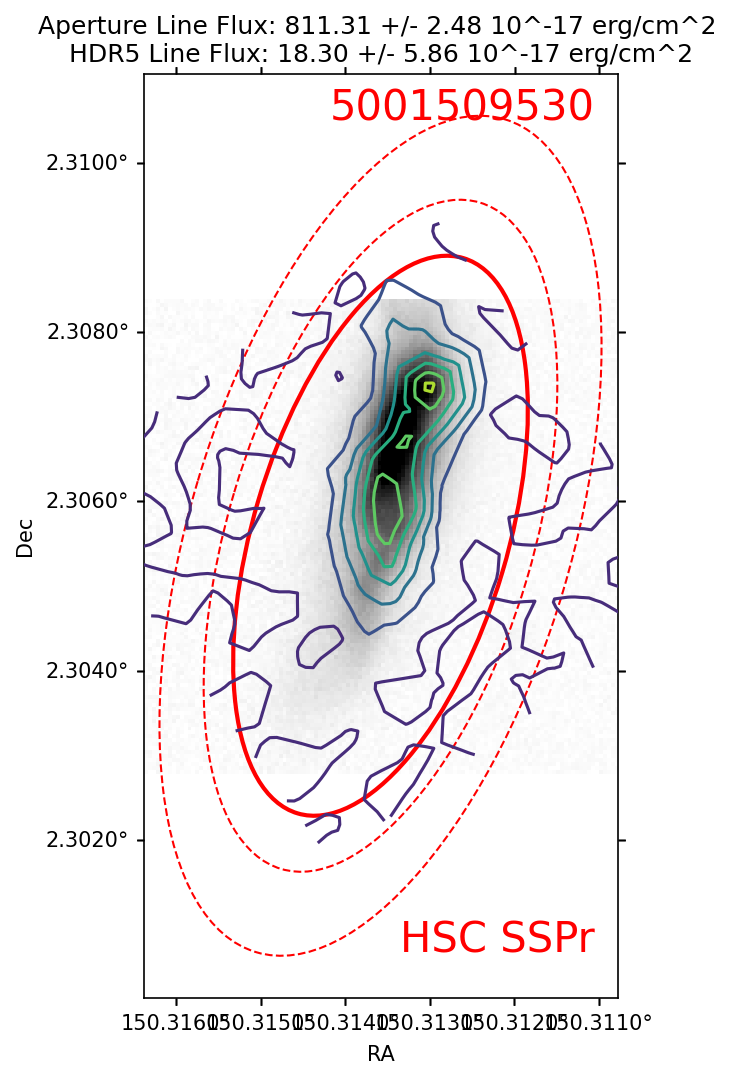}
    \caption{Examples of \OII\ line-flux maps (shown in contours) that are used to measure continuum-subtracted \OII\ fluxes for spatially resolved low-redshift galaxies. Elliptical apertures are defined using imaging data from HSC or DECALs as measured in the \elixer\ catalog \citep{Davis2023}. In total, \nlzgtooiiswitch\ galaxies are found to have \OII\ line emission that was not initially measured by the HETDEX reduction pipeline.}
\label{fig:resolveoii}
\end{figure*}

The HETDEX emission-line detection algorithm produces PSF-weighted line-flux values by assuming the object being measured is a point source.  However, many nearby [\ion{O}{2}] line emitters are extended, resulting in underestimated measurements of the objects' fluxes. In addition, the emission-line detection pipeline contains an upper limit on the continuum value, so very bright emission-line galaxies are completely missing from the emission-line database, although they are found in the HETDEX continuum catalog and are assigned a redshift. Without an associated emission line, they are initially classified as a low-redshift galaxy (an `lzg'). Figure~\ref{fig:resolveoii} illustrates this issue. The two leftmost sources are only found in the continuum search method with no reported \OII\ line flux; in the third galaxy, the HETDEX-report line flux severely underestimates the total [\ion{O}{2}] emission from the galaxy.

For these resolved galaxies, we supplement the HETDEX line-flux measurements with estimates of galaxies' total line emission.  A major strength of the wide-IFU (dithered) coverage provided by HETDEX is that the observations automatically produce an emission-line map of resolved galaxies.  However, due to the grid-like distribution of IFUs in the HET's focal plane, many of these systems have incomplete coverage, as their light extends beyond their IFU's limits. We deal with this issue by using shape information from ancillary imaging observations either provided by HSC or DECam Legacy Survey (DECALs; \citealt{Dey2019}). 

Isophotal elliptical apertures based on broadband imaging are included in HETDEX's \elixer\ classification tool \citep{Davis2023}. The program applies \textsc{Source Extraction and Photometry} \citep[\textsc{SEP};][]{sep_pkg} to all available broadband imaging at the location of each HETDEX detection. We use the \elixer\ catalog \texttt{selected=True} option (indicating this was the imaging counterpart of choice) and preferentially choose $r$-band measurements over $g$-band measurements to define each galaxy's elliptical aperture. In general, the $r$-band imaging has a fainter limiting magnitude, and better image quality.

Elliptical parameters for each low-$z$ galaxy are provided in HPSC2 (columns described in Table~\ref{tab:column_info}) under the columns \texttt{major}, \texttt{minor}, and \texttt{theta}. Additionally the aperture center and the measured continuum aperture magnitude are in the expanded \textit{Detection Information Table} in Appendix~\ref{appendix:DetInfoTable} in the columns \texttt{ra\_aper}, \texttt{dec\_aper}, \texttt{mag\_aper}, and \texttt{mag\_aper\_err}. 

For every galaxy (both `lzg' and `oii') at \texttt{z\_hetdex}~$<0.5$, we first construct a sky-subtracted narrowband image centered on the observed \OII\ wavelength (given by \texttt{z\_hetdex}). The cube is collapsed spectrally over a $\pm15$\,\AA\ window to form the line-flux map. To remove continuum emission, we create two additional 50\,\AA\ wide collapsed images placed $\pm10$\,\AA\ outside the line-map limits. The average of these two continuum images is subtracted from the narrowband image, producing a continuum-subtracted \OII\ emission-line flux map.

Because many resolved galaxies extend beyond the HETDEX IFU footprint, only part of the elliptical aperture defined from broadband imaging is actually covered by IFU data. After generating the continuum-subtracted map, we therefore measure the \OII\ flux within the portion of the aperture that has fiber coverage, compute the fraction of the aperture that is sampled, and scale the measured flux by the inverse of this coverage fraction. This yields an aperture-corrected estimate of each galaxy’s total \OII\ flux assuming the flux distribution is smooth and evenly distributed within the aperture. In practice, emission-line regions can exhibit radial structure and clumpy morphology, so this correction should be regarded as an approximate estimate of the total flux. The aperture-corrected fluxes and their uncertainties are provided in the catalog under the columns \texttt{flux\_aper} and \texttt{flux\_aper\_err}, respectively. 

A significant number of low-$z$ galaxies exhibit a positive \OII\ line-flux value within the resolved galaxy aperture, despite an absence of an \OII\ flux detection in the initial pipeline database. While the redshift of these systems remains unchanged, their classification is revised from `lzg' to `oii'. Using this method, \nlzgtooiiswitch\ sources in the HPSC2 catalog have been reclassified as \oiis\ from an initial classification as a low-z passive galaxy. 

\subsection{\OII\ and Ly$\alpha$ Line Fluxes and Luminosities}
\label{sec:fluxes}
For the two main emission-line datasets, we include columns for their continuum-subtracted, extinction-corrected line fluxes (\texttt{flux\_oii, flux\_lya}) and corresponding luminosities (\texttt{logL\_oii, logL\_lya}), along with their respective uncertainties marked by the \texttt{\_err} subscript. This information is only relevant to sources with \OII\ and \lya\ emission lines and pertains exclusively to objects categorized with \texttt{source\_type} as `oii', `lae', or `agn'. Only the AGN with \lya-emission within the HETDEX spectral range will have reported flux values. For accessing other emission-line-flux data, users are directed to Appendix~\ref{appendix:DetInfoTable}.

The native source detection pipeline as described in \citet{Gebhardt2021} is designed to deliver continuum-subtracted line fluxes under the assumption that the source is unresolved. However, many HETDEX sources are resolved. Moreover, an additional shortcoming in the HETDEX detection procedure is that an upper bound on continuum counts is set during the search for emission lines. As a result, the line flux emitted at the central location of a bright \OII\ emitter or AGN is not reported.

As outlined in Section~\ref{sec:aperflux}, the issue is addressed for \oiis\ by fitting spatially resolved line fluxes for each system with  \texttt{z\_hetdex}$<0.5$. When the aperture flux, denoted as \texttt{flux\_aper}, exceeds the pipeline-reported \texttt{flux}, this spatially integrated flux measurement value is assigned to \texttt{flux\_oii}, and the \texttt{flag\_aper} is set to 1. Conversely, if the HETDEX pipeline's estimate of the [\ion{O}{2}] surpasses the aperture flux, the pipeline flux is preserved in \texttt{flux\_oii}, and \texttt{flag\_aper} is set to 0. If no \OII\ flux is associated with the source, the \texttt{flag\_aper} is assigned a value of $-1$ and the source is not to be classified as an `oii'.

For LAEs, it is generally acceptable to assume that their morphology resembles that of a point source, especially considering the relatively poor image quality delivered by the HET. Research focused on quantifying surface-brightness profiles and determining isophotal radii suggests that for a subset of LAEs with a signal-to-noise ratio greater than six \citep{MentuchCooper2026_HETDEX_LAN}, an exponential model with a scale length of approximately 10 kiloparsecs is a more accurate representation of LAE morphology. This improved model increases the reported Ly$\alpha$ flux by a factor of $\sim 1.3$ over the point-source calculation as reported in \citet{MentuchCooper2026_HETDEX_LAN}. However, we have chosen not to implement any corrections to adjust for this discrepancy. We also have opted to not provide spatially resolved flux values for extended \lya\ emitters; these measurements can be found in \citet{MentuchCooper2026_HETDEX_LAN}.

As mentioned above, some AGNs are often bright enough so that their continuum counts are above the maximum allowed for an emission-line search.  In these objects, the Ly$\alpha$ line flux is measured by subtracting the surrounding continuum and fitting a multicomponent Gaussian model to the emission line \citep{liu2025}. These line-flux values are reported in \texttt{flux\_lya} in HPSC2. For AGNs whose Ly$\alpha$ emission lies outside the spectral range of the VIRUS spectrographs, no flux is reported. Readers that desire more advanced line-flux values are referred to the AGN catalogs created from HETDEX data release HDR4 \citep{liu2025}.

\subsubsection{Extinction Correction Due to Dust Attenuation}
All flux values, unless designated with an \texttt{\_obs} subscript have been extinction corrected. The Python software package \textsc{dustmaps} \citep{dustmaps} provides local Milky Way Dust reddening values for each source's coordinates as measured by \citet{Schlegel1998}. The software returns the locally measured color excess value, $E$(\bv), based on a source coordinate. We assume the ratio of $V$-band extinction, $A_V$, to color excess, $E$(\bv), to be $R_V = 3.1$ and apply a factor of 2.742 to measure the local $V$-band extinction as $A_V = 2.742\times E(\bv)$ according to the re-calibration using \sdss\ stars of the \citet{Schlegel1998} maps by \citet{Schlafly2011}. However, we note that large variations of this re-calibration are seen in the Milky Way interstellar medium \citep{2025Sci...387.1209Z, lee2025}. $A_V$ values range from 0.01 to 1.44 with a median value of 0.04.  Extinction corrections of line fluxes are applied at the central wavelength of each catalogued emission line according to the $R_V=3.1$ extinction curve of \citet{Fitzpatrick1999}, implemented using the open source Python software \textsc{extinction}\footnote{https://github.com/kbarbary/extinction}.

\subsection{ML/AI Assisted Classification}
\label{sec:MLAI}

Machine Learning (ML) and Artificial Intelligence (AI) tools provide efficient methods for data classification, and the HETDEX collaboration has explored several such approaches. Unsupervised learning has proven valuable for identifying supernovae \citep{Vinko2022} and AGN \citep{valentina2023} within the rich HETDEX dataset, while also offering interactive methods for artifact removal and sample cleaning.

For HPSC2, ML/AI methods have been most successful in identifying and removing artifacts from the raw detection database and improving the low-$S/N$ emission-line sample. In this section, we describe three approaches that have proven particularly effective and that are incorporated into HPSC2 and the PDR1 data cube mask model.

\subsubsection{Rapid Automatic Image Categorization}
\label{sec:raic}
Every emission-line and continuum detection in the HETDEX catalog is passed through a classification workflow using the RAIC Labs\footnote{\url{https://raiclabs.com} classification platform (RAIC: Rapid Automatic Image Categorization), developed as a scalable visual AI system for analyzing large image datasets using semi-supervised learning and human-in-the-loop interaction. This system provides a user-friendly interface for semi-automated labeling of astronomical detections, enabling rapid and scalable classification of large survey datasets without requiring fully labeled training sets.}

For each emission-line detection, we generate a narrowband cutout, $20\arcsec$ on a side, with the wavelength dimension collapsed over $\pm 2\sigma$ around the fitted central wavelength of the line, where $\sigma$ is taken from the Gaussian profile fit produced by the HETDEX pipeline. No continuum subtraction is performed. For continuum detections, cutouts of the same size are produced by collapsing the spectrum over the rest-frame wavelength range $4000$--$5000$\,\AA. In both cases, the detection of interest is centered within the cutout. These narrowband cutouts are uploaded to the RAIC Labs classification platform, which projects the image set into a low-dimensional feature space where unsupervised learning naturally groups detections with similar morphology. The tool provides an interactive GUI in which users can visually explore this feature space and interactively choose selections to assign labels to groups of sources simultaneously. This approach enables efficient labeling of large numbers of detections with consistent morphological classifications.

\begin{figure*}[t]
    \centering
    \begin{tabular}{cc}
        \includegraphics[width=\textwidth]{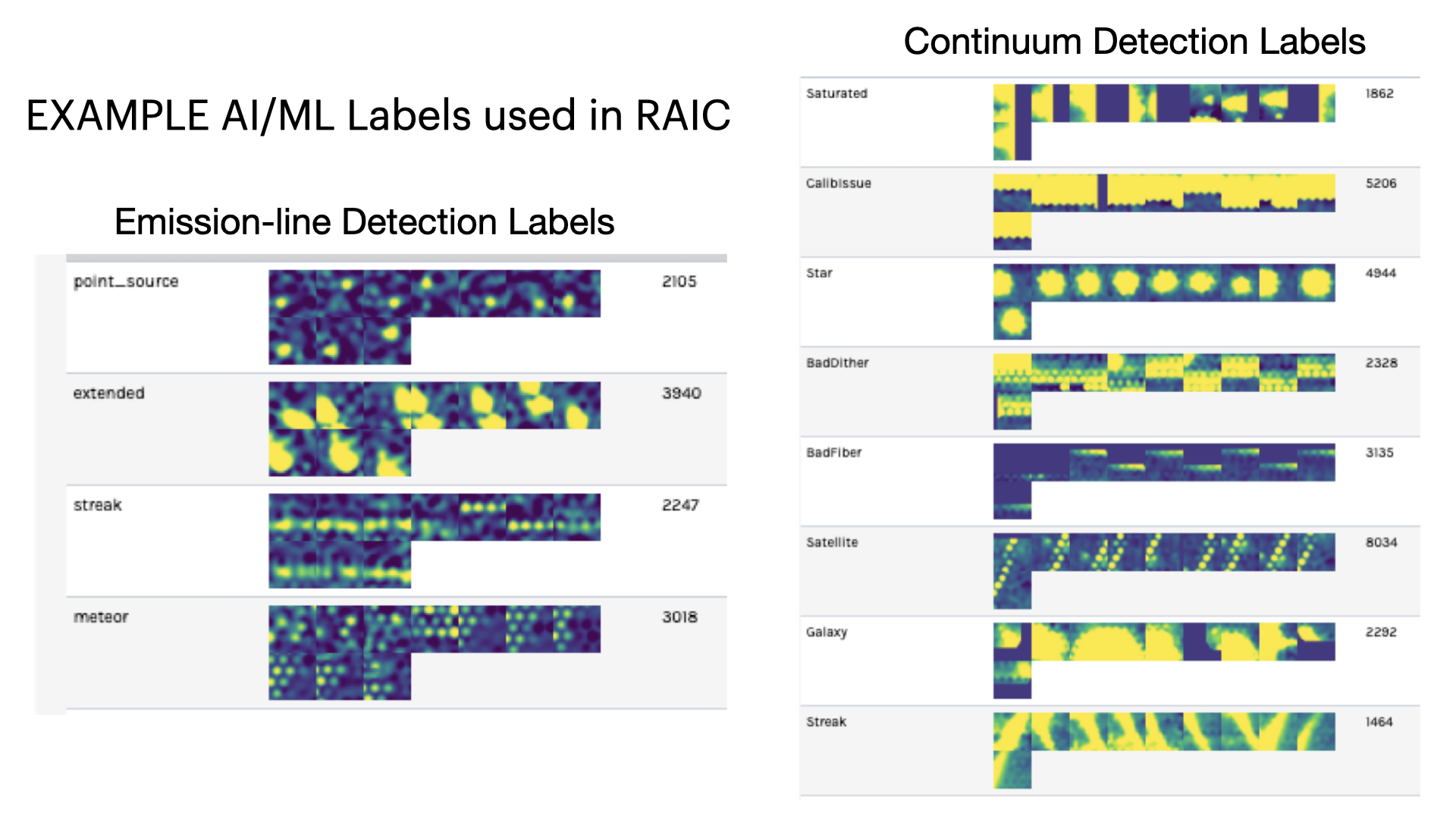}\\
%        \includegraphics[width=0.48\textwidth]{line_labels.png} &
%        \includegraphics[width=0.48\textwidth]{continuum_labels.png} \\
    
%    (a) Emission-line detection labels & (b) Continuum detection labels
    \end{tabular}
    \caption{RAIC Labs classification label sets used for HETDEX detections. 
    Panel (a) shows the classes applied to emission-line cutouts, while panel (b) 
    shows the classes for continuum cutouts.}
    \label{fig:raic_labels}
\end{figure*}

The resulting label sets are shown in Figure~\ref{fig:raic_labels}. For emission-line detections, the labeled sets are point source, extended object, streak, and meteor. For continuum detections, the categories include saturated object, calibration issue, star, bad dither, bad fiber, satellite, galaxy, streak, low counts, and faint star. Each category is built up to include at least 1000 example labels.

Once a sufficient number of labels has been collected via the RAIC Labs interactive GUI, the system trains its internal classification model to apply those labels to previously unlabeled detections. Each source in the HETDEX catalog (emission‐line or continuum) is passed through the relevant classification model, and is assigned a likelihood score, set between 0.0 and 1.0, for each label for each detection. This process means that any HETDEX detection can be assigned multiple labels. Those that are above the threshold classification score of 0.5 are assigned that label. 

The labels for ``meteor'' and ``satellite'' are used to help identify contaminants in the catalog and create the masking for each category, respectively (see Sections~\ref{sec:meteor} \& \ref{sec:sat}.) From the continuum labels, we flag the labels ``calibissue'', ``baddither'', ``badfiber'' from the catalog when their label score is greater than 0.3. From the emission-line labels, we use the flag and remove any detections with the ``streak'' label and a score greater than 0.35. 

From our internal quality-controlled catalog 113,954 detections are flagged by RAIC. We do not include these in any public catalogs. While many of these artifacts are found through our other automated flagging procedures, 41,002 are only found using the RAICLabs AI classification platform; demonstrating the success of this software as a tool for efficiently labeling and classifying HETDEX detections. These detections are masked in the data cubes under the BADDET bitmask described in Section\,\ref{sec:baddet}.

\subsubsection{Citizen Science + Unsupervised Learning}
\label{sec:dee_tsne}

Another method to reduce false positives due to noise and artifacts is to combine large-scale citizen science vetting from the NASA Zooniverse project \textit{Dark Energy Explorers}\footnote{\url{https://www.zooniverse.org/projects/erinmc/dark-energy-explorers}} (DEE) with unsupervised learning on the detection spectra \citep{house2023}. DEE presents compact “mini” views of each detection produced by HETDEX custom software \elixer\ and collects a minimum of ten binary classifications per source. This result is a per-object DEE probability, $p_{\rm DEE}$, that a candidate is real, as opposed to an artifact. Since its launch in February 2021, DEE has amassed over 4M classifications from over $\sim$22,000 volunteers.

As described in \citet{house2023}, we then map spectra into a low-dimensional space using t-distributed stochastic neighbour embedding (t-SNE), operating on a $\pm50$\,\AA\ window centered on the emission-line wavelength. We found that the labels alone lead to too many real sources being assigned low $p_{\rm DEE}$ scores; however the combination an object's $p_{\rm DEE}$ and its location in t-SNE space -- particularly in regions where the majority of detections also have low $p_{\rm DEE}$ values -- was very successful at labeling artifacts with $S/N>5.5$. 

As of this writing, the final set of HDR5 detections with $S/N>5.5$ is still being evaluated, but we have flagged all candidates that satisfy the exclusion conditions described in \citet{house2023} and removed them from the public catalog. These objects are marked \texttt{flag\_best==0} in our internal catalog but are not retained in the public release. In total, 28,841 detections are flagged and removed. Many of these are also identified by the masking procedure described in Section\,\ref{sec:bitmask}. For those missed by our primary masking methods, the remaining sources are masked in the data cubes with the cubic BADDET mask.

\subsubsection{CNN Classifier: \texttt{p\_cnn}}
\label{sec:cnn}

The largest challenge for HETDEX and its untargeted emission-line search is validating the low $S/N$ emission-line detections found by the HETDEX reduction pipeline. Standard quality flags and conservative $S/N$ thresholds ($>5.5$) are effective at reducing contamination but also exclude a large fraction of genuine LAEs, limiting the source density needed for cosmological analyses. 

Extending the threshold down to the pipeline limit ($S/N=4.8$) requires an additional layer of classification to suppress residual noise and artifacts. To address this issue, \citet{Mukae2025} developed a convolutional neural network (CNN) tailored for the HETDEX two-dimensional spectra. The model was trained on $\sim4300$ emission-line candidates from the COSMOS field, with validation from HETDEX repeat observations, ancillary spectroscopy, and multiple narrowband LAE surveys and citizen science classifications. By incorporating both the spatial and spectral structure of detections, the CNN achieves $\sim90\%$ accuracy overall and maintains strong performance ($86\%$) in the low-S/N regime. Visualization methods confirm that the network identifies smooth, centrally concentrated features as true LAEs, while rejecting irregular or noisy patterns.  

Applied to the full catalog, the classifier extends the usable $S/N$ range, suppresses spurious sky-line–induced redshift spikes in the blue region of the spectrum, and recovers the target LAE density required for cosmological measurements. This approach demonstrates that domain-specific deep learning methods can enhance the reliability of blind spectroscopic surveys and improve the scientific yield of HETDEX.  

The CNN score, \texttt{p\_cnn}, is provided in HPSC2 and the supplemental \textit{Detection Information Table} in Appendix~\ref{appendix:DetInfoTable}. It is measured for the full LAE sample of $\sim1.6$\, million LAE candidates as well as a subset of $\sim400,000$ faint \oiis\ with \hetg$>22$ which are found to be well described by this CNN classifying approach.

\subsubsection{Random Forest Classifier: \texttt{p\_conf}}
\label{sec:rf}

We train a single, compact Random Forest (RF) classifier to distinguish high-confidence LAEs from low-confidence detections in the COSMOS Legacy field. This field benefits from multiple HETDEX observations, which allow for confirmation of genuine LAEs, and provides a set of high-confidence LAEs suitable for use in supervised machine learning methods.

Training is restricted to COSMOS LAE emission-line detections and excludes any objects previously classified as AGNs\null. It is further limited to faint objects with \hetg$>22$. Classification labels are binary, with 1 denoting a high-confidence LAE candidate and 0 denoting a low-confidence LAE candidate. It is important to note that verifying that a HETDEX detection is real is much more straightforward than proving the candidate is noise, unless the object exhibits characteristics of an obvious artifact. Because HETDEX is dominated by low-$S/N$ emission lines, uncertainties in the noise model lead to detections with questionable validity.

High-confidence classifications originate from several sources. HETDEX observed some regions of the COSMOS Legacy field multiple times; any LAE detected in a second observation beyond its original detection is considered a validated source. We also include objects confirmed through positional and redshift and/or spectral matches to 
other spectroscopic programs \citep[via the compilation by][]{khostovan2025}, or deep narrowband surveys. Included in the latter category are SILVERRUSH \citep[$z=2.2$ and $3.3$;][]{Kikuta2023}, ODIN \citep[$z=2.4$ and $3.1$;][]{Firestone2024}, and the SC4K narrow- and medium-band surveys \citep[$z\sim2.2$–$3.4$;][]{Sobral2018}. To increase sample size and better match the distribution of HETDEX LAEs, we also include sources classified as “Real” in the citizen science project {\it Dark Energy Explorers} \citep{house2023}\footnote{\url{https://www.zooniverse.org/projects/erinmc/dark-energy-explorers}} with a confidence score of $p_{\rm DEE}\geq0.8$. In total, 2935 sources are assigned real. The 4,582 low-confidence sources come exclusively from DEE with $p_{DEE}\le0.1$.

We explore a number of catalog columns as possible feature vectors for our LAE training set. Ultimately, we find that a compact feature set performs well, relying only on parameters measured directly from the detection pipeline:

\[
\mathbf{x} = \{\texttt{wave},\, \texttt{sn},\, \texttt{chi2},\, \texttt{linewidth},\, \texttt{continuum}\}.
\]

where \texttt{wave} is the wavelength of the line, \texttt{sn} is the pipeline's signal-to-noise, \texttt{chi2} is the reduced $\chi^2$ value of the Gaussian line-fits to the possible feature, \texttt{linewidth} is the standard deviation for the line-width Gaussian, and \texttt{continuum} is the pipeline's fitted value for the continuum underlying the emission line.

We use \textsc{scikit-learn}’s \texttt{RandomForestClassifier} (version~1.5.2) with standard settings \citep{scikit-learn}, including \texttt{n\_estimators=100}, \texttt{criterion=`gini'}, and \texttt{max\_features=`sqrt'}. The classifier’s output is a probability-like score, \texttt{p\_conf}, assigned to each detection in our catalog. As with the CNN classifier, \texttt{p\_conf} is computed for both the full LAE sample and for a subset of fainter \oiis. Our Random Forest model achieves an accuracy of 0.89 and an area under the receiver operating characteristic curve (AUC) of 0.95 when trained on COSMOS repeat detections and ancillary spectroscopic matches. It reliably separates high-confidence LAEs from low-confidence detections, with most predictive power supplied by the \texttt{sn} (0.303) and \texttt{continuum} (0.286) features. The remaining features—\texttt{wave}, \texttt{linewidth}, and \texttt{chi2}—provide smaller but meaningful contributions, with relative importances of 0.189, 0.115, and 0.107, respectively.

\subsection{LAE Confirmation and Best LAE Selection}

\begin{figure}[t]
\centering
% done in stampede2/cosmos/calibration/hypothesis-testing-cosmos20251006.ipynb
\includegraphics[width=0.5\textwidth]{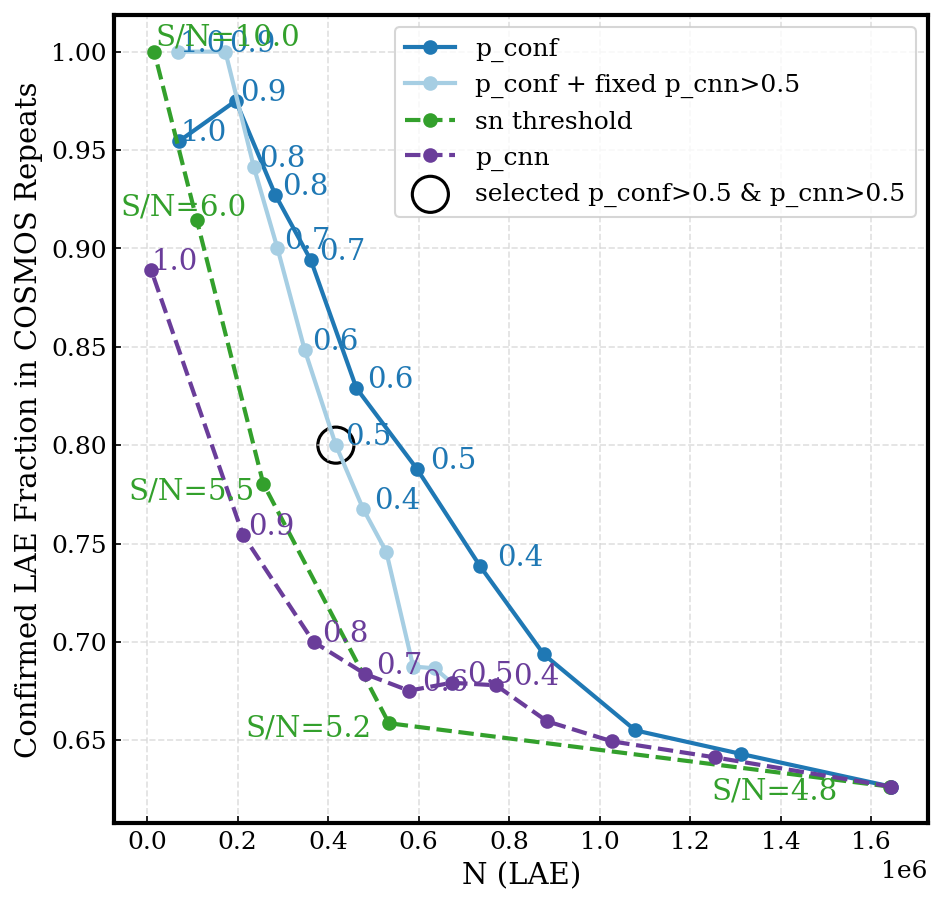}
\caption{The $y$-axis represents the fraction of LAEs that are confirmed from multiple methods (repeat HETDEX observations, ancillary spectroscopic and narrowband confirmation). The $x$-axis indicates total LAE count from the full LAE catalog based on the plotted thresholds. We consider different confirmation rates for various LAE selection cuts based on a $S/N$ threshold cut (green curve), and threshold cuts for our ML/AI classifiers based on a convolution neural network (CNN) (\texttt{p\_cnn}; see Section\,\ref{sec:cnn}) and a random forest (RF) classifier (\texttt{p\_conf}; see Section\,\ref{sec:rf}). This figure visualizes the trade-off between purity (confirmation rate) and yield (total LAEs) for RF-only, RF+CNN, S/N-only, and CNN-only operating strategies. The supplemental table in Appendix\,\ref{appendix:DetInfoTable}  contains the full sample, while the main catalog, HPSC2, contains a limited more robust combined sample selection of \texttt{p\_conf}$>0.5$ and \texttt{p\_cnn}$>0.5$ indicated in the black open circle selected. 
}
\label{fig:pconf_vs_nlae}
\end{figure}

To determine the optimal thresholds that maximize the LAE sample size and minimize false-positive contamination, we take a systematic evaluation approach that is demonstrated in Figure\,\ref{fig:pconf_vs_nlae}. This figure shows how the confirmation rate of a subset of LAEs from the COSMOS Legacy field varies with a number of different selection criteria thresholds. As stated above, it is difficult to prove with certainty that a low $S/N$ emission line detection is not real unless it is visually verified as an artifact: when the completeness fraction is below 50\%, it can take a prohibitively large number of observations to say with $>90$\% confidence that a candidate flagged by the HETDEX pipeline is a false detection. Conversely, it is relatively straightforward to prove an object is real.  If the object is found in more than one HETDEX observation, or it has already been cataloged by an independent spectroscopic or narrowband survey, then there is no question that the object exists. Thus, instead of plotting the false-positive rate on the $y$-axis, we plot the confirmation rate, which can be considered an upper limit on the inverse of the false-positive rate.

The key to measuring the confirmation rate is to select a sample in which confirmation is possible. While the entire COSMOS field has a large number of spectroscopic redshift sources, provided by the compilation by \citet{khostovan2025}, and narrowband sources that overlap with the HETDEX LAE population, the majority of HETDEX objects will not be found in ancillary studies. Conventional spectroscopic surveys, which require a broadband detection to identify candidates for follow-up redshift measurements, generally target brighter galaxies (with consequently larger stellar masses) than the average HETDEX LAE\null.  Moreover, while the physical properties of the LAEs identified in deep narrowband surveys, such as SILVERRUSH, ODIN, and SC4K, are similar to those of HETDEX LAEs, these programs target very narrow redshift slices of the Universe. The parameter space explored by HETDEX is quite unlike any other LAE survey except for the small coverage ($\sim 20\arcmin$) provided by the MUSE COSMOS sample (i.e. \citealt{Urrutia2019}). Operating with longer exposure times and reaching deeper flux sensitivity, this program is too limited in size (and redshift overlap) to provide useful comparison.

The most effective method to confirm a HETDEX LAE is by using HETDEX itself.  Unfortunately, due to the large variation in observing conditions, the flux sensitivity of any given observation can be vastly better than another. One repeat observation alone is generally not enough to confirm. Most HETDEX LAEs straddle the 50\% completeness limit in a typical HETDEX observation and in that regime, it takes at least five independent observations to excluded the reality of a detection with 95\% confidence.  By design, we limit the COSMOS confirmation sample to those observations that contain at least eight overlap apertures. This results in a small set of 414 LAEs which have the potential to be confirmed by repeat HETDEX observations as well as ancillary spectroscopic and narrowband surveys. Future dedicated HETDEX validation observations at the HET/VIRUS (obtained by PI Laurel Weiss in 2025) will expand this statistical baseline.

For any threshold on a classifier score, we define the \emph{Confirmed LAE Fraction} as
\begin{equation}
    f_{\rm conf}(\mathrm{threshold}) =
    \frac{N_{\rm confirmed}(> \mathrm{threshold})}
         {N_{\rm total}(> \mathrm{threshold})},
\end{equation}

where the denominator is the number of LAEs in the confirmation sample whose classifier scores exceed the threshold (either a selection-cut based on $S/N$, \texttt{p\_cnn} or \texttt{p\_conf} or a combination of these), and the numerator is the subset of those LAEs that are confirmed by repeat HETDEX observation or by ancillary COSMOS catalogs in that same sample.

In Figure~\ref{fig:pconf_vs_nlae}, the solid dark blue curve shows the confirmation fraction for different \texttt{p\_conf} thresholds, described in Section\,\ref{sec:rf}, as a function of the total number of LAEs included in the full HETDEX catalog sample above the indicated threshold; this is annotated with text for clarity. As expected, for all metrics, we see a trend where higher thresholds correspond to an increased \emph{Confirmed LAE Fraction} (and consequently less contamination from false positives), indicating that the use of more stringent criteria results in a more reliable selection of LAEs. However, this comes at the cost of decreasing the overall sample size, as fewer candidates meet the stricter selection criteria. For example, a threshold of \texttt{p\_conf}=0.5 increases the confirmation fraction from 62\% to 80\% but reduces the sample of LAE candidates from 1.6\,million candidates to 600\,K LAE candidates. We therefore adopt \texttt{p\_conf}$=0.5$ as a practical compromise between confirmation fraction and sample size, as higher thresholds yield only modest additional improvements in confirmation while rapidly reducing the number of LAE candidates available for statistical studies.

Variation based on thresholds for the CNN classifier, \texttt{p\_cnn}, described in Section\,\ref{sec:cnn}, is shown in the dashed purple line. Confirmation of the sample increases as the threshold for \texttt{p\_cnn} is raised, but the sample size decreases significantly. As the curve plateaus at \texttt{p\_cnn}$\sim0.3$--$0.7$, the confirmation rate does not indicate a single obvious threshold from this metric alone. In \citet{Mukae2025}, additional criteria are explored, including the recovery of known LAEs in the DESI–HETDEX validation sample \citep{landriau2025} and ancillary COSMOS catalogs. These analyses show that adopting a CNN threshold of \texttt{p\_cnn}$=0.5$ provides a practical balance between completeness and contamination, recovering the majority of confirmed LAEs while maintaining a catalog size close to the survey’s target LAE number density. Because the CNN score uncertainties are typically $\sim0.2$, sources with scores in the range $\sim0.3$--$0.7$ are intrinsically ambiguous, making a threshold near 0.5 a natural dividing point between high- and low-confidence classifications.

To compare against alternative selection criteria, we also compute (i) a combined RF + fixed value CNN sequence requiring a fixed $\texttt{p\_cnn}=0.5$ alongside varying \texttt{p\_conf} (light blue), (ii) pure $S/N$ cuts at $S/N\in\{4.8,5.2,5.5,6.0,10.0\}$, and (iii) pure CNN cuts at thresholds from $0$ to $1$.

It is crucial to strike a balance between maximizing the confirmation rate and preserving a sufficiently large sample for statistical analysis. In this release, we adopt thresholds of \texttt{p\_conf}$\ge0.5$ and \texttt{p\_cnn}$\ge0.5$ to provide a robust sample while maintaining a sufficiently large catalog for statistical analyses. The choice of \texttt{p\_conf}$=0.5$ reflects a balance between confirmation fraction and sample size, as higher thresholds yield only modest improvements in confirmation while rapidly reducing the number of LAE candidates. The CNN threshold of \texttt{p\_cnn}$=0.5$ is motivated by the analysis of \citet{Mukae2025}, which shows that this value provides a practical balance between completeness and contamination while preserving the survey’s target LAE number density. These values reduce the LAE sample from 1,632,604 to \nlae\ and the \OII\ sample from 559,544 to \noii\null.

Future work will involve exploring alternative selection methods and thresholds to optimize these criteria. For users who are interested in exploring a less stringent selection criteria, we provide the full suite of LAE candidates in the \textit{Supplemental Detection Information Table} in Appendix~\ref{appendix:DetInfoTable}. Candidates at lower \texttt{p\_conf} and \texttt{p\_cnn} values are included. This flexibility allows future users to tailor their analyses based on specific scientific goals.

\section{Public Data Model and Access}
\label{sec:datamodel}

HETDEX public data products are found at the data mount hosted by the Texas Advanced Computing Center (TACC) at The University of Texas at Austin at \url{https://web.corral.tacc.utexas.edu/hetdex/HETDEX/pdr/}. Additional information and data hosting mirrors are provided at \url{https://hetdex.org/data-results/}.

\begin{figure*}[t]
    \centering
    \includegraphics[width=\linewidth]{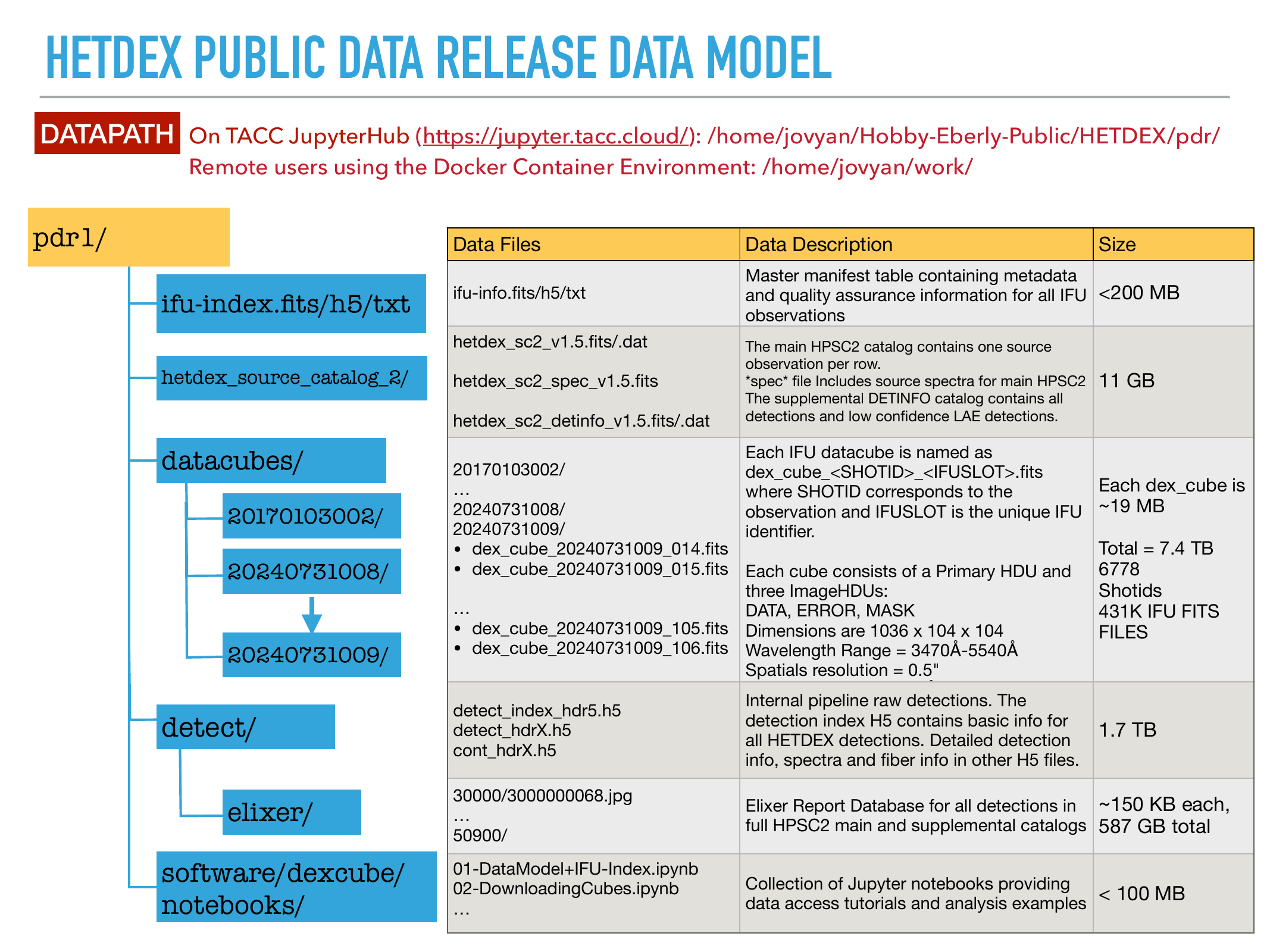}
    \caption{Structure of the HETDEX Public Data Release~1 (PDR1) IFU data cubes. 
Each cube is stored in a FITS file named \texttt{dex\_cube\_<SHOTID>\_<IFUSLOT>.fits}, 
where \texttt{SHOTID} identifies the observation and \texttt{IFUSLOT} the IFU. 
Cubes have dimensions of $1036 \times 104 \times 104$ (wavelength $\times$ spatial $\times$ spatial), 
with a spectral resolution of 2\,\AA\ over the range $3470$–$5540$\,\AA\ and a spatial sampling of $0\farcs5$. 
Each file contains a Primary HDU and three Image HDUs (\texttt{DATA}, \texttt{ERROR}, \texttt{MASK}), 
with a typical size of 19\,MB, totaling $\sim$7.4\,TB for \nifu\ IFU data cubes across \nhetdexobs\ observations.}
    \label{fig:datamodel}
\end{figure*}

A schematic of the PDR1 data model is shown in Figure~\ref{fig:datamodel}. The primary data products are:

\vspace{10pt}

\begin{itemize}
    \item \textbf{IFU metadata table}: The \texttt{ifu-index} file (FITS/HDF5/ASCII) provides the master manifest with astrometric, observational, and quality assurance information for every IFU.
    \item \textbf{IFU data cubes}: \nifu\ FITS files, each corresponding to a single IFU observation (\texttt{dex\_cube\_<SHOTID>\_<IFUSLOT>.fits}), with dimensions $1036 \times 104 \times 104$ (wavelength $\times$ spatial $\times$ spatial). Each cube covers 3470--5540\,\AA\ at 2\,\AA\ sampling, with $0\farcs5$ spatial resolution, and contains \texttt{DATA}, \texttt{ERROR}, and \texttt{MASK} extensions. The WCS header information must be used for orientation as the cubes are on a grid that is aligned with the original telescope rotation angle.
    \item \textbf{Source catalogs}: .fits/.dat files containing spectra and properties for all detected sources, described in detail in Section~\ref{sec:catalog} and Appendix~\ref{appendix:DetInfoTable}
    \item \textbf{Raw Detection Databases}: A record of the raw emission line and continuum detection search output. A detailed description is given in Appendix~\ref{appendix:detections}.
\end{itemize}

\vspace{10pt}

The data release is 9.8\,Tb in size with the data cubes making up the bulk of storage size at 7.4\,Tb. For the typical user, downloading the full data release is not practical so we provide example notebooks on how to query the database based on coordinates or observing quality and download only the data cubes of interest. %These data can then be accessed and analyzed on a local machine.
We first describe the data products that make up the PDR1 data release and then provide instructions on how to access the data.
%The most efficient way to access the data is through the HETDEX JupyterLab cloud computing environment offered by TACC\null. Data products can be accessed through Jupyter notebook tutorials and downloaded to a local machine for further analysis. We describe the system below in Section\,\ref{sec:jupyter}.

\subsection{IFU Index Table}
\label{sec:index_table}

The file \texttt{ifu-index.fits} (also available in .h5/.txt format) is a master look up table for each IFU data cube observation made by HETDEX. 

Table~\ref{tab:ifu_schema} describes the columns from the IFU Index table. Each row in the table represents a single IFU-observation, with the parent observation labeled as \texttt{shotid}. %The observation ID, labeled \texttt{shotid}, identifies the observation.
A HETDEX observation will have between 14 and 78 IFU observations, identified by column \texttt{ifuslot}. The \texttt{shotid}/\texttt{ifuslot} combination provides the unique identifier of an HETDEX IFU data cube.

The central sky coordinate of each IFU is listed in Table~\ref{tab:ifu_schema}. Users can search for data cubes that overlap within an arcminute of this coordinate and then check the WCS of the data cube for exact coverage. An example query is given in Section~\ref{sec:query}. Observational quality assessment values for seeing (FWHM), and transparency (\texttt{response\_4540}) are tabulated to allow for querying of desired shots. We include the \texttt{fwhm\_virus} value, which is produced by the HETDEX pipeline based on VIRUS observations.

For each IFU, the fraction of the data cube flagged due to bad amplifiers, bad fibers, meteors, and satellites is reported in the aptly named columns. Of the full set of IFU data cubes, $91.6\%$ of the data remain unflagged, meaning that $8.4\%$ of the data cubes are flagged at all wavelengths. Breaking this down by source of contamination, $7.8\%$ of the total data cube volume are flagged due to bad amplifiers, $0.2\%$ due to bad fibers, $0.12\%$ due to satellites, and $0.04\%$ due to meteors. Additional wavelength-dependent masking can occur; this mask is contained in the BITMASK HDU. 

\begin{table*}[t]
    \centering
    \caption{IFU Metadata Schema}
    \begin{tabular}{llll}
    \hline
    \hline
    Column Name & Description & Data Type & Units \\
    \hline
    shotid & Shot ID integer representing a shot: DATEOBS & int64 & — \\
    ifuslot & IFU slot number (e.g., 046, 063, etc.) & str & — \\
    ra\_cen & Right Ascension of IFU center (ICRS J2000) & float32 & deg \\
    dec\_cen & Declination of IFU center (ICRS J2000) & float32 & deg \\
    flag & Fraction of IFU with useable data (1\,=\,fully useable, 0\,=\,fully flagged) & float32 & — \\
    flag\_badamp & Fraction of IFU unaffected by bad amplifier(s) & float32 & — \\
    flag\_badfib & Fraction of IFU unaffected by bad fiber(s) & float32 & — \\
    flag\_meteor & Fraction of IFU unaffected by meteor trail contamination & float32 & — \\
    flag\_satellite & Fraction of IFU unaffected by satellite trail contamination & float32 & — \\
    flag\_shot & Fraction of IFU unaffected by shot-level flagging & float32 & — \\
    flag\_throughput & Fraction of IFU unaffected by low throughput & float32 & — \\
    field & Field identifier (e.g., `dex-spring') & str & — \\
    objid & Object ID (string identifier) & str & — \\
    date & Date of the observation (YYYYMMDD) & int32 & — \\
    obsid & Observation index & int32 & — \\
    ra\_shot & RA of shot center (ICRS J2000) & float64 & deg \\
    dec\_shot & Dec of shot center (ICRS J2000) & float64 & deg \\
    pa$^{1}$ & Position angle of focal plane & float64 & deg \\
    n\_ifu & Number of active IFUs in the shot & int32 & — \\
    fwhm\_virus & Seeing FWHM from VIRUS & float32 & arcseconds \\
    fwhm\_virus\_err & Error in seeing measurement & float32 & arcseconds \\
    response\_4540 & Effective Normalized System Throughput at 4540\,\AA\ in 360\,s & float32 & — \\
    ambtemp & Ambient temperature & float32 & °C \\
    datevobs & Observation date string & str & — \\
    dewpoint & Dew point & float32 & °C \\
    exptime & Exposure time & float32 & s \\
    humidity & Relative humidity & float32 & \% \\
    mjd & Modified Julian Date & float32 & d \\
    nstars\_fit\_fwhm & Number of stars used for seeing estimate & int32 & — \\
    obsind & Observation index (unused) & int32 & — \\
    pressure & Barometric pressure & float32 & Torr \\
    structaz & Telescope azimuth & float32 & deg \\
    time & Start time of exposure & str & UTC \\
    trajcdec & TRAJDEC from telescope header & float32 & deg \\
    trajcpa & TRAJPA (position angle) from telescope header& float32 & deg \\
    trajcra & TRAJRA from telescope header & float32 & deg \\
    \hline
    \multicolumn{4}{l}{$^{1}$\,The \texttt{pa} column is the HET tracker rotation angle. The position angle east of north on the celestial sphere }\\
    \multicolumn{4}{l}{\phantom{$^{1}$\,}is given by $\theta = 360\degr - (90\degr + \mathtt{pa} + 1\fdg55)$.}\\
    \multicolumn{4}{l}{\footnotesize Note: This table is available in its entirety in machine-readable form at \url{https://hetdex.org/data-results/} and in the online journal.} \\
    \end{tabular}
    \label{tab:ifu_schema}
\end{table*}

\subsection{Catalog Data Products}

The creation of the primary catalog delivered in this release is the HETDEX Public Source Catalog 2 (HPSC2) was described in Section\,\ref{sec:catalog}. Here we describe its data model. HPSC2 provides one row per source observation and summarizes the key physical and observational properties for each object. HPSC2 includes the source position, redshift, physical size when relevant, and the dust– and aperture–corrected \OII\ or \lya\ fluxes and luminosities. It contains all necessary information for most scientific analyses. To support more specialized use cases, we also provide the full \textit{Detection Information Table}, described in Appendix~\ref{appendix:DetInfoTable}, which lists every individual HETDEX line-emission and continuum detection that passes the quality checks and line-parameter criteria in \S\ref{sec:bitmask} and \S\ref{sec:detection_cleaning}. This expanded table contains per-detection measurements and metadata for all HETDEX observations of a source and can be used to trace each HPSC2 entry back to its constituent detections. While the detailed \textit{Detection Information Table} enables advanced analyses, HPSC2 provides a streamlined and user-friendly summary appropriate for most users.

%The information in this release is presented in two separate catalog formats: the full detection catalog \textit{Detection Information Table}, with columns described in Table\,\ref{tab:det_col_info} in Appendix~\ref{appendix:DetInfoTable}, which contains information about every HETDEX line emission and continuum detection that has passed the quality checks and line parameter criteria described in \S\,\ref{sec:bitmask} and \ref{sec:detection_cleaning} respectively; and the main HETDEX Public Source Catalog 2 (HPSC2 hereafter) which contains aggregate information from the more detailed \textit{Detection Information Table} for each source observation. Both contain fundamental information on a source (position, redshift, physical size if relevant, \OII\ or \lya\ flux and luminosity where appropriate) and is repeated for each separate HETDEX observation of the source. For most users, HPSC2 will be sufficient and it is a limited, easier-to-parse summary of the expanded table of full detections.

A HETDEX source, identified by \texttt{source\_id}, is a collection of all detections at the same on-sky position combined through the detection grouping method described in Section~\ref{sec:detgroup}. If the source is observed more than once, its \texttt{source\_name} will be the same, but the \texttt{source\_id} will be different as will the reported catalog measurements. For each \texttt{source\_id} row, we report a single representative detection identifier, \texttt{detectid}, which may be matched to the \textit{Detection Information Table} for each source observation in the \texttt{detectid} column; this column corresponds to the detection member with the brightest (i.e., smallest) \hetg\ value for all sources that are not LAEs. For LAEs, the highest $S/N$ \lya\ line detection is the selected representative \texttt{detectid}. A user may search the \textit{Detection Information Table} for this representative detectid by the selecting the column \texttt{selected\_det==True}.

For sources identified as \texttt{lae} or \texttt{oii}, we provide the objects' Ly$\alpha$ or [\ion{O}{2}] line-flux in the columns \texttt{flux\_lya} and \texttt{flux\_oii}, along with the corresponding errors on these quantities. As discussed in Section~\ref{sec:aperflux}, for each low-$z$ galaxy, an aperture \OII\ line flux is measured: \texttt{flux\_aper} at \zhet. This flux is assigned as the source's \texttt{flux\_oii} if it is a positive value and the major-axis of the galaxy, as measured on broadband imaging is greater than 2\arcsec, otherwise the measured line flux comes from the native pipeline \texttt{flux} value obtained from the the line fit to the extracted spectrum of the brightest \texttt{detectid} in the source group.  Line fluxes and associated errors are converted to intrinsic \OII\ and \lya\ line luminosities (denoted \texttt{logL\_lya} and \texttt{logL\_oii}) using our best measured redshift, \zhet, and the cosmological model defined by \citet{Planck2018}.

The following files are included in this release:
\begin{itemize}
    
    \item \textit{HETDEX Public Source Catalog 2} (HPSC2, columns described in Table~\ref{tab:column_info}):
\begin{verbatim}
hetdex_sc2_vX.X.dat/.fits
\end{verbatim}
    This table consists of one row per source observation. For each source observation, it provides the source's J2000 equatorial coordinates, and redshift (\zhet). Every source is classified into one of the following source\_type options: \texttt{lae}, \texttt{oii}, \texttt{agn}, \texttt{lzg}, or \texttt{star} as described in Section~\ref{sec:classification}. For sources with either \lya\ or \OII\ line emission, the table provides the optimal measurement for the dust-corrected, aperture-corrected flux and luminosity in \texttt{flux\_lya}, \texttt{flux\_oii}, \texttt{logL\_lya}, and \texttt{logL\_oii}. 
    
    \item \textit{HPSC2 Spectra} in FITS format:
\begin{verbatim}
hetdex_sc2_spec_vX.X.fits
\end{verbatim}
    Each row in HPSC2 provides the corresponding 1D extracted spectra in a FITS file format consisting of 4 Header Data Units (HDUs). Multiple HDUs are included as listed in Table\,\ref{tab:spec_format}. The primary HDU is empty. HDU1:INFO contains a copy of the main HPSC2 catalog. At the same row index for each source in this table, HDU2:SPEC and HDU3:SPEC\_ERR contain the aperture corrected, 1D PSF-weighted extinction-corrected spectra and their associated uncertainties in \fluxden, computed according to the procedure outlined in Section~\ref{sec:fluxes}. The final HDU4:WAVELENGTH is a 1036 array corresponding to the spectral dimension in \AA\null. All spectra have the same spectral range from 3470\,\AA\ to 5540\,\AA\ in steps of 2\,\AA.
    
    \item \textit{Supplemental Detection Information Table} (columns described in Table~\ref{tab:det_col_info}): 
\begin{verbatim}
hetdex_sc2_detinfo_vX.X.dat/.fits 
\end{verbatim}
    This table contains specific information for every quality-controlled line emission and continuum detection. Every emission-line detection row contains all parameter information, including the emission-line's observed wavelength, its fitted parameters and measured flux. If the observed wavelength corresponds to a commonly found spectral species\footnote{\url{http://classic.sdss.org/dr6/algorithms/linestable.html}} at redshift \zhet, the species is indicated in the column \texttt{line\_id}. There are also several columns related to imaging counterpart matches, redshift assignments, and emission-line classification as found by \elixer\ \citep{Davis2023}. Also included are a number of columns containing details about the specific observation, the instrument, and the detection grouping parameters. A full column description is provided in Table~\ref{tab:det_col_info}. Detailed information concerning this catalog is provided in Appendix\,\ref{appendix:DetInfoTable}.

\end{itemize}

\begin{table*}[t]
    \centering
    \caption{Format of the HETDEX Public Source Catalog 2 (HPSC2) (columns described in Table\,\ref{tab:column_info}) Spectra FITS file}
    \label{tab:spec_format}
    \begin{tabular}{llcll}
    \hline
    HDU Name &   Type  &   Dimensions &  Description \\
    \hline
    0:PRIMARY       & PrimaryHDU       &    &  \\     
    1:INFO       &    BinTableHDU     &   7367R x 27C   & Source information for each catalog source, \\
    & & & one row per source observation. \\
    2:SPEC        &   ImageHDU         &   ($1036\times$\nsource) &  Extinction- and aperture-corrected, PSF-weighted 1D spectrum \\
    & & & at the source's representative \texttt{detectid} \\
    & & & at position \texttt{RA\_det}, \texttt{DEC\_det} in units of \fluxden. \\
    3:SPEC\_ERR     & ImageHDU         &  ($1036\times$\nsource)   &  Uncertainty in SPEC.   \\
%    4:SPEC\_OBS      & ImageHDU         &   ($1036\times$\nsource)  & Aperture-corrected, 1D PSF-weighted spectrum \\
%    & & & at \texttt{detectid} in units of \fluxden.  \\
%    5:SPEC\_OBS\_ERR   & ImageHDU         &   ($1036\times$\nsource) &  Uncertainty in SPEC\_OBS in \fluxden.   \\
%    6:APCOR         & ImageHDU         &   ($1036\times$\nsource)  &  Aperture correction applied to spectrum and catalogue flux values\\
    4:WAVELENGTH & ImageHDU & (1036,) & Wavelength array from 3470\,\AA\ to 5540\,\AA\ in 2\,\AA\ bins. \\
    \hline
    \end{tabular}
    \par\vspace{2pt}
    \raggedright \textbf{Note.} This file contains the full HPSC2 table (Table~\ref{tab:column_info}) in HDU 1 (also available in a simple .dat ASCII format). The spectrum provided is a 1D PSF-extracted spectrum at the representative \texttt{detectid} location. Different spectral extractions can be obtained from HETDEX data cubes directly.
\end{table*}

\subsection{ELiXer Reports}

An essential diagnostic for HETDEX detections is provided by the \elixer\ reports that are produced for every detection in the quality-controlled HETDEX catalog. A full description of an \elixer\ detection report is provided in the Appendix of \citet{Davis2023}. These reports provide a summary of ancillary imaging data at the location of the detection. Small cutouts of the amplifier array containing the four highest weighted fibers that contribute to the PSF-extraction of the detection. Also included is a full plot of the HETDEX PSF-extracted spectrum, associated error with an expanded view of the spectral region of the detection. These reports are used to visually validate HETDEX detections. We include in this release \ndettable\ \elixer\ reports for all detections in full supplemental HETDEX catalog described in Appendix~\ref{appendix:DetInfoTable}. This includes every detection in the main HPSC2 catalog. 

They can be accessed individually through the HETDEX data mount at \url{https://web.corral.tacc.utexas.edu/hetdex/HETDEX/pdr/pdr1/detect/elixer/}. The files are organized based on the first five digits of their detectid into directories to allow for easier navigation. For example, \texttt{detectid}=4022202460 is located at: \url{https://web.corral.tacc.utexas.edu/hetdex/HETDEX/pdr/pdr1/detect/elixer/40222/4022202460.jpg} 

\subsection{Access Options}
\label{sec:access}

There are two primary ways a user can access HETDEX public data. (1) Use the IFU index table to determine dexcubes of interest and then download these cubes locally for personal analysis, or (2) utilize a public JupyterHub access point. In this section we outline these two points of access and present some basic data access examples including making line flux maps at a spectral region of interest and doing 1D spectral extractions.

\subsubsection{Remote Access}
\label{sec:remote}

For users without direct HETDEX JupyterHub access, data cubes may be downloaded manually. First you will need to download the IFU index file from \url{https://web.corral.tacc.utexas.edu/hetdex/HETDEX/pdr/pdr1/ifu-index.fits}. Then files can be downloaded through a terminal using \texttt{wget}.

\begin{lstlisting}[style=arxivcode]
$ wget --cut-dirs=4 -nH -x -i {url}
\end{lstlisting}

Where url = \url{http://web.corral.tacc.utexas.edu/hetdex/HETDEX/pdr/pdr1/data cubes/20190405020/dex_cube_20190405020_034.fits} for example. The options here ensure the same file structure as is located on the host directory system so that only minor edits of the base path are require in the tutorial notebooks.

The script can be easily adapted to download many cubes by looping
over lists of \texttt{shotid} and \texttt{ifuslot} values. For examples of downloading data cubes, including batch parallel downloading options, see \url{https://github.com/HETDEX/dexcube/blob/main/notebooks/02-DownloadingCubes.ipynb} and \url{https://github.com/HETDEX/dexcube/blob/main/notebooks/12-BatchDownloads-ForRemoteUsers.ipynb}.

\subsubsection{JupyterHub Access}
\label{sec:jupyter}

\begin{figure*}[t]
\centering
    \includegraphics[width=0.9\textwidth]{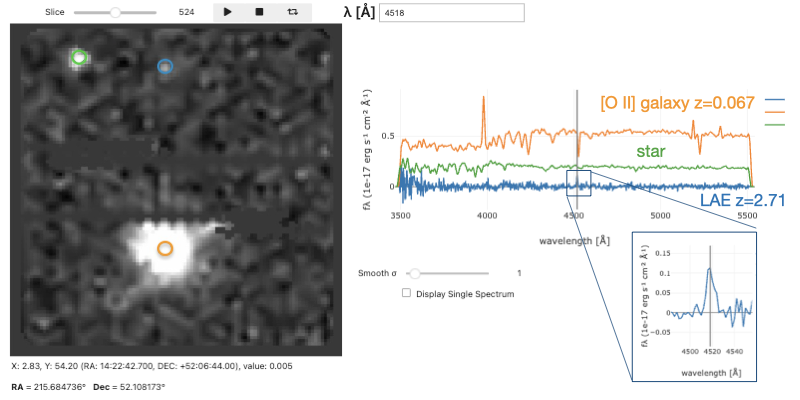}
\caption{Example HETDEX IFU data cube visualization using the \texttt{CubeWidget} 
tool for \texttt{shotid = 20190405020}, \texttt{ifuslot = 034}. 
The left panel shows a 2\,\AA\ wide spectral region at $\lambda=4518$\,\AA, overlaid with circular 
apertures marking three sources. The right panel displays the extracted spectra: 
HETDEX\_J142245.70+520636.2, an LAE at $z=2.717$ 
(blue; detectid 3002870541, $S/N$=8.7); HETDEX\_J142244.63+520702.8, an [O\,\textsc{ii}] galaxy at $z=0.067$ (orange; detectid 3002870503, $S/N$=64.5); and HETDEX\_J142244.36+520629.4, a foreground star (green; detectid 3090069113). A zoom-in cutout panel highlights the \lya-emission line feature at $\lambda=4517.5$\,\AA. There is a noteable masked region in this wavelength slice, caused by a known artifact that occurs at 4500\,\AA\ due to CCD readout present in many HETDEX data cubes. Despite this issue, the LAE is clearly detected in both the slice and its spectrum.}
\label{fig:cubewidget}
\end{figure*}

A public JupyterHub is available for all public users through the Texas Advanced Computing Center at \url{https://jupyter.tacc.cloud/}. The JupyterHub provides access to the full data release and is the most efficient way to perform bulk actions on the survey data. For example, one can perform searches on a catalog of coordinates and extract spectra, create narrowband like images and run other analysis packages. We also include an interactive widget, CubeWidget, to explore HETDEX IFU cubes. An example of CubeWidget is shown in Figure~\ref{fig:cubewidget}.

A TACC account is required for access but users do not need to be associated with a project allocation. More details on JupyterHub access is found at \url{https://hetdex.org/data-results/}. No persistent storage is offered to public users. Tutorial notebooks can be run and edited but when the user's server is shut down, the server will reboot back to the original container. It is strongly recommended to download your work and any data products that you produce.

\subsection{Examples}

The best approach for a user to become familiar with the data model is to run the notebooks found on the \texttt{dexcube} repository\footnote{\url{https://github.com/HETDEX/dexcube}}. These tutorials introduce the data model and provide simple examples of querying the survey and downloading data cubes of interest for remote users. Users will generally want to do two things with HETDEX data cubes: (1) Extract spectra at a list of catalog coordinates and (2) create line-flux maps at specific wavelength and coordinate locations. In this section we demonstrate, using Python, how to query, access and interpolate HETDEX data cubes.

\subsubsection{Coordinate Query Example}
\label{sec:query}

The following Python snippet demonstrates how to query the HETDEX survey for data cubes near a given sky coordinate. This example searches for IFU observations within $37\arcsec$ of a Ly$\alpha$ nebula at $z=2.53$.

\begin{samepage}
\begin{lstlisting}[style=arxivcode,language=Python]
import os.path as op
from astropy.table import Table
from astropy.coordinates import SkyCoord
import astropy.units as u

# Load IFU index
pdr_dir = '/pathto/pdr1/'
ifu_data = Table.read(op.join(pdr_dir, 
    'ifu-index.fits'))

# Create SkyCoord array of IFU centers
ifu_coords = 
    SkyCoord(
        ra=ifu_data['ra_cen']*u.deg,
        dec=ifu_data['dec_cen']*u.deg
    )

# Target coordinate (a LAB at z=2.53)
coord = 
    SkyCoord(
        ra=228.78581*u.deg, 
        dec=51.268036*u.deg
    )

# Select IFUs within 37 arcseconds
sel = coord.separation(ifu_coords) < 37*u.arcsec

# Print matching data cube path(s)
for row in ifu_data[sel]:
    shotid  = row['shotid']
    ifuslot = row['ifuslot']
    path = 
        op.join(
            pdr_dir, 
            'data cubes', str(shotid),
            f'dex_cube_{shotid}_{ifuslot}.fits')
    print(path)
\end{lstlisting}
\end{samepage}

\subsubsection{Spectral Extractions}

\begin{figure}[t]
    \centering
    \includegraphics[width=\linewidth]{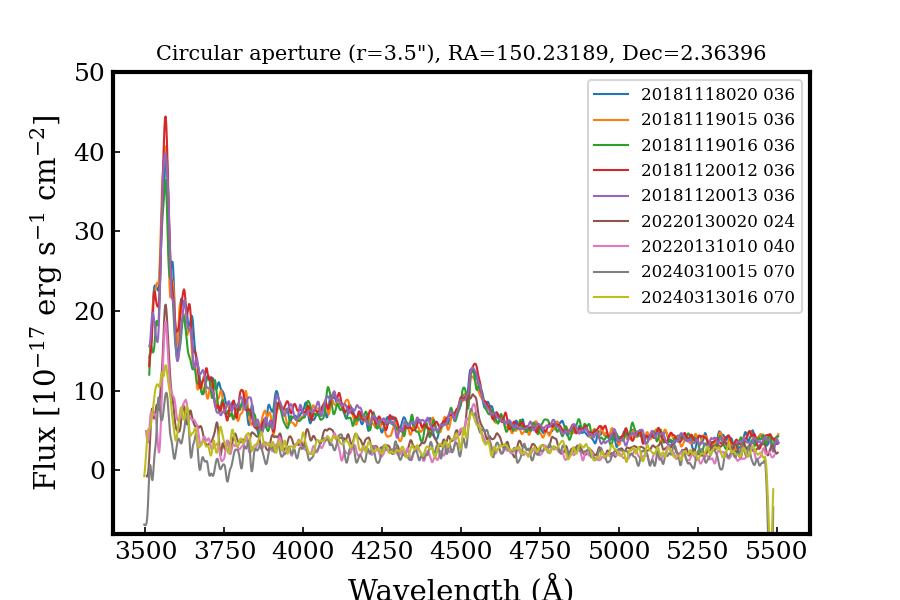}
    \caption{Example aperture-extracted spectra of an AGN in COSMOS at R.A.=150.23189$^\circ$, decl.=2.36396$^\circ$ using a $3\farcs5$ circular aperture. The source is observed multiple times across HETDEX, showing clear evidence of variability. In early observations from 2018, strong \lya\ emission is present, but the feature diminishes in later data from 2022 and 2024. This “changing-look” AGN demonstrates the time-domain potential of repeated spectroscopic coverage within the survey.}

    \label{fig:changingagn}
\end{figure}

Spectra can be extracted from HETDEX data cubes by applying a circular aperture at a given sky position and summing the flux across spatial pixels within the aperture. This process is the most direct way to obtain 1D spectra from the 3D cubes, and it can be adapted depending on the science case. A simple example is shown in Figure~\ref{fig:changingagn}. Additional examples are provided on the \texttt{dexcube} repository\footnote{\url{https://github.com/HETDEX/dexcube/blob/main/notebooks/08-ExtractingSpectra.ipynb}}. 

For more robust use cases, we provide a dedicated spectral extraction script, \texttt{get\_spectra}\footnote{
\url{https://github.com/HETDEX/dexcube/blob/main/notebooks/dexcube/get_spec.py}}. This tool performs proper aperture corrections and allows for a variety of masking options. For example, in Figure~\ref{fig:changingagn} we removed the \texttt{BADCAL} mask flag so that the AGN’s \lya-emission line at $\lambda=3563$\,\AA\ could be recovered, even though it overlaps with a region that is typically masked. Users are encouraged to adapt the example code to their own needs and apply masking strategies appropriate for their targets.

Finally, while spectra can be extracted directly from the full data cube, efficiency can often be improved by slicing the cube along the spectral axis to limit the wavelength range of interest. This reduces memory load and speeds up the extraction when only a narrow spectral region is required.

\begin{lstlisting}[style=arxivcode,language=Python]
from astropy.io import fits
from astropy.wcs import WCS
from astropy.coordinates import SkyCoord
import astropy.units as u
import numpy as np, matplotlib.pyplot as plt

# Open data cube
hdul = fits.open("dex_cube_20181118020_036.fits")
flux, hdr = hdul["DATA"].data, hdul["DATA"].header
wcs, coord = WCS(hdr), SkyCoord(150.23189*u.deg, 2.363963*u.deg)

# Pixel center + radius in pixels
x,y = wcs.celestial.world_to_pixel(coord)
r_pix = 2.0 / wcs.proj_plane_pixel_scales()[0].to(u.arcsec).value

# Circular aperture mask
yy,xx = np.indices(flux.shape[1:])
mask = (xx-x)**2 + (yy-y)**2 < r_pix**2

# Extract spectrum
spec = np.nansum(flux[:,mask],axis=1)
wave = hdr['CRVAL3'] + (np.arange(flux.shape[0]) - hdr['CRPIX3'] + 1) * hdr['CDELT3']

plt.plot(wave*1.e10, spec); plt.xlabel("Wavelength (AA)")
\end{lstlisting}

\subsubsection{Collapsing Cubes to Make Line-Flux Maps}
\label{sec:narrowband}

To visualize emission at a known wavelength, data cubes can be collapsed along the spectral axis across a small wavelength window around the targeted line. This process produces a pseudo–narrowband (NB) “line-flux” image. 
A similar procedure can be performed on a adjacent /wide-pass set of wavelengths to build a “continuum” image, which can be scaled and subtracted to remove underlying continuum. During this interpolate process, masked pixels (bitmasks or zeros) should be set to NaN before summation so they don’t bias the collapse. 

Below is an example of collapsing in a $\pm\Delta\lambda$ window around \lya\ at a known observed wavelength. In this case, we collapse $\pm10$\,\AA\ at $\lambda=4295.9$\,\AA\null.  No continuum subtraction is done here, but an example is available online\footnote{\url{https://github.com/HETDEX/dexcube/blob/main/notebooks/07-CollapsingCubes.ipynb}}.

\begin{lstlisting}[style=arxivcode,language=Python]
from astropy.io import fits
from astropy.wcs import WCS
import numpy as np

hdul = fits.open("dex_cube_20180618017_024.fits")

F, M = hdul["DATA"].data, hdul["MASK"].data

# mask data cube
F[(M>0) | (F==0)] = np.nan

wave = hdr['CRVAL3'] + (np.arange(flux.shape[0]) - hdr['CRPIX3'] + 1) * hdr['CDELT3']

lam0, dlam = 4295.9, 10.0
sel = np.abs(wave - lam0) <= dlam

#collapse along spectral dim
img_nb = np.nansum(F[sel], axis=0)
\end{lstlisting}

\begin{figure*}[t]
    \centering
    \includegraphics[width=0.8\linewidth]{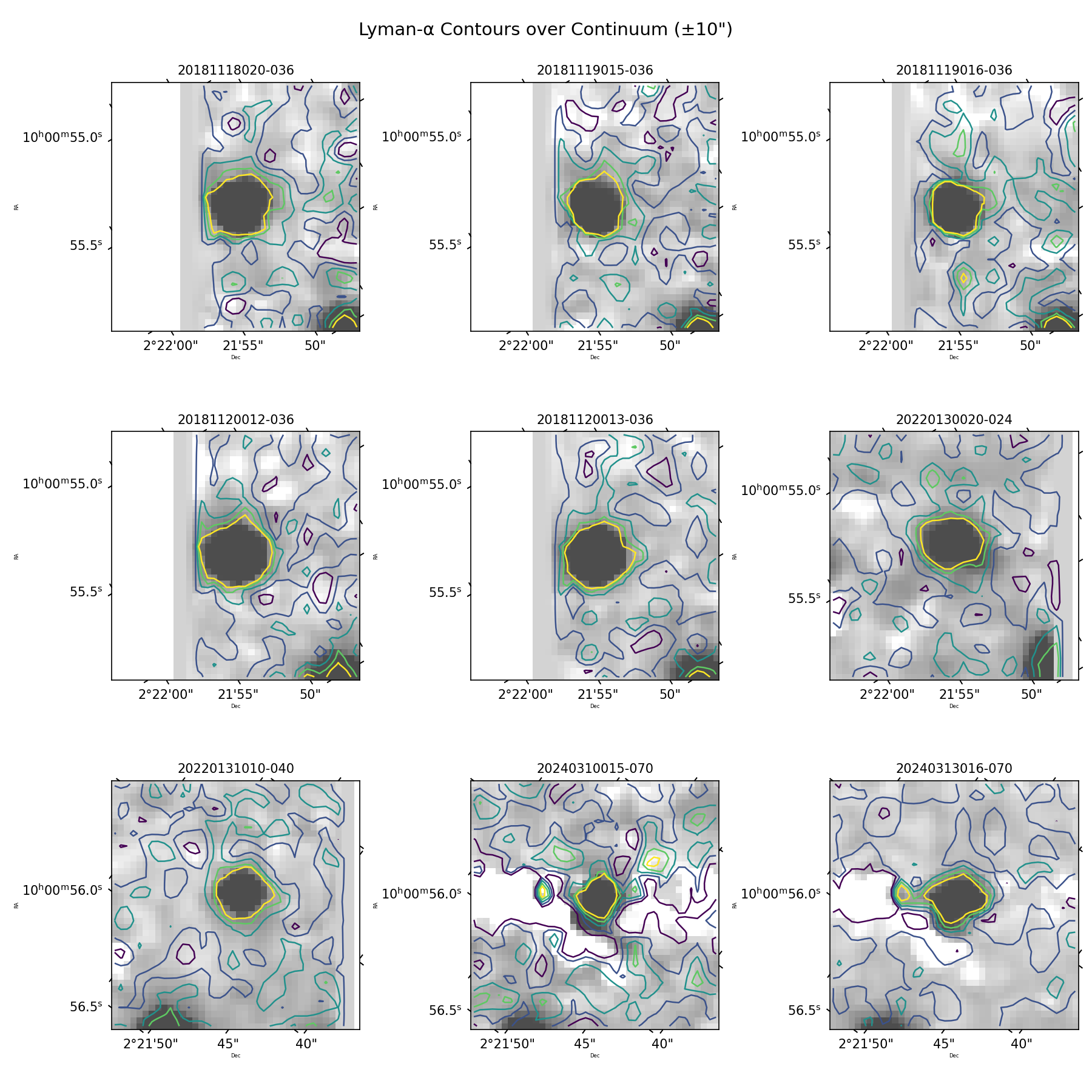}
    \caption{Collapsing the HETDEX data cubes into pseudo-narrowband (Ly$\alpha$, $\lambda=3563$\,\AA\ $\pm50$\,\AA) and continuum (3800–5200\,\AA) images for the example changing-look AGN in COSMOS shown in Figure~\ref{fig:changingagn}. Each panel represents a $20''\times20''$ cutout, with greyscale continuum flux calculated from the HETDEX data cubes and Ly$\alpha$ contours overlaid. These maps are not projected to a common WCS orientation but are in the orientation native to the data cubes. These maps demonstrate both the temporal and spatial information available from repeated IFU coverage in HETDEX.}
    \label{fig:changingagn_lineflux}
\end{figure*}

In Figure~\ref{fig:changingagn_lineflux}, we extend this simple example by collapsing the data cube into narrowband and continuum images and then visualizing their spatial distribution across multiple epochs for the same changing-look AGN shown in Figure~\ref{fig:changingagn}. For each observation covering the AGN, we build (i) a line-flux map by collapsing the cube within $\pm50$\,\AA\ of \lya\ at 3563\,\AA\, and (ii) a continuum map from 3800–5200\,\AA\null. Square cutouts centered on the AGN are extracted with consistent physical size, but the orientation here is in the IFU plane, so they are not matched in WCS. \lya\ narrowband images are overlaid as contours on top of the greyscale continuum maps made from the HETDEX data cubes. This approach highlights the combined spatial and spectral advantage offered by HETDEX observations.

\section{Caveats: How you can use these data and how you should not}
\label{sec:caveats}

%\textcolor{red}{Please add more caveats that you can think of or expand on what I've already write.}

The optimal use of HETDEX data cubes is to retrieve data related to sources that are in the public source catalog provided with this release (Section\,\ref{sec:catalog}) or extract spectra or data cube cutouts on your own list of known sources. The HSPC2 spectral FITS file only provides simple 1D, PSF-weighted spectra for the representative \texttt{detectid} of the source. Many of the objects, however, are resolved and the HETDEX data cubes provide the spatial information needed to explore their physical properties. For example, \citet{MentuchCooper2026_HETDEX_LAN} reported surface-brightness profile measurements and isophotal sizes for a sample of high $S/N$ LAEs revealing that over 50\% of this population has extended emission.

Extracting spectra at the positions of known objects is also an ideal way to access HETDEX data. Objects with $g < 22$ are sufficiently bright for radial-velocity measurements of Milky Way stars \citep[$\sigma_v < 30~km~s^{-1}$;][]{hawkins2021}, or redshift estimates based on a galaxy's absorption lines (see, for example, \citealt{Cooper2023, Zeimann2024, debski2025}).

\subsection{Beware of Early Data}

During its first year (2017-01-01 to 2018-01-01), VIRUS operated with fewer than 20 IFUs. The data quality was poor, with about half of the amplifiers flagged as bad, and detectors were plagued by pox and charge traps. Over time, IFU units were replaced, and new generations of detectors proved to be more stable and clean of artifacts. As a result, early catalog sources and IFU data cubes are of subpar quality. Early observations can be identified using the \texttt{shotid} variable. Values beginning with 2017 correspond to early data and should be used with caution. The \texttt{shotid} appears in the naming structure of each data cube and is included in all source catalogs.  We do not recommend the use of LAE samples from this era without manual inspection. Users should be cautious; these data are included mainly for follow-up studies of external samples.

\subsection{Flux Sensitivity and Luminosity Functions}

HETDEX observations are very heterogeneous. The sensitivity of each spaxel varies considerably depending on wavelength, detector/amplifier, and observing conditions. The flux sensitivity of HETDEX is quantified through extensive source injection simulations calibrated against the survey noise properties \citep{Gebhardt2021}. In these simulations, artificial emission lines are added directly into the fiber spectra across the full wavelength range of VIRUS, and recovery rates are measured using the same detection algorithms applied to the real data. 

The survey reaches $\sim$50\% completeness at a line flux of $1.1\times10^{-16}$ erg s$^{-1}$ cm$^{-2}$, depending on wavelength and seeing conditions, based on simulations. Sensitivity decreases toward the blue end of the spectrograph, but remains relatively uniform redward of $\lambda \simeq 4700$\,\AA. 

We do not include completeness corrections as they need to be appropriately adapted and tested with the interpolated data cube model presented in PDR1.  As a result, PDR1 is not suited for direct luminosity function or number density measurements of faint line emitters without additional modeling. While PDR1 provides the essential raw data products to do so, the completeness models needed to translate catalog detections into population statistics is beyond the scope of the release.

\subsection{Source Detection}

Users may attempt their own source detection algorithms on the released data cubes; however, this should be done with caution. The primary targets of HETDEX, LAEs, are extremely faint emission-line sources that typically occupy only a couple spatial and spectral resolution elements in the VIRUS IFU data. As a result, noise fluctuations and instrumental artifacts can occasionally mimic emission-line detections.

We have made a concerted effort to provide systematic masking of non-astronomical signals through the mask model described in Section~\ref{sec:bitmask}. In addition, some detections were removed during catalog construction using automated machine learning classifiers and manual vetting procedures that operate at the detection level rather than directly within the cube masks. To improve reproducibility for users performing independent analyses, we introduce an additional bitmask flag (see Section\,\ref{sec:baddet}), BADDET, which masks the locations of detections that were rejected by these procedures but were not captured by the previous noise model. This mask is applied as a $5\times5\times5$ pixel cube (corresponding to $2$\farcs$5 \times2$\farcs$5\times10$\,\AA) centered on the emission-line detection and is included in the MASK array HDU of the data cube FITS files.

Despite these masking efforts, some artifacts may still remain and may be difficult to identify using the data cubes alone. Intermediate data processing products, particularly the 2D spectral images used in \citet{Mukae2025}, provide additional diagnostic power for distinguishing real emission lines from artifacts. Users developing independent detection algorithms should therefore consider using the provided mask extensions and, where possible, auxiliary data products to minimize contamination from non-astronomical features.

Source detection is performed on the internal fiber database using a PSF-extracted grid search, with noise estimated by propagating fiber uncertainties into PSF-weighted spectral extractions. All detections are assumed to be point sources. Due to the interpolated nature of the public IFU data cubes, one-to-one recovery is not expected, even with masking applied.

To quantify agreement between independent detections on the public data cubes and the released catalog, we re-measured S/N for 1000 faint emission-line sources drawn from the expanded source catalog (see Appendix\,\ref{appendix:DetInfoTable}), including those below the primary selection threshold (\texttt{p\_conf}$<0.5$ or \texttt{p\_cnn}$<0.5$). Using the same PSF-weighted Moffat extraction ($\beta = 3.0$, FWHM from the IFU index) and S/N estimator as the HETDEX pipeline \citep{Gebhardt2021}, we find a median recovered S/N of $0.75 \pm 0.23$ relative to catalog values.

Agreement improves at higher significance: sources with $\mathrm{S/N} \gtrsim 6$ approach unity with low scatter, while dispersion increases near the detection threshold ($\mathrm{S/N} \sim 5$). The $\sim$25\% deficit arises from correlated noise introduced by spatial interpolation onto the $0\farcs5$ grid, unlike the uncorrelated noise in native fiber spectra. For high-confidence detections (\texttt{p\_conf}~$>0.5$ and \texttt{p\_cnn}~$>0.5$), 95\% of sources with $\mathrm{S/N} \geq 6$ are recovered at $\mathrm{S/N} > 4$, decreasing to 81\% for $\mathrm{S/N} \geq 5.5$ and 5\% near $\mathrm{S/N} \sim 5$.

%We recommend that users do \textit{not} attempt to write their own source detection algorithm, as a number of artifacts still remain in the dataset and these can masquerade as real signal in the data cubes. We have made an earnest effort to provide systematic masking of as many non-astronomical signals as possible within the construction of our mask model as described previously in Section~\ref{sec:bitmask}. But a number of artifacts were removed at a detection level based on automated techniques.  These isolated calibration issues are not directly included in the data cube masks.

%Visually, these artifacts can be difficult to identify in the data cubes alone. Intermediate data processing products, particularly the 2D Spectra images, as is used in \citet{Mukae2025}, are most effective at identifying these features.

\subsection{Local Sky Subtraction and Extended Flux Calibration Limitations}

Flux calibration in HETDEX is performed independently within each IFU. While this approach works well for detecting the low-$S/N$ emission lines that are the primary targets of HETDEX, it can lead to inaccurate flux measurements near bright or extended sources such as large galaxies, bright stars, or the occasional planetary nebula that would benefit from larger-scale modeling of the sky.

To address this, we also generate a full–focal-plane sky estimate using the entire VIRUS array for each observation. In practice, however, the current implementation of the full-frame sky model introduces low-level spectral features that degrade the default flux calibration. For this reason, fiber spectra calibrated with the full-frame sky model are not included in this release, although this may be revisited in future data products. Despite these limitations, \citet{maja2025} successfully used full-frame sky–subtracted data to perform \lya\ intensity mapping by statistically removing contaminating components.

Because of these issues, we caution users against performing intensity-mapping analyses (e.g., \citealt{maja2025}) or absolute sky-brightness measurements \citep{laurel2025} directly from the public PDR1 data cubes. The local sky subtraction provided here is optimized for source detection but is not suitable for analyses requiring a globally uniform sky model. Users studying very extended sources should also note that the local sky subtraction can suppress low surface-brightness emission on spatial scales comparable to the IFU footprint.

\subsection{Stacking}

One of the biggest strengths of HETDEX is its potential for studying the properties of faint targets via spectral stacking. By combining thousands to hundreds of thousands of spectra, the faint spectral continuum signal from LAEs is boosted revealing details about its stellar population \citep{davisstack2023} and signatures of neutral hydrogen in the circumgalactic medium \citep{weiss2024, mahan2025}. 

When stacking HETDEX spectra, we account for faint sky residuals by constructing a correction spectrum from stacks of 200 randomly selected empty-sky apertures per field \citep{davisstack2023}. This residual stack, typically less than 1\% of the sky level in individual spectra, is subtracted from source stacks to remove small calibration offsets that only become significant when combining large numbers of spectra.  This measured offset, which is applied in published HETDEX stacking analyses, such as \citet{davisstack2023,weiss2024,mahan2025}, is not included in the HETDEX data cubes.

\section{Summary}

HETDEX is a medium-wide area, IFU spectroscopic survey that covers the wavelength range 3470--5540\,\AA\ at a resolving power of $750 < R < 950$. Main survey operations are now complete with \skycoverage\,deg$^2$ of noncontiguous sky coverage mapped across $\sim540$\,deg$^2$.

This data release provides access to all science quality IFU data cubes from the main HETDEX survey. It consists of \nifu\ data cubes, reduced from $>600$\,million fiber spectra obtained by the VIRUS instrument on the HET\null. The data cubes are offered at a spatial resolution of 0\farcs5, and a spectral resolution of 2\AA\null. Because the data cubes use a local sky-subtraction procedure optimized for detecting faint emission-line sources, they are not designed for analyses requiring absolute surface-brightness measurements (e.g., intensity mapping) or for studies of very extended nearby galaxies.

This paper describes the public release of HETDEX Public Source Catalog 2 (HPSC2) which contains \nlae\ LAEs, \noii\ \OII-emitting galaxies, \nstar\ stars, \nlzg\ low-$z$ continuum selected galaxies, and \nagn\ AGNs. By utilizing a three-pronged classification approach, we provide robust spectroscopic redshifts and classifications for the entire catalog. When compared to spectroscopic redshifts from external catalogs, 94.1\% of the sources are within $|\Delta z| < 0.02$.

Data access and details about the catalog can be found online at \url{http://hetdex.org}. The link to the data at TACC is at \url{https://web.corral.tacc.utexas.edu/hetdex/HETDEX/pdr/pdr1/}.  A copy of the HETDEX Public Source Catalog 2 is also publicly available via Zenodo (\href{https://doi.org/10.5281/zenodo.19581262}{DOI: 10.5281/zenodo.19581262}), ensuring long-term access and reproducibility.

Papers making use of the HETDEX dataset should include citations for the HET (\citealt{Ramsey1998, Hill2021}), the VIRUS instrument (\citealt{Hill2021}), and the HETDEX survey (\citealt{Gebhardt2021}), and acknowledgments as specified at https://hetdex.org/papers/.

\section*{Acknowledgements}

HETDEX is led by the University of Texas at Austin McDonald Observatory and Department of Astronomy with participation from the Ludwig-Maximilians-Universit\"at M\"unchen, Max-Planck-Institut f\"ur Extraterrestrische Physik (MPE), Leibniz-Institut f\"ur Astrophysik Potsdam (AIP), Texas A\&M University, The Pennsylvania State University, Institut f\"ur Astrophysik G\"ottingen, The University of Oxford, Max-Planck-Institut f\"ur Astrophysik (MPA), The University of Tokyo, and Missouri University of Science and Technology.

Observations for HETDEX were obtained with the Hobby-Eberly Telescope (HET), which is a joint project of the University of Texas at Austin, the Pennsylvania State University, Ludwig-Maximilians-Universit\"at M\"unchen, and Georg-August-Universit\"at G\"ottingen. The HET is named in honor of its principal benefactors, William P. Hobby and Robert E. Eberly. 

The Visible Integral-field Replicable Unit Spectrograph (VIRUS) was used for HETDEX observations. VIRUS is a joint project of the University of Texas at Austin,
Leibniz-Institut f\"ur Astrophysik Potsdam (AIP), Texas A\&M University
(TAMU), Max-Planck-Institut f\"ur Extraterrestrische Physik (MPE),
Ludwig-Maximilians-Universit\"at Muenchen, Pennsylvania State
University, Institut f\"ur Astrophysik G\"ottingen, University of Oxford,
and the Max-Planck-Institut f\"ur Astrophysik (MPA). In addition to
Institutional support, VIRUS was partially funded by the National
Science Foundation, the State of Texas, and generous support from
private individuals and foundations.

The authors acknowledge the Texas Advanced Computing Center (TACC) at The University of Texas at Austin for providing high performance computing, visualization, and storage resources that have contributed to the research results reported within this paper. URL: http://www.tacc.utexas.edu

We acknowledge RAIC Labs (\url{https://raiclabs.com}) for providing access to the RAIC platform, which enabled the efficient classification and identification of artifacts, meteor tracks, and satellite streaks within the HETDEX dataset.

Dark Energy Explorers is recognized as an official NASA Citizen Science partner. This publication utilizes data generated through the Zooniverse.org platform, the development of which is supported by generous funding, including a Global Impact Award from Google and a grant from the Alfred P. Sloan Foundation.

The Institute for Gravitation and the Cosmos is supported by the Eberly College of Science and the Office of the Senior Vice President for Research at the Pennsylvania State University. The Kavli IPMU is supported by World Premier International Research Center Initiative (WPI), MEXT, Japan. 

This work makes use of the Sloan Digital Sky Survey IV, with funding provided by the Alfred P. Sloan Foundation, the U.S. Department of Energy Office of Science, and the Participating Institutions. \sdss-IV acknowledges support and resources from the Center for High-Performance Computing at the University of Utah. The \sdss\ website is www.sdss.org.

%This work makes use of the Pan-STARRS1 Surveys (PS1) and the PS1 public science archive, which have been made possible through contributions by the Institute for Astronomy, the University of Hawaii, the Pan-STARRS Project Office, the Max-Planck Society and its participating institutes.

%This work makes use of data from the European Space Agency (ESA) mission {\it Gaia} (\url{https://www.cosmos.esa.int/gaia}), processed by the {\it Gaia} Data Processing and Analysis Consortium (DPAC, \url{https://www.cosmos.esa.int/web/gaia/dpac/consortium}). Funding for the DPAC has been provided by national institutions, in particular the institutions participating in the {\it Gaia} Multilateral Agreement.

%This work makes use of observations made with the NASA/ESA Hubble Space Telescope obtained from the Space Telescope Science Institute, which is operated by the Association of Universities for Research in Astronomy, Inc., under NASA contract NAS 5–26555. 

In addition to Institutional support, HETDEX is funded by the National Science Foundation (grant AST-0926815), the State of Texas, the US Air Force (AFRL FA9451-04-2-0355), and generous support from private individuals and foundations. K.G. acknowledges support from NSF-2008793. 
S.S. and H.K. acknowledge support from the National Science Foundation under grants NSF-2219212 and NSF-2511145.

\software{
\textsc{Astropy} \citep{astropy:2018},
\textsc{Python} \citep{Pythonref},
\textsc{NumPy} \citep{harris2020array},
\textsc{SciPy} \citep{scipy},
\textsc{Scikit-Learn} \citep{scikit-learn},
\textsc{hetdex-api} (\url{https://github.com/HETDEX/hetdex_api}),
\textsc{Elixer} \citep[\url{https://github.com/HETDEX/elixer};][]{Davis2023},
\textsc{Diagnose} \citep[\url{https://github.com/grzeimann/Diagnose};][]{diagnose},
\textsc{Photutils} \citep{photutils_1.3.0},
\textsc{Dustmaps} \citep{dustmaps},
\textsc{Extinction} (\url{https://github.com/kbarbary/extinction})
}

%Software: This research was made possible by the open-source projects astropy \citep{astropy:2018}, Python \citep{Pythonref}, numpy \citep{harris2020array}, Scipy \citep{scipy}, hetdex-api (\url{https://github.com/HETDEX/hetdex_api}), elixer \citep[\url{https://github.com/HETDEX/elixer};][]{Davis2021}, diagnose \citep[\url{https://github.com/grzeimann/Diagnose};][]{diagnose}, photutils \citep{photutils_1.3.0}, dustmaps \citep{dustmaps}, extinction (\url{https://github.com/kbarbary/extinction}) 

\bibliography{hetdex}
\bibliographystyle{aasjournal}

\newpage
\appendix

\section{Raw Emission-Line and Continuum Detections}\label{appendix:detections}

In addition to the IFU data cubes, this data release includes the raw databases of all emission-line and continuum detections from the HETDEX internal data pipeline. Emission lines capture discrete spectral features such as Ly$\alpha$ and [O~II], while continuum detections identify bright broadband sources such as galaxies and stars. Both are distributed as HDF5 files under \texttt{pdr1/detect/}. 
Each file contains three main tables: \texttt{Detections}, \texttt{Spectra}, and \texttt{Fibers}. 
Continuum catalogs use the same structure, though many line-specific columns are not applicable and are filled with 0.0 values. 

The raw detection HDF5 files provide a structured record of every emission-line 
or continuum detection found by the HETDEX pipeline, along with its associated spectra and fiber-level extractions. 
Each file is organized into three linked groups: \textbf{Detections}, \textbf{Spectra}, 
and \textbf{Fibers}. 

The \textbf{Detections} group contains one row per candidate detection, including positional information (RA, decl., IFU coordinates), best-fit line measurements 
(central wavelength, flux, line width, continuum), and quality metrics ($S/N$, $\chi^2$ values, noise estimates). Instrumental identifiers (e.g., \texttt{shotid}, \texttt{specid}, \texttt{ifuslot}) and fiber metadata (brightest fiber ID, aperture correction factors) are also stored here.  

The \textbf{Spectra} group links to each detection via \texttt{detectid} and 
records the 1D extracted spectrum. Both aperture-corrected and 
uncorrected versions are provided, with associated errors. Additional arrays 
include the wavelength grid, raw counts, PSF-weighted counts, and the 
wavelength-dependent aperture correction. No dust-extinction correction is applied to the raw database spectral data. 

The \textbf{Fibers} group stores the fiber-level information for each fiber contained in a single detection. This dataset includes fiber coordinates (on-sky and within the IFU), CCD positions, weights, 
flags, time stamps, and instrument identifiers. Each fiber entry is linked back to 
its parent detection, allowing reconstruction of the full fiber ensemble 
contributing to a given source.  

Together, these three groups allow users to trace each detection from the raw 
fiber spectra, through PSF-weighted extraction, to the final line and continuum 
measurements.

\begin{table}[h]
\centering
\caption{Detection Catalog Files in PDR1}
\label{tab:detect_files}
\begin{tabular}{lllr}
\hline
File & Description & Date Range & Number of Detections \\
\hline
\texttt{cont\_hdr3.h5} & Continuum detections & 2017-01-01 to 2021-08-31 & 297{,}877 \\
\texttt{cont\_hdr4.h5} & Continuum detections & 2021-09-01 to 2023-08-31 & 276{,}465 \\
\texttt{cont\_hdr5.h5} & Continuum detections & 2023-09-01 to 2024-07-31 & 107{,}725 \\
\texttt{detect\_hdr3.h5} & Emission-line detections & 2017-01-01 to 2021-08-31 & 13{,}843{,}051 \\
\texttt{detect\_hdr4.h5} & Emission-line detections & 2021-09-01 to 2023-08-31 & 14{,}180{,}087 \\
\texttt{detect\_hdr5.h5} & Emission-line detections & 2023-09-01 to 2024-07-31 & 3{,}938{,}578 \\
\texttt{detect\_index\_hdr5.h5} & Index for HDR5 detections & 2023-09-01 to 2024-07-31 & 32{,}643{,}783 \\
\texttt{elixer\_hdr345\_cluster\_cat.h5} & Filtered subset of clustered detections & 2017-01-01 to 2024-07-31 & 4,710,195 \\
\hline
\end{tabular}
\end{table}

Two additional tables are contained in this directory. The Detection Index table provides a look-up reference so that the full collection of raw detection databases may be queried. The column names for the Detection Index table are given in Table~\ref{tab:detectindex}. Querying may be done using healpix with a resolution of $N_{\rm side}=2^{15}$.

\newcommand{\notes}[1]{\begin{minipage}[t]{0.58\columnwidth}\raggedright #1\end{minipage}}

\begin{table}[t]
\centering
\caption{Schema of the DetectIndex table (file: detect\_index\_hdr5.h5). The table contains 32{,}643{,}783 rows.}
\label{tab:detectindex}
\small
\setlength{\tabcolsep}{4pt}
\begin{tabular}{lll}
\hline\hline
\textbf{Column} & \textbf{Type} & \textbf{Notes} \\
\hline
\texttt{detectid} & Int64    & \notes{Unique detection identifier. Indexed (CSI)} \\
\texttt{shotid}   & Int64    & \notes{Observation identifier. Indexed (CSI)} \\
\texttt{ra}       & Float32  & \notes{R.A. (ICRS J2000 deg). Indexed (CSI)} \\
\texttt{dec}      & Float32  & \notes{Decl. (ICRS J2000 deg)} \\
\texttt{wave}     & Float32  & \notes{Central wavelength (\AA)} \\
\texttt{sn}       & Float32  & \notes{Signal-to-noise ratio of the detection.} \\
\texttt{healpix}  & Int64    & \notes{HEALPix index of detection location. Indexed (CSI)} \\
\texttt{det\_type} & String(4)  & \notes{Detection type (\texttt{line} or \texttt{cont})} \\
\texttt{survey}   & String(4)  & \notes{Survey flag (e.g., HDR2, HDR3, HDR4, HDR5)} \\
\texttt{fiber\_id} & String(38) & \notes{Unique fiber identifier string} \\
\hline
\end{tabular}
\end{table}

\subsection{Detection Methods}

Emission-line detections are identified through a two-stage grid search of the data cubes. 
An initial coarse search is performed in $0\farcs5$ spatial steps and 8\,\AA\ spectral steps using a Gaussian line profile with the instrumental resolution ($\sigma=1.7$\,\AA). 
Continuum emission is subtracted locally, and candidate lines with $S/N > 4$ and $\chi^2 < 3$ are retained. 
A refined raster with $0\farcs15$ spatial steps and an unconstrained $\sigma$ then optimizes the fit, with duplicates within $3\arcsec$ and 3\,\AA\ merged. 
Continuum detections are identified by searching for fibers with more than 50 counts in either a blue (3700--3900\,\AA) or red (5100--5300\,\AA) window (corresponding to $g \sim 22.5$). 
Candidate positions are refined with a PSF raster fit, and spectra are extracted at the best-fit location. 

Each row in the catalogs is assigned a unique integer \texttt{detectid}. 
A single astronomical source may have multiple detection entries if it is spatially extended, has multiple emission lines, or is bright enough to be detected in via its line and continuum emission. 
Additionally, the same emission line observed in separate exposures will receive independent \texttt{detectid} identifiers. 
Users should therefore treat the detection catalogs as a superset of measurements from which higher-level source catalogs can be constructed. 

\subsection{Access Example}

The catalogs are stored in HDF5 format and can be accessed directly with \textsc{PyTables} \citep{pytables}. Each internal data release has its own set of detections for the date range as outlined in Table~\ref{tab:detect_files}. To streamline querying, a master index file \texttt{detect\_index\_hdr5.h5} can first be queried either by coordinates, detectid, or other variables (see Table~\ref{tab:detectindex}) to determine the detection's survey and detection type (\texttt{line} vs. \texttt{cont}). This query indicates which survey HDF5 file to open (e.g., \texttt{detect/detect\_\{survey\}.h5} for line detections, or \texttt{cont/cont\_\{survey\}.h5} for continuum detections). The matching rows in the \texttt{/Detections} and \texttt{/Spectra} tables are then read to obtain sky coordinates, \texttt{shotid}/\texttt{ifuslot}, the central wavelength for line detections, and the pipeline 1D spectrum and errors.

\begin{lstlisting}[style=arxivcode,language=Python]
import tables as tb
import os.path as op
import numpy as np
from astropy.coordinates import SkyCoord

# --- Inputs ---
detectid_obj = 3002870541   # <- example detectid (int)

pdr_dir = '/home/jovyan/Hobby-Eberly-Public/HETDEX/pdr/pdr1/'

# 1) Open detection index to access pipeline detection index for the full survey
DI = tb.open_file(op.join(pdr_dir, 'detect/detect_index_hdr5.h5'), 'r')

det_file_info = DI.root.DetectIndex.read_where(f'detectid == {detectid_obj}')[0]
# PyTables string columns are byte strings; we use \texttt{.decode()} when needed.
survey = det_file_info['survey'].decode()
det_type = det_file_info['det_type'].decode()

# 2) Choose file family based on detection type
if det_type == 'line':
    det_file_type = 'detect'
else:
    det_file_type = 'cont'

# 3) Spectra and IFU/shotid information are accessed from H5 files for
#    each survey/detection-type combination
det_file = tb.open_file(op.join(pdr_dir, f'{det_file_type}/{det_file_type}_{survey}.h5'), 'r')

# Pull basic detection info
det_info = det_file.root.Detections.read_where(f'detectid == {detectid_obj}')[0]
coord = SkyCoord(ra=det_info['ra'], dec=det_info['dec'], unit='deg')
shotid = det_info['shotid']
ifuslot = det_info['ifuslot'].decode()

cw = None
if det_type == 'line':
    cw = det_info['wave']   # central wavelength [Angstrom] for line detections

# 4) Load the pipeline 1D spectrum row for this detection
spec_pipeline = det_file.root.Spectra.read_where(f'detectid == {detectid_obj}')[0]

# 5) Fix for calibration issue in pipeline spectra

corr_file = np.loadtxt(op.join(pdr_dir, 'detect/wdcor.txt'))
spectrum = spec_pipeline['spec1d']/corr_file[:,1]

# 6) If desired convert to a flux density
spectrum /= 2.0

# (Optional) clean-up
# DI.close()
# det_file.close()
\end{lstlisting}

For emission-line detections \texttt{cw} stores the fitted line center in \AA\null.
The pipeline spectrum (column \texttt{spec1d}, more information in Table~\ref{tab:raw_h5_columns}) contains PSF-weighted flux in units of $10^{-17}$\,erg\,s$^{-1}$\,cm$^{-2}$ in 2\,\AA\ bins (users should divide by 2 to obtain conventional flux density units of
$10^{-17}$\,erg\,s$^{-1}$\,cm$^{-2}$\,\AA$^{-1}$ as is provided in the 1D spectra provided in HPSC2). A calibration error is present in the raw H5 detection spectra. Output spectra should be normalized by the spectrum `/pdr1/detect/wdcor.txt' as shown in the example above.

Continuum catalogs (\texttt{cont\_hdr*}) have the same structure, although many line-specific 
columns (e.g., \texttt{wave}, \texttt{flux}, \texttt{sn}) are zero.

\begin{table}[t]
\centering
\caption{Column Descriptions in the Raw Detections HDF5 Table Groups}
\label{tab:raw_h5_columns}
\small
\setlength{\tabcolsep}{4pt}
\renewcommand{\arraystretch}{1.05}
\begin{tabular}{lll}
\hline\hline
Table & Column & Description \\
\hline
\multirow{24}{*}{Detections}
 & \texttt{detectid} & Unique integer detection ID \\
 & \texttt{inputid} & Input extraction ID (\texttt{DATEvOBS\_inputname}) \\
 & \texttt{shotid} & Shot identifier (\texttt{YYYYMMDDnnn}) \\
 & \texttt{date} & Observation date (YYYYMMDD) \\
 & \texttt{obsid} & Observation number on that date \\
 & \texttt{ra}, \texttt{dec} & R.A. and decl. (ICRS J2000 deg) \\
 & \texttt{wave}, \texttt{wave\_err} & Line central wavelength \AA and uncertainty \\
 & \texttt{flux}, \texttt{flux\_err} & Line flux $10^{-17}$ erg s$^{-1}$ cm$^{-2}$ and uncertainty \\
 & \texttt{linewidth}, \texttt{linewidth\_err} & Gaussian $\sigma$ \AA and uncertainty \\
 & \texttt{continuum}, \texttt{continuum\_err} & Continuum flux density \fluxden\ and uncertainty \\
 & \texttt{sn}, \texttt{sn\_err} & Emission-line signal-to-noise ratio and uncertainty \\
 & \texttt{chi2}, \texttt{chi2\_err} & $\chi^2$ of line fit and uncertainty \\
 & \texttt{fiber\_id} & Brightest fiber identifier \\
 & \texttt{multiframe} & \texttt{specid/ifuslot/ifuid/amp} identifier \\
 & \texttt{fibnum} & Fiber number \\
 & \texttt{expnum} & Dither exposure number \\
 & \texttt{x\_raw}, \texttt{y\_raw} & CCD coordinates of detection \\
 & \texttt{x\_ifu}, \texttt{y\_ifu} & IFU position (arcsec) \\
 & \texttt{specid}, \texttt{ifuslot}, \texttt{ifuid}, \texttt{amp} & Instrument identifiers \\
 & \texttt{weight} & Weight of brightest fiber \\
 & \texttt{apcor} & Aperture correction factor \\
 & \texttt{sn\_cen} & $S/N$ summed across 7 pixels at line center \\
 & \texttt{flux\_noise\_1sigma} & 1$\sigma$ noise estimate on flux \\
 & \texttt{sn\_3fib}, \texttt{sn\_3fib\_cen} & $S/N$ for 3-fiber extractions \\
\hline
\multirow{7}{*}{Spectra}
 & \texttt{detectid} & Unique integer detection ID (links to Detections) \\
 & \texttt{wave1d} & Wavelength array \AA \\
& \texttt{spec1d}, \texttt{spec1d\_err} 
& Flux-calibrated, PSF-weighted, aperture-corrected spectrum\\
&  & and error in $10^{-17}$ erg s$^{-1}$ cm$^{-2}$ in 2\,\AA\ bins \\
 & \texttt{counts1d}, \texttt{counts\_err} & Counts and uncertainty \\
 & \texttt{apsum\_counts}, \texttt{apsum\_counts\_err} & PSF-weighted counts and uncertainty \\
 & \texttt{spec1d\_nc}, \texttt{spec1d\_nc\_err} & Spectrum without aperture correction and uncertainty \\
 & \texttt{apcor} & Applied aperture correction (wavelength dependent) \\
\hline
\multirow{16}{*}{Fibers}
 & \texttt{detectid} & Unique integer detection ID (links to Detections) \\
 & \texttt{ra}, \texttt{dec} & Fiber RA, Dec (ICRS J2000 deg) \\
 & \texttt{fiber\_id} & Fiber identifier string \\
 & \texttt{x\_ifu}, \texttt{y\_ifu} & Fiber position in IFU (arcsec) \\
 & \texttt{multiframe} & \texttt{specid/ifuslot/ifuid/amp} identifier \\
 & \texttt{fibnum} & Fiber number \\
 & \texttt{expnum} & Dither exposure number \\
 & \texttt{distance} & Distance to requested RA,Dec (arcsec) \\
 & \texttt{wavein} & Input wavelength for extraction \\
 & \texttt{timestamp} & Exposure time stamp \\
 & \texttt{date}, \texttt{obsid} & Observation date and number \\
 & \texttt{flag} & Fiber flag \\
 & \texttt{weight} & Weight applied in PSF sum \\
 & \texttt{ADC} & Five-element array of ADC values \\
 & \texttt{specid}, \texttt{ifuslot}, \texttt{ifuid}, \texttt{amp} & Instrument identifiers \\
 & \texttt{x\_raw}, \texttt{y\_raw} & CCD coordinates of fiber \\
\hline
\end{tabular}
\end{table}

\clearpage

\section{HPSC2 Supplemental Detection Information Table}
\label{appendix:DetInfoTable}

This appendix describes the supplemental \textit{Detection Information Table}, noted by the \texttt{hetdex\_sc2\_detinfo\_vX.X.fits/.dat} file name, which contains information for every emission-line and continuum detection found by the HETDEX pipeline (see in Section~\ref{sec:detection}). As described in Section~\ref{sec:detgroup}, a HETDEX source can be composed of a collection of line-emission and continuum-emission detections. The \textit{HETDEX Public Source Catalog 2} (HPSC2), outlined in Table\,\ref{tab:column_info}, provides a simplified version of the \textit{Detection Information Table} with one row per source observation; it provides basic information about a source such as coordinates, redshift, \hetg\ magnitude, and the [\ion{O}{2}]/Ly$\alpha$ line flux and luminosity, where applicable. The \textit{Detection Information Table} presented in this appendix is expanded to provide additional information for every detection in a source. Many columns are the same to those in HPSC2, e.g.  \texttt{source\_id}, \texttt{source\_name}, \texttt{RA}, \texttt{DEC}, \zhet. Additional information is provided regarding line fit parameter information, including the specific position of the detection (\texttt{RA\_det}, \texttt{Dec\_det}) and wavelength (\texttt{wave}) for the detection, the detection's line width, ($\sigma$: \texttt{sigma}), continuum-subtracted line flux and the local continuum measurement. Each observed wavelength is checked for a restframe match to a common line species at \zhet. Specifically, we consider \ion{C}{3}] $\lambda 1909$, \ion{C}{4} $\lambda 1550$,
\hbeta, \hdelta, \hgamma, \HeII, \lya, [\ion{O}{2}] $\lambda 3727$, and [\ion{O}{3}] $\lambda\lambda 4959,5007$ \footnote{\url{http://classic.sdss.org/dr6/algorithms/linestable.html}}. If a match is found, it is listed in \texttt{line\_id}. Not all detections have a \texttt{line\_id} as some HETDEX detections can result from discontinuities in a spectrum or calibration issues. We attempt to mitigate these by excluding high line width sources that are not selected as the main detection (ie., \texttt{selected\_det}==True) of a source. Other information as described in the text is also provided, including (1) detection group information from 3D and 2D FOF detection clustering, (2) \elixer\ imaging counterpart information, (3) specific observation parameters such as image quality (e.g., \texttt{fwhm}), \texttt{shotid}, \texttt{date}, \texttt{obsid}, \texttt{field}), and (4) specific information related to the highest weight fiber in the spectral extraction (such as \texttt{multiframe}, \texttt{fiber\_id}, \texttt{weight} and others). The detectid whose spectrum is included in HPSC2 for the source (typically the brightest magnitude detection) is identified by \texttt{selected\_det==True}. The description of all parameters is provided in Table~\ref{tab:det_col_info}.

\input{column_info.tex}\label{tab:det_col_info}

\facilities{HET}

\end{document}

%% file: HETDEX_pn.tex
\begin{table}[t]
\centering
\small
\setlength{\tabcolsep}{1pt} % reduce column separation (default 6pt)
\caption{Planetary Nebulae in HETDEX PDR1\label{tab:pn_cat}}
\begin{tabular}{cccc}
\hline \hline
PNG ID & Coordinates (RA, Decl.) & shotid & r (arcmin) \\
\hline
PNG 085.3+52.3 & $(230.444^\circ,\ 52.368^\circ)$ & 20210510014 & 4.08 \\
PNG 085.3+52.3 & $(230.444^\circ,\ 52.368^\circ)$ & 20220131013 & 4.08 \\
PNG 144.8+65.8 & $(179.437^\circ,\ 48.938^\circ)$ & 20240314023 & 3.00 \\
PNG 144.8+65.8 & $(179.437^\circ,\ 48.938^\circ)$ & 20240403008 & 3.00 \\
PNG 136.6+61.9 & $(181.868^\circ,\ 54.025^\circ)$ & 20210309019 & 2.83 \\
PNG 136.6+61.9 & $(181.868^\circ,\ 54.025^\circ)$ & 20210401013 & 2.83 \\
\hline
\end{tabular}
\end{table}

%% file: obs_column_info.tex
% Wrap long text in the 2nd column while keeping {ll}
%\newcommand{\desc}[1]{\begin{minipage}[t]{0.8\columnwidth}\raggedright #1\end{minipage}}
\begin{table*}[!h]
\centering
\caption{\textit{Source Observation Table} Column Descriptions\label{tab:column_info}}
\footnotesize
\setlength{\tabcolsep}{2pt} % tweak if you need tighter columns
\begin{tabular}{ll}
\hline\hline
\textbf{Name} & \textbf{Description} \\
\hline
source\_name & HETDEX IAU designation (e.g., \texttt{HETDEX~J123449.19+511733.7}).\\
source\_id & HETDEX Source Identifier.\\
shotid & Integer observation ID: \texttt{int(date+obsid)}.\\
ifuslot & String identifier of IFU location in focal plane.\\ 
RA & Right ascension for source (ICRS deg).\\
DEC & Declination for source (ICRS deg).\\
RA\_det & representative detectid R.A. (ICRS J2000 deg)\\
DEC\_det & representative detectid decl. (ICRS J2000 deg)\\
gmag & SDSS $g$ magnitude measured in the HETDEX spectrum.\\
Av & Applied extinction correction in $V$ band.\\
z\_hetdex & HETDEX spectroscopic redshift.\\
z\_hetdex\_src & Source/method for HETDEX spectroscopic redshift.\\
z\_hetdex\_conf & Confidence (0--1) in HETDEX spectroscopic redshift source.\\
source\_type & One of \texttt{star}, \texttt{lae}, \texttt{agn}, \texttt{lzg}, \texttt{oii}\\
detectid & Detection ID of representative detection (where \texttt{selected\_det = True} in the \textit{Detection Info Table}).\\
field & Survey field: \texttt{dex-fall}, \texttt{dex-spring}, \texttt{cosmos}, \texttt{goods-n}, \texttt{nep}, or \texttt{ssa22}.\\
flux & extinction corrected line flux in $10^{-17}\,\mathrm{erg\,s^{-1}\,cm^{-2}}$ at \texttt{wave}. -999.0 for continuum sources but is  \\
 & correctly assigned in \texttt{flux\_lya} or \texttt{flux\_oii}. \\
flux\_err  & Mcmc error in extinction corrected line flux.\\
flux\_aper & Extinction-corrected [O\,\textsc{ii}] line flux in $10^{-17}\,\mathrm{erg\,s^{-1}\,cm^{-2}}$ measured in the resolved galaxy aperture.\\
flux\_aper\_err & Uncertainty in \texttt{flux\_aper}.\\
flag\_aper & Aperture-flux usage flag for \texttt{lum\_oii}: 1 = use \texttt{flux\_aper}; 0 = use PSF line flux \texttt{flux}.\\
major & Major axis (arcsec) of the resolved [O\,\textsc{ii}] aperture ellipse (from imaging).\\
minor & Minor axis (arcsec) of the resolved [O\,\textsc{ii}] aperture ellipse (from imaging).\\
theta & Position angle (deg) of the aperture ellipse.\\
logL\_lya & $\log_{10}$ Ly$\alpha$ luminosity erg\,s$^{-1}$ from extinction-corrected \texttt{flux}.\\
logL\_lya\_err & Uncertainty in \texttt{logL\_lya}.\\
logL\_oii & $\log_{10}$ [O\,\textsc{ii}] luminosity erg\,s$^{-1}$ from \texttt{flux} if \texttt{flag\_aper=0} or from \texttt{flux\_aper} if \texttt{flag\_aper=1}.\\
logL\_oii\_err & Uncertainty in \texttt{logL\_oii}.\\
flux\_lya & Ly$\alpha$ flux in $10^{-17}\,\mathrm{erg\,s^{-1}\,cm^{-2}}$ from extinction-corrected \texttt{flux}.\\
flux\_lya\_err & Uncertainty in \texttt{flux\_lya}.\\
flux\_oii & [O\,\textsc{ii}] flux in $10^{-17}\,\mathrm{erg\,s^{-1}\,cm^{-2}}$ from \texttt{flux} if \texttt{flag\_aper=0} or \texttt{flux\_aper} if \texttt{flag\_aper=1}.\\
flux\_oii\_err & Uncertainty in \texttt{flux\_oii}.\\
sn & Signal-to-noise ratio for line emission. Note that continuum sources (\texttt{det\_type}==`cont') will have \texttt{sn}=-999.0.\\
det\_type & detection type of the main detectid for the source: `line' or `cont'.  \\
apcor & Aperture correction applied to the spectrum at 4500\,\AA. \\
p\_conf & LAE/Faint \OII\ emitter confidence score from RF Classifier. 1=high confidence, 0=low confidence.\\
p\_cnn & LAE/Faint \OII\ emitter confidence score based on CNN Classifier. 1=high confidence, 0=low confidence.\\
\\
\hline
\multicolumn{2}{l}{\parbox{\linewidth}{\footnotesize Note: Bad values are $-999.0$ for floating-point columns and \texttt{n/a} for string columns.}} \\
\multicolumn{2}{l}{\parbox{\linewidth}{\footnotesize This table is available in its entirety in machine-readable form at \url{https://hetdex.org/data-results/} and in the online journal.}} \\
\end{tabular}
\end{table*}

%% file: column_info.tex
\begin{longtable}{ll}
\caption{\textit{Detection Information Table} Column Descriptions}\\ 
\toprule
\hline %\hline
Name & Description \\
\hline
source\_id & HETDEX Source Identifier \\
source\_name & HETDEX IAU designation \\
RA & source\_id R.A. (ICRS J2000 deg) \\
DEC & source\_id decl. (ICRS J2000 deg) \\
z\_hetdex & HETDEX spectroscopic redshift \\
z\_hetdex\_src & HETDEX spectroscopic redshift source \\
z\_hetdex\_conf & 0 to 1 confidence HETDEX spectroscopic redshift source. Not well calibrated. \\
source\_type & options are \texttt{star}, \texttt{lae}, \texttt{agn}, \texttt{lzg}, \texttt{oii}, and \texttt{none} \\
detectid & emission-line or continuum detection ID \\
selected\_det & best detectid for Ly$\alpha$ flux or \OII\ line flux \\
det\_type & detection type: `line' or `cont' \\
shotid & integer represent observation ID: int( date+obsid) \\
ifuslot & string identifier of IFU location in focal plane \\ 
line\_id & line identification at observed wavelength (wave) assuming redshift of \zhet\ \\
p\_conf & LAE/Faint OII emitter confidence score from RF Classifier. 1=high confidence, 0=low confidence. \\
p\_cnn & LAE/Faint OII emitter confidence score based on CNN Classifier. 1=high confidence, 0=low confidence. \\
RA\_det & detectid R.A. (ICRS J2000 deg) \\
DEC\_det & detectid decl. (ICRS J2000 deg) \\
src\_separation & separation in arcsec between the detectid (RA\_det, DEC\_det) and the source\_id center (RA, DEC) \\
n\_members & number of detections in the source group. Note some may be missing from final catalog due to selections. \\
gmag & \sdss-$g$ magnitude measured in HETDEX spectrum \\
Av & applied extinction correction in the $V$-band \\
ebv & applied selective extinction \\
wave & central wavelength of line emission (\AA) \\
wave\_err & MCMC error in fitted central wavelength (\AA) \\
flux & extinction corrected line flux $10^{-17}\,\mathrm{erg\,s^{-1}\,cm^{-2}}$\\
flux\_err & MCMC error in extinction corrected line flux \\
flux\_obs & observed line flux $10^{-17}\,\mathrm{erg\,s^{-1}\,cm^{-2}}$ \\
flux\_obs\_err & MCMC error in observed line flux \\
flux\_aper & extinction corrected, [\ion{O}{2}] line flux measured in elliptical galaxy aperture in $10^{-17}\,\mathrm{erg\,s^{-1}\,cm^{-2}}$ \\
flux\_aper\_err & error in flux\_aper \\
flux\_aper\_obs & [\ion{O}{2}] line flux measured in elliptical galaxy aperture in $10^{-17}\,\mathrm{erg\,s^{-1}\,cm^{-2}}$  \\
flux\_aper\_obs\_err & error in flag\_aper\_obs \\
flag\_aper & 1 = aperture line flux used for lum\_oii, 0 = PSF line flux used from "flux" column \\
flux\_lya & Ly$\alpha$ flux in $10^{-17}\,\mathrm{erg\,s^{-1}\,cm^{-2}}$ from dust-corrected \texttt{flux}. \\
flux\_lya\_err & Uncertainty in \texttt{flux\_lya}. \\
flux\_oii & [O\,\textsc{ii}] flux in $10^{-17}\,\mathrm{erg\,s^{-1}\,cm^{-2}}$ from \texttt{flux} if \texttt{flag\_aper=0} or \texttt{flux\_aper} if \texttt{flag\_aper=1}. \\
flux\_oii\_err & Uncertainty in \texttt{flux\_oii}. \\
logL\_lya & $\log_{10}$ Ly$\alpha$ luminosity (erg\,s$^{-1}$) from extinction-corrected \texttt{flux}. \\
logL\_lya\_err & Uncertainty in \texttt{logL\_lya}. \\
logL\_oii & $\log_{10}$ [O\,\textsc{ii}] luminosity (erg\,s$^{-1}$) from \texttt{flux} if \texttt{flag\_aper=0} or from \texttt{flux\_aper} if \texttt{flag\_aper=1}. \\
logL\_oii\_err & Uncertainty in \texttt{logL\_oii}. \\
sigma & sigma linewidth in gaussian line fit (\AA) \\
sigma\_err & MCMC error in sigma linewidth (\AA) \\
continuum & local fitted extinction-corrected continuum in \fluxden \\
continuum\_err & MCMC error in continuum in \fluxden \\
continuum\_obs & local fitted observed continuum in \fluxden \\
continuum\_obs\_err & MCMC error in continuum in \fluxden \\
sn & signal-to-noise ratio for line emission \\
sn\_err & MCMC error in signal-to-noise \\
chi2 & reduced $\chi^2$ quality of line fit \\
chi2\_err & MCMC uncertainty in reduced $\chi^2$ \\
flux\_noise\_1sigma\_obs & observed 1 sigma flux sensitivity in $10^{-17}\,\mathrm{erg\,s^{-1}\,cm^{-2}}$ \\
flux\_noise\_1sigma & extinction corrected 1 sigma flux sensitivity in $10^{-17}\,\mathrm{erg\,s^{-1}\,cm^{-2}}$ \\
apcor & aperture correction applied to spectrum at 4500\AA \\
counterpart\_mag & selected closest counterpart mag from source extracting on image data \\
counterpart\_mag\_err & uncertainty in counterpart\_mag \\
counterpart\_dist & distance to closest counterpart \\
counterpart\_catalog\_name & image catalog source of counterpart \\
counterpart\_filter\_name & image filter of counterpart \\
plya\_classification & \elixer\ confidence line is Ly$\alpha$ ranges 0 to 1 (1=high probility line is \lya) \\
z\_elixer & \elixer\ best redshift \\
best\_pz & confidence in best\_z \\
z\_diagnose & best fit redshift from \texttt{Diagnose} \\
cls\_diagnose & best classification from \texttt{Diagnose}. Options are `STAR', `GALAXY', `QSO', `UNKNOWN' \\
stellartype & \texttt{Diagnose} spectral type classification for stars \\
agn\_flag & $-1$ not an AGN, 0 broad line source but not confirmed AGN, 1 confident AGN \\
wave\_group\_id & id for 3D Friend-of-Friends (FOF) clustering at common ra, dec, wave \\
wave\_group\_a & semi-major axis in arcseconds from 3D FOF clustering \\
wave\_group\_b & semi-minor axis in arcseconds from 3D FOF clustering \\
wave\_group\_pa & positional angle from 3D FOF clustering \\
wave\_group\_ra & mean ra from 3D FOF clustering (ICRS J2000 deg) \\
wave\_group\_dec & mean dec from 3D FOF clustering  (ICRS J2000 deg)\\
wave\_group\_wave & mean wavelength from 3D FOF clustering \\
fwhm & measured seeing of the observation in arcseconds \\
throughput & relative spectral response at 4540 assuming a 360 s nominal exposure \\
field & field ID: cosmos, goods-n, dex-fall, dex-spring, nep, ssa22 \\
date & date \\
obsid & observation number \\
multiframe & string identifier for the ifuslot/specid/ifuid/amp combination \\
fiber\_id & string identifier for the highest weight fiber \\
weight & flux weight of the highest weight fiber \\
x\_raw & $x$ value on the CCD of the detection (ds9 x value) \\
y\_raw & $y$ value on the CCD of the detection (ds9 y value) \\
x\_ifu & $x$ position in the ifu in arcseconds \\
y\_ifu & $y$ position in the ifu in arcseconds \\
ra\_aper & Right Ascension of aperture center of imaging counterpart (ICRS J2000 deg)  \\
dec\_aper & Declination of aperture center of imaging counterpart (ICRS J2000 deg)  \\
catalog\_name\_aper & imaging source for measuring \OII\ resolved apertures \\
filter\_name\_aper & filter of imaging used for measuring \OII\ resolved apertures \\
dist\_aper & distance between aperture center and detectid position in arcseconds \\
mag\_aper & photometric magnitude in aperture in imaging source \\
mag\_aper\_err & photometric magnitude error in aperture in imaging source \\
major & major axis of aperture ellipse of resolved \OII\ galaxy defined by imaging \\
minor & minor axis of aperture ellipse of resolved \OII\ galaxy defined by imaging \\
theta & angle in aperture ellipse \\
 &  \\
 \hline
 \multicolumn{2}{l}{\footnotesize Note: Bad values are $-999.0$ for floating-point columns and \texttt{n/a} for string columns.} \\
\multicolumn{2}{l}{\footnotesize This table is available in its entirety in machine-readable form at \url{https://hetdex.org/data-results/} and in the online journal.} \\
\end{longtable}